\documentclass[prd,tightenlines,nofootinbib,superscriptaddress]{revtex4}%{revtex4-2}

\usepackage{pre_all}
\usepackage{pre_QH}

\begin{document}

\title{Spinor Representation of the Hamiltonian Constraint in 3D LQG with a Non-zero Cosmological Constant}

\author{{\bf Valentin Bonzom}}\email{bonzom@lipn.univ-paris13.fr}
\affiliation{LIPN, UMR CNRS 7030, Institut Galil\'ee, Universit\'e Paris 13, Sorbonne Paris Cit\'e, 99, avenue Jean-Baptiste Cl\'ement, 93430 Villetaneuse, France, EU}

\author{{\bf Ma\"it\'e Dupuis}}\email{mdupuis@perimeterinstitute.ca}
\affiliation{Perimeter Institute, 31 Caroline St North, Waterloo N2L 2Y5, Ontario, Canada}
\affiliation{Department of Applied Mathematics, University of Waterloo, Waterloo, Ontario, Canada}

\author{{\bf Qiaoyin Pan}}\email{qpan@fau.edu}
\affiliation{Department of Physics, Florida Atlantic University, 777 Glades Road, Boca Raton, FL 33431, USA}
\affiliation{Perimeter Institute, 31 Caroline St North, Waterloo N2L 2Y5, Ontario, Canada}
\affiliation{Department of Applied Mathematics, University of Waterloo, Waterloo, Ontario, Canada}

\date{\today}

\begin{abstract}
We develop in a companion paper the kinematics of three-dimensional loop quantum gravity in Euclidean signature and with a negative cosmological constant, focusing in particular on the spinorial representation which is well-known at zero cosmological constant. In this paper, we put this formalism to the test by quantizing the Hamiltonian constraint on the dual of a triangulation. The Hamiltonian constraints are obtained by projecting the flatness constraints onto spinors, as done in the flat case by the first author and Livine. Quantization then relies on $q$-deformed spinors. The quantum Hamiltonian constraint acts in the $q$-deformed spin network basis as difference equations on physical states, which are thus the Wheeler-DeWitt equations in this framework. Moreover, we study how physical states transform under Pachner moves of the canonical surface. We find that those transformations are in fact $q$-deformations of the transition amplitudes of the flat case as found by Noui and Perez. Our quantum Hamiltonian constraints, therefore, build a Turaev-Viro model at real $q$.
\end{abstract}

\maketitle

\section*{Introduction}

Three-dimensional gravity is often taken as a testing ground for new approaches to quantum gravity as it is much better understood compared to the four-dimensional case. Here we consider 3D gravity as a topological model (meaning the triad can degenerate) through the BF formulation. A criterion for the validity of novel approaches to their quantization is that they reproduce the results obtained via previous approaches, such as its topological invariance. In the absence of a cosmological constant, non-perturbative quantization in the canonical approach, \ie the loop quantum gravity (LQG) based on the BF formulation of gravity \cite{Rovelli:2007quantum,Thiemann:2007zz}, and the covariant approach, based on spin foams and more precisely the Ponzano-Regge model \cite{Ponzano:1969semi,Baez:1997zt,Baez:1999sr,Oriti:2001qu,Livine:2010zx}, give consistent results. In particular, it has been shown by Noui and Perez in \cite{Noui:2004iy} that the scalar products of physical states defined in LQG are given by the Ponzano-Regge amplitudes.

A more recent approach has emerged that aims at quantizing the Hamiltonian constraint instead of the flatness constraint derived from the BF formulation. In the case of vanishing cosmological constant (which we will often refer to as the flat case), the Hamiltonian constraint translates in the spin network basis to \emph{difference equations} on the coefficients of the physical states. These difference equations should really be seen as Wheeler-DeWitt equations for 3D LQG. In particular, they are solved by spin network evaluations, as expected from the Ponzano-Regge model \cite{Bonzom:2011hm,Bonzom:2011nv}. 

When the cosmological constant is non-zero, the connection between LQG and spin foams is less clear. On the spin foam side, the Turaev-Viro model \cite{Turaev:1992hq} is known to provide the partition function of 3D gravity in Euclidean signature with a positive cosmological constant \cite{Archer:1991rz}. It is a sum of states in $\SU_q(2)$ representation, with $q$ a root of unity encoding the cosmological constant. It is thus a $q$-deformation of the Ponzano-Regge model, further providing a regularization through a natural cut-off on representations when replacing $\SU(2)$ with $\SU_q(2)$. The large spin limit of the $q$-$6j$ symbol matches Regge calculus for curved tetrahedra \cite{Taylor:2006j}. The Turaev-Viro model thus provides an example of the interplay between the cosmological constant, curved geometries and the quantum group deformation of Lie groups.

On the LQG side, the Hamiltonian takes a more complicated form when the cosmological constant is non-zero, so much so that even how to discretize it has been unclear and it seems to evade traditional LQG methods. It has nevertheless been conjectured for a long time that the quantum theory ought to be described by quantum groups, as expected from the spin foam model \cite{Smolin:1994qb,Smolin:1995vq,Borissov:1996ge,Major:1995yz,Smolin:2002sz}. 
One (indirect) way to relate 3D LQG with a non-vanishing cosmological constant to the Turaev-Viro model is to take the Chern-Simons formulation of 3D gravity and consider the Witten-Reshetikhin-Turaev path integral ${\mathcal Z}_{WRT}(\cM)$ on a 3-manifold $\cM$ with the Chern-Simons action with opposite levels, say $k$ and $-k$. It has been well-known that the Turaev-Viro state sum matches such path integral as ${\mathcal Z}_{TV}(\cM)=|{\mathcal Z}_{WRT}(\cM)|^2$ \cite{Walker:1991wi,Turaev:2010tw}. 

A more direct approach for bridging the two quantum gravity approaches would be to work on the BF formulation with a cosmological constant term itself. Canonical analysis for the BF action written with the standard triad and connection variables leads to a torsion equation independent of the cosmological constant so that the kinematical Hilbert space upon quantization is spanned by the $\SU(2)$ spin network as in the case with a zero cosmological constant. In this setting, one can expect, through the connection of cosmological constant and quantum group deformation, that the quantum group structure would only appear at the level of the physical Hilbert space since only the curvature equation depends on the cosmological constant. One proposal to realize this and thereby connect LQG directly to the Turaev-Viro model was given in \cite{Noui:2011im,Noui:2011aa,Pranzetti:2014xva}. There, a new curvature constraint was defined via a new (Poisson non-commutative) connection, leading to a redefinition of the physical scalar product and recovering the Turaev-Viro amplitude. 

From the geometrical point of view, the non-deformed kinematical structures given by imposing the torsion-free, or Gauss, constraints represent discrete flat 2D geometries. Then the deformed dynamical structures are expected to describe the gluing of these flat 2D geometries to approximate the curved 3D geometries, as a deformed version of the case with a zero cosmological constant \cite{Bonzom:2011hm}. Indeed, one can approximate a curved 3D geometry by gluing flat 2D pieces and take the limit as the sizes of these pieces approach zero. It was moreover argued in \cite{Dittrich:2008pw,Bahr:2009ku,Bahr:2009mc} that the continuous symmetries survive at the discrete level when one uses curved 2D pieces instead of flat ones.

This suggests we rethink the definition of the kinematics and dynamics in the BF formulation with a non-vanishing cosmological constant. The kinematical phase space defined with the Gauss constraint can in fact be deformed so that its quantization naturally leads to a quantum group deformation. In particular, this kinematical phase space describes 2D \emph{curved} geometries \cite{Bonzom:2014wva}. Then the physical phase space, defined by imposing the flatness constraint, describes the gluing of these 2D curved geometries into 3D curved geometries. 
The gap between the discrete, classical theory and the continuous action was further filled recently in \cite{Dupuis:2020ndx}. 

This programme was carried out in \cite{Bonzom:2014wva, Bonzom:2014bua} to a large degree. There, the phase space is defined in terms of deformed fluxes and holonomies and the Poisson structure is based on the Heisenberg double of $\SU(2)$. In \cite{Bonzom:2014bua} we have investigated the quantization, using the same techniques as in \cite{Bonzom:2011hm}, \ie by building a Hamiltonian constraint out of the flatness constraints. It can be classically interpreted as generating displacements of the vertices of the triangulation \cite{BonzomDittrich}. At the quantum level, the Hamiltonian constraints give rise to \emph{difference equations}, which can therefore be considered as the Wheeler-DeWitt equations in the spin network basis. We considered in \cite{Bonzom:2014bua} the (simple) case of the boundary of the tetrahedron and showed that the solution to those difference equations is the $q$-$6j$ symbol.

Here we are interested in using the spinorial formalism for LQG instead of holonomies and fluxes, and in further extracting all building blocks for the transition amplitudes, \ie to go beyond the case of the tetrahedron from \cite{Bonzom:2014bua}. In a companion paper \cite{Bonzom:2022bpv}, we revisit all kinematical aspects of this $q$-deformed LQG model in more detail, and in the spinor representation. (This was initiated in \cite{Dupuis:2014fya}.) In particular, the quantization of the deformed spinors can be performed in terms of $q$-bosons. We then use those $q$-bosons to define the invariant operators which are needed for the quantization of the Hamiltonian constraint in spinor variables.

In this paper, we describe the dynamics using the deformed spinors and $q$-bosons of \cite{Bonzom:2022bpv}. The Hamiltonian constraint built from the deformed spinors is a direct generalization of the non-deformed version given in \cite{Bonzom:2011nv}. At the quantum level, the Hamiltonian constraints also give rise to difference equations which are direct $q$-deformed generalizations of those of \cite{Bonzom:2011nv}. Here we go further to provide the transformations of the physical states (in the spin network basis) under Pachner moves of the canonical surface. This is equivalent to finding the building blocks for spin foams, or for the transition amplitudes, as emphasized by Noui and Perez \cite{Noui:2004iy}. In particular, we find that those building blocks are exactly those of the Turaev-Viro model in a version with $q$ real (note that this version suffers from the same finiteness issues as the Ponzano-Regge model).

This paper is organized as follows. In Section \ref{sec:phase_space}, we concisely recall the discrete classical phase space in terms of the holonomies and (deformed) fluxes introduced in \cite{Bonzom:2014wva}, as well as the Gauss constraints (used to define the kinematical phase space) and the flatness constraints (used to define the dynamical phase spaces). In Section \ref{sec:spinor_rep} we move on to the construction of the deformed spinors, following \cite{Bonzom:2022bpv}, and of the Hamiltonian constraints. The quantization is performed in Section \ref{sec:quantum_hamiltonian}, again following the prescriptions of \cite{Bonzom:2022bpv}. This is where in particular we find the difference equations encoding the Wheeler-DeWitt equations in the spin network basis. Then in Section \ref{sec:Pachner}, we study how solutions to the difference equations are related under Pachner moves, thereby providing the building blocks for the transition amplitudes {\it \`a la} Noui-Perez.

\section{Classical phase space and constraints}
\label{sec:phase_space}

We start by recalling the main ingredients of the classical phase space for 3D loop gravity with a negative cosmological constant $\Lambda$ in the Euclidean signature. More details for the mathematical setup can be found in \cite{Bonzom:2014wva,Bonzom:2022bpv}. The phase space is associated to a graph $\Gamma$ which is dual to a cellular decomposition of the canonical surface $\Sigma$. It has $V$ vertices, $E$ edges and $F$ faces (the connected components of $\Sigma\setminus\Gamma$).

Here and throughout the paper, we use $\kappa := \f{G\sqrt{-\Lambda}}{c}$, for a cosmological constant $\Lambda<0$. It is a parameter which deforms the Poisson structure with respect to the case of vanishing cosmological constant.

%%%%%%%%%%%%%
\subsection{Phase space for a single edge}
%%%%%%%%%%%%%

We first consider a single edge and associate to it a phase space: the Heisenberg double $(\cD(\SU(2)),\pi_H)$ of $\SU(2)$. It is the group $\cD(\SU(2))=\SL(2,\bC)\cong \SU(2)\bowtie \AN(2)$ with Poisson structure $\pi_H$ fully determined by a classical $r$-matrix $r\in \sl(2,\bC)\otimes \sl(2,\bC)$. The Poisson brackets can be compactly written as
\be
\{d_1,d_2\}=-\rT d_1d_2+d_1d_2 r=r d_1d_2-d_1d_2\rT\,,\quad 
\forall d\in \SL(2,\bC)\,,
\label{eq:bivector_SL2C}
\ee
where $d_1=d\otimes \id\,,d_2=\id\otimes d$. The $r$-matrix is chosen as
\be
r= \f{i\ka}{4} \sum_{i=1}^3 \sigma_i\otimes \rho_i =
\f{i\ka}{4}\mat{cccc}{
1 & 0 & 0 & 0 \\ 0 & -1 & 4 & 0 \\ 0 & 0 & -1 & 0 \\ 0 & 0 & 0 & 1}\,.
\label{eq:r_4X4}
\ee 
Here $\sigma_{1,2,3}$ are the Pauli matrices while $\rho_{i} = \sigma_i + \tfrac{1}{2}[\sigma_3,\sigma_i]$. Finally, $r_{21}$ is given by the permutation of the two vector space components of $r$ (in the above $4\times 4$ representation, $r_{21}$ is simply the matrix transpose of $r$). The equality of the last two expressions in \eqref{eq:bivector_SL2C} is guaranteed by the property that $r_s:=\f12(r+r_{21})$ is the Casimir thus $[r_s,d_1d_2]=0$.

It is important for loop gravity to split an $\SL(2,\bC)$ element via the Iwasawa decomposition into the product of an $\AN(2)$ element and an $\SU(2)$ element. One can write $d\in\SL(2,\bC)$ in exactly two ways as 
\be
d=\ell u=\ut\lt\,,\quad
\ell,\lt\in \AN(2)\,,\quad
u,\ut\in \SU(2)\,.
\label{eq:Iwasawa}
\ee
This phase space can be seen as a deformation of the holonomy-flux phase space at $\Lambda=0$ \cite{Bonzom:2014wva,Pan:2022the}. In the flat/non-deformed ($\Lambda=0$) case, the phase space of an edge is described by $\ISU(2)$, the holonomies are described by $\SU(2)$ and the fluxes are described by $\R^3$. Here, in the deformed phase space $\SL(2,\bC)$, we also let the $\SU(2)$ subgroup describes the holonomies while the (deformed) fluxes correspond to an $\AN(2)$ subgroup. That is, for each phase space variable $d$, we perform the Iwasawa decomposition \eqref{eq:Iwasawa} then $u$ and $\tilde{u}$ are holonomies, while $\ell$ and $\tilde{\ell}$ are fluxes. 

We call the constraint,
\begin{equation} \label{RibbonConstraintOneEdge}
\cC=\mathbbm{1} \in\SL(2,\bC) \quad \text{for} \quad \cC:=\ell u\lt^{-1}\ut^{-1}
\end{equation}
the \emph{ribbon constraint}, associated to every edge of $\Gamma$. It has six real components and forms a set of second-class constraints with respect to the Poisson brackets \eqref{eq:bivector_SL2C} (meaning that the brackets between the components do not close).

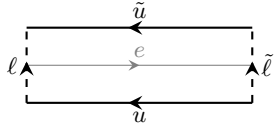
\begin{figure}[h!]
	\begin{tikzpicture}[scale=1]
		
	\coordinate (O) at (0,0);
	\coordinate (A) at (3,0);
	\coordinate (B) at (3,1);
	\coordinate (C) at (0,1);
	
	\coordinate (l) at (0,0.5);
	\coordinate (r) at (3,0.5);
		
	\draw[gray,decoration={markings,mark=at position 0.5 with {\arrow[scale=1.3,>=stealth]{>}}},postaction={decorate}] (l)-- node[midway,above]{$e$} (r);

	\draw[thick,dashed,decoration={markings,mark=at position 0.55 with {\arrow[scale=1.3,>=stealth]{>}}},postaction={decorate}] (O) -- node[midway, left]{$\ell$}(C); 
	\draw[thick,dashed,decoration={markings,mark=at position 0.55 with {\arrow[scale=1.3,>=stealth]{>}}},postaction={decorate}] (A) -- node[midway, right]{$\lt$}(B);
	\draw[thick,decoration={markings,mark=at position 0.55 with {\arrow[scale=1.3,>=stealth]{>}}},postaction={decorate}] (B) -- node[midway, above]{$\ut$}(C);
	\draw[thick,decoration={markings,mark=at position 0.55 with {\arrow[scale=1.3,>=stealth]{>}}},postaction={decorate}] (A) -- node[midway, below]{$u$}(O);
	
	\end{tikzpicture}
\caption{A ribbon edge $R(e)$. The variables $\ell,u,\lt,\ut$ are assigned to the four sides of the ribbon edge and they are subject to the ribbon constraint $\ell u\lt^{-1}\ut^{-1}$ represented as the trivialization of the loop around $R(e)$. The positions of these variables are fixed such that the directions of $u$ and $\ut$ are opposite to that of the edge $e$ ({\it in gray}).}
\label{fig:box}
\end{figure}

The ribbon constraint has a natural graphical interpretation. Since the edges of $\Gamma$ are embedded in a surface, there is a natural clockwise walk around each of them. It goes $i)$ along the edge on one side, $ii)$ crosses it at its end, $iii)$ goes back along the edge on its other side and $iv)$ finally crosses it again to close the walk. Equivalently, one thickens the edge by taking a tubular neighbourhood in $\Sigma$, as in Figure \ref{fig:box}. The boundary has four pieces which naturally correspond to the four parts of the walk above.  

If $e$ is an edge in $\Gamma$, then we denote $R(e)$ its thickening, called \emph{ribbon edge}. The boundary pieces parallel to $e$ will be called the \emph{long edges} of $R(e)$ and the boundary pieces which cross $e$ at its ends will be called the \emph{short edges} of $R(e)$. We can orient the long and short edges clockwise around $R(e)$. The matrices $u, \ell, \tu^{-1}, \tell^{-1}$ are then assigned in this order and as pictured in Figure \ref{fig:box}. In particular, $u$ and $\tu$ are assigned to the long edges of $R(e)$, while $\ell$ and $\tell$ are assigned to its short edges. Equivalently, we can think of $u$ and $\tu$ as associated to $e$ itself and $\ell$ and $\tell$ to each \emph{half-edge}, \ie a pair of an edge and an incident vertex.

To fix the position of the variables around the ribbon, one can use the orientation on $\Gamma$ and decide for instance that $u$ is oriented opposite to $e$. This is the convention we will use. The ribbon constraint $\cC=\ell u\lt^{-1}\ut^{-1}\cong \id$ is then a flatness constraint around $R(e)$.

The Poisson brackets \eqref{eq:bivector_SL2C} can be equivalently written as brackets between holonomies and fluxes
\be\ba{llll}
\{\ell_1,\ell_2\}=-[\rT,\ell_1\ell_2]\,,&\{\ell_1,u_2\}=-\ell_1\rT u_2\,,& \{u_1,\ell_2\}=\ell_2r u_1\,,& \{u_1,u_2\}=-[r,u_1u_2]\,, \\
\{\lt_1,\lt_2\}=[\rT,\lt_1\lt_2]\,,& \{\lt_1,\ut_2\}=-\ut_2 \rT \lt_1\,,& \{\ut_1,\lt_2\}=\ut_1 r \lt_2\,,& \{\ut_1,\ut_2\}=[r,\ut_1\ut_2]\,.
\ea
\label{eq:poisson_Iwasawa}
\ee
All other Poisson brackets \eg $\{\ell_1,\ut_2\}$ can also be obtained by combining \eqref{eq:poisson_Iwasawa} and \eqref{RibbonConstraintOneEdge}.

The $\AN(2)$ elements $\ell$ and $\lt$ can be parametrized as $2\times 2$ lower triangular matrices
\be
\ell = \mat{cc}{\lambda & 0 \\ z & \lambda^{-1} }\,,\quad
\lt=\mat{cc}{\tlambda & 0 \\ \tz & \tlambda^{-1}}\,,\quad
\lambda,\tlambda \in \R^+\,,\quad z,\tz\in \bC\,.
\ee 
By also writing the holonomies $u$ and $\ut$ in the fundamental representation, one can write down all the Poisson brackets between the matrix elements of $u, \ut$ and $\lambda, \tlambda, z, \tz, \bar{z}, \bar{\tz}$ (See \cite{Bonzom:2014wva,Bonzom:2022bpv} for details). 

%%%%%%%%%%%%%
\subsection{Ribbon graph phase space}
%%%%%%%%%%%%%

We extend the phase space defined above for a single edge to the whole graph $\Gamma$, by taking the product of $\SL(2,\bC)$ over the edges. Graphically, one thickens every edge of $\Gamma$ as before. However, this is not enough since there will be interactions between the group elements of different edges which meet at vertices of $\Gamma$. An advantageous graphical representation is to thicken $\Gamma$ itself. Each vertex $v$ of valency $d$ is fattened to a \emph{ribbon vertex} $R(v)$, \ie a $d$-gon whose boundary edges correspond to the edges incident to $v$, and are glued to the short edges of the ribbon edges (the boundary vertices of $R(v)$ correspond to the ``corners'' at $v$ between adjacent edges).
In other words, a ribbon vertex is a face whose boundary sides are dressed with fluxes $\ell$s and $\tell$s.  

Overall, the holonomies $u_e, \tu_e$ are labelled by the edges of $\Gamma$ and assigned to the long edges of the ribbon edges. There are two fluxes associated to every edge of $\Gamma$, denoted $\ell_{e}\in\AN(2)$ if $e$ is outgoing at $v$, $\tell_{e}\in \AN(2)$ if $e$ is incoming at $v$. Since each is in fact assigned to a half-edge $(e,v)$ (equivalently a short edge of $R(e)$), we will use the generic notation $\ell_{ev}$ for either one of them.

Since there are also two holonomies along $e$, it is tempting to distinguish them in terms of half-edges. This is possible using orientations. We denote $u_{ev}$ the $\SU(2)$ matrix which points towards $\ell_{ev}$ (so that if $\ell_{ev}=\tell_e^{-1}$ then $u_{ev} = \tu_e^{-1}$ and else $\ell_{ev} = \ell_e$ and $u_{ev}=u_e$).

%%%%%%%%%%%%%
\subsection{Gauss and flatness constraints}
%%%%%%%%%%%%%

The phase space for $\Gamma$ described above is constrained by the ribbon constraints $\mathcal{C}_e=\mathbbm{1}$ on every edge. Gravity further imposes two additional sets of constraints, namely the Gauss and flatness constraints. Gauss constraints are associated to vertices and impose that the ordered products of the fluxes along the short edges of every $R(v)$ are trivial. Flatness constraints are associated to the faces and impose that the ordered products of the holonomies along the long edges which border every face is trivial. Those two sets of constraints are first class.

To write the Gauss constraints explicitly, choose (randomly) one edge of reference at each vertex of $\Gamma$ and call it $e_1$, then order the edges from 1 to $n$, \ie $e_1, \dotsc, e_n$ by going counter-clockwise around $v$. Notice that all the $\AN(2)$ matrices $\ell_{e_i v}$ on the boundary of the ribbon vertex $R(v)$ are oriented counter-clockwise, as shown in fig.\ref{fig:ribbon_node}, for any choice of orientations of the edges incident to $v$. The Gauss constraint is then simply the flatness around $R(v)$. It reads
\be
\ell_{e_n v} \dotsm \ell_{e_1 v} = \mathbbm{1}\,.
\label{eq:Gauss}
\ee
We repeat this construction on faces instead of vertices: choose a random edge of reference around each face $f$ and denote it $e_1$, then $e_2, \dotsc, e_d$ are the edges encountered counter-clockwise around $f$. For all possible orientations of the edges $e_1, \dotsc, e_d$ on the boundary of $f$ are, the $\SU(2)$ matrices $u_{e_i v}$ are all counter-clockwise. The flatness constraint on $f$ reads
\be
u_{e_d v_1} \dotsm u_{e_2 v_3} u_{e_1 v_2} = \mathbbm{1}\,,
\label{eq:flatness}
\ee
as pictured in Figure \ref{fig:ribbon_face}.
\begin{figure}[h!]
\centering
\begin{tikzpicture}
\coordinate (o) at (0,0);
\def\x{60};
\coordinate (a1) at ([shift=(30:1cm)]o);
\coordinate (a2) at ([shift=({30+\x}:1cm)]o);
\coordinate (a3) at ([shift=({30+2*\x}:1cm)]o);
\coordinate (a4) at ([shift=({30+3*\x}:1cm)]o);
\coordinate (a5) at ([shift=({30+4*\x}:1cm)]o);
\coordinate (a6) at ([shift=({30+5*\x}:1cm)]o); 

\draw (o) -- node[at end, above right]{$e_1$} (a1);
\draw (o) -- node[at end, above]{$e_2$} (a2);
\draw (o) -- node[at end, above left]{$e_3$} (a3);
\draw (o) -- node[at end, below left]{$e_4$} (a4);
\draw (o) -- node[at end, below ]{$e_5$} (a5);
\draw (o) -- node[at end, below right]{$e_6$} (a6);

\draw (o) node[left]{$v$};

\draw[<->,thick] (2,0) -- (3,0);

\coordinate (O) at (6,0);
\def\y{0.8};
\coordinate (A1) at ([shift=(0:\y cm)]O);
\coordinate (A2) at ([shift=({\x}:\y cm)]O);
\coordinate (A3) at ([shift=({2*\x}:\y cm)]O);
\coordinate (A4) at ([shift=({3*\x}:\y cm)]O);
\coordinate (A5) at ([shift=({4*\x}:\y cm)]O);
\coordinate (A6) at ([shift=({5*\x}:\y cm)]O);

\draw[dashed,decoration={markings,mark=at position 0.55 with {\arrow[scale=1.3,>=stealth]{>}}},postaction={decorate}] (A1) -- node[pos=0.7,right]{$\ell_{e_1v}$} (A2);
\draw[dashed,decoration={markings,mark=at position 0.55 with {\arrow[scale=1.3,>=stealth]{>}}},postaction={decorate}] (A2) -- node[midway,above]{$\ell_{e_2v}$} (A3);
\draw[dashed,decoration={markings,mark=at position 0.55 with {\arrow[scale=1.3,>=stealth]{>}}},postaction={decorate}] (A3) -- node[pos=0.3,left]{$\ell_{e_3v}$} (A4);
\draw[dashed,decoration={markings,mark=at position 0.55 with {\arrow[scale=1.3,>=stealth]{>}}},postaction={decorate}] (A4) -- node[pos=0.7,left]{$\ell_{e_4v}$} (A5);
\draw[dashed,decoration={markings,mark=at position 0.55 with {\arrow[scale=1.3,>=stealth]{>}}},postaction={decorate}] (A5) -- node[midway,below]{$\ell_{e_5v}$} (A6);
\draw[dashed,decoration={markings,mark=at position 0.55 with {\arrow[scale=1.3,>=stealth]{>}}},postaction={decorate}] (A6) -- node[pos=0.3,right]{$\ell_{e_6v}$} (A1);

\def\z{30};
\coordinate (B1) at ([shift=({\z}:\y cm)]A1);
\coordinate (C1) at ([shift=({-\z}:\y cm)]A1);
\coordinate (B2) at ([shift=({\z}:\y cm)]A2);
\coordinate (C2) at ([shift=({\z+\x}:\y cm)]A2);
\coordinate (B3) at ([shift=({\z+\x}:\y cm)]A3);
\coordinate (C3) at ([shift=({\z+2*\x}:\y cm)]A3);
\coordinate (B4) at ([shift=({\z+2*\x}:\y cm)]A4);
\coordinate (C4) at ([shift=({\z+3*\x}:\y cm)]A4);
\coordinate (B5) at ([shift=({\z+3*\x}:\y cm)]A5);
\coordinate (C5) at ([shift=({\z+4*\x}:\y cm)]A5);
\coordinate (B6) at ([shift=({\z+4*\x}:\y cm)]A6);
\coordinate (C6) at ([shift=({\z+5*\x}:\y cm)]A6);

\draw (A1) -- (B1);
\draw (A1) -- (C1);
\draw (A2) -- (B2);
\draw (A2) -- (C2);
\draw (A3) -- (B3);
\draw (A3) -- (C3);
\draw (A4) -- (B4);
\draw (A4) -- (C4);
\draw (A5) -- (B5);
\draw (A5) -- (C5);
\draw (A6) -- (B6);
\draw (A6) -- (C6);

\end{tikzpicture}
\caption{A vertex $v$ on which edges meet becomes a ribbon vertex $R(v)$ incident to ribbon edges. The ribbon vertex is here depicted with a dashed boundary. Due to the clockwise orientation of the short edges of every ribbon edge, the matrix $\ell_{ev}$ around $R(v)$ are all oriented counter-clockwise.}
\label{fig:ribbon_node}
\end{figure}
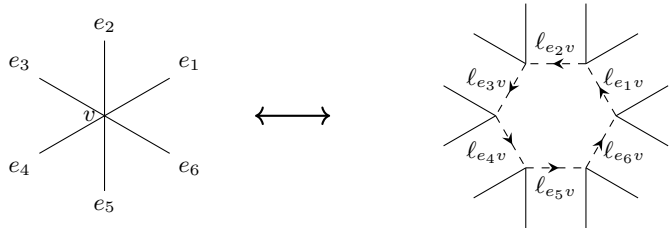
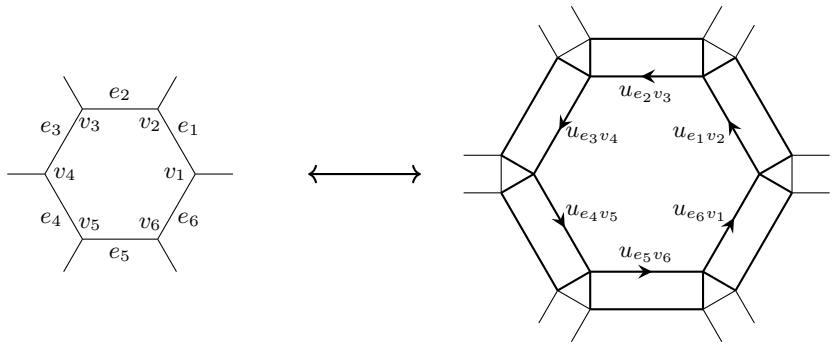
\begin{figure}[h!]
\centering
\begin{tikzpicture}
\coordinate (O) at (0,0);
\def\x{60};
\def\y{1};
\def\z{1.5};
\coordinate (A1) at ([shift=(0:\y cm)]O);
\coordinate (A2) at ([shift=({\x}:\y cm)]O);
\coordinate (A3) at ([shift=({2*\x}:\y cm)]O);
\coordinate (A4) at ([shift=({3*\x}:\y cm)]O);
\coordinate (A5) at ([shift=({4*\x}:\y cm)]O);
\coordinate (A6) at ([shift=({5*\x}:\y cm)]O);

\coordinate (B1) at ([shift=(0:\z cm)]O);
\coordinate (B2) at ([shift=({\x}:\z cm)]O);
\coordinate (B3) at ([shift=({2*\x}:\z cm)]O);
\coordinate (B4) at ([shift=({3*\x}:\z cm)]O);
\coordinate (B5) at ([shift=({4*\x}:\z cm)]O);
\coordinate (B6) at ([shift=({5*\x}:\z cm)]O);

\draw (A1) -- node[pos=0.7,right]{$e_1$} (A2);
\draw (A2) -- node[midway,above]{$e_2$} node[pos=0.1,below]{$v_2$}  node[pos=0.9,below]{$v_3$} 
(A3);
\draw (A3) -- node[pos=0.3,left]{$e_3$} (A4);
\draw (A4) -- node[pos=0.7,left]{$e_4$} (A5);
\draw (A5) -- node[midway,below]{$e_5$} node[pos=0.1,above]{$v_5$}  node[pos=0.9,above]{$v_6$} (A6);
\draw (A6) -- node[pos=0.3,right]{$e_6$} (A1);
\draw (A1) node[left]{$v_1$};
\draw (A4) node[right]{$v_4$};

\draw (A1) -- (B1);
\draw (A2) -- (B2);
\draw (A3) -- (B3);
\draw (A4) -- (B4);
\draw (A5) -- (B5);
\draw (A6) -- (B6);

\draw[<->,thick] (2.5,0) -- (4,0);

\coordinate (O) at (7,0);
\def\y{1.5};
\coordinate (A1) at ([shift=(0:\y cm)]O);
\coordinate (A2) at ([shift=({\x}:\y cm)]O);
\coordinate (A3) at ([shift=({2*\x}:\y cm)]O);
\coordinate (A4) at ([shift=({3*\x}:\y cm)]O);
\coordinate (A5) at ([shift=({4*\x}:\y cm)]O);
\coordinate (A6) at ([shift=({5*\x}:\y cm)]O);

\draw[thick,decoration={markings,mark=at position 0.55 with {\arrow[scale=1.3,>=stealth]{>}}},postaction={decorate}] (A1) -- node[pos=0.4,left]{$u_{e_1v_2}$} (A2);
\draw[thick,decoration={markings,mark=at position 0.55 with {\arrow[scale=1.3,>=stealth]{>}}},postaction={decorate}] (A2) -- node[midway,below]{$u_{e_2v_3}$} (A3);
\draw[thick,decoration={markings,mark=at position 0.55 with {\arrow[scale=1.3,>=stealth]{>}}},postaction={decorate}] (A3) -- node[pos=0.6,right]{$u_{e_3v_4}$} (A4);
\draw[thick,decoration={markings,mark=at position 0.55 with {\arrow[scale=1.3,>=stealth]{>}}},postaction={decorate}] (A4) -- node[pos=0.4,right]{$u_{e_4v_5}$} (A5);
\draw[thick,decoration={markings,mark=at position 0.55 with {\arrow[scale=1.3,>=stealth]{>}}},postaction={decorate}] (A5) -- node[midway,above]{$u_{e_5v_6}$} (A6);
\draw[thick,decoration={markings,mark=at position 0.55 with {\arrow[scale=1.3,>=stealth]{>}}},postaction={decorate}] (A6) -- node[pos=0.6,left]{$u_{e_6v_1}$} (A1);

\def\z{30};
\def\y{0.5};
\coordinate (B1) at ([shift=({\z}:\y cm)]A1);
\coordinate (C1) at ([shift=({-\z}:\y cm)]A1);
\coordinate (B2) at ([shift=({\z+\x}:\y cm)]A2);
\coordinate (C2) at ([shift=({\z}:\y cm)]A2);
\coordinate (B3) at ([shift=({\z+2*\x}:\y cm)]A3);
\coordinate (C3) at ([shift=({\z+\x}:\y cm)]A3);
\coordinate (B4) at ([shift=({\z+3*\x}:\y cm)]A4);
\coordinate (C4) at ([shift=({\z+2*\x}:\y cm)]A4);
\coordinate (B5) at ([shift=({\z+4*\x}:\y cm)]A5);
\coordinate (C5) at ([shift=({\z+3*\x}:\y cm)]A5);
\coordinate (B6) at ([shift=({\z+5*\x}:\y cm)]A6);
\coordinate (C6) at ([shift=({\z+4*\x}:\y cm)]A6);

\draw[thick] (A1) -- (B1);
\draw[thick] (A1) -- (C1);
\draw[thick] (A2) -- (B2);
\draw[thick] (A2) -- (C2);
\draw[thick] (A3) -- (B3);
\draw[thick] (A3) -- (C3);
\draw[thick] (A4) -- (B4);
\draw[thick] (A4) -- (C4);
\draw[thick] (A5) -- (B5);
\draw[thick] (A5) -- (C5);
\draw[thick] (A6) -- (B6);
\draw[thick] (A6) -- (C6);

\draw[thick] (B1) -- (C2);
\draw[thick] (B2) -- (C3);
\draw[thick] (B3) -- (C4);
\draw[thick] (B4) -- (C5);
\draw[thick] (B5) -- (C6);
\draw[thick] (B6) -- (C1);

\draw (B1) -- (C1);
\draw (B2) -- (C2);
\draw (B3) -- (C3);
\draw (B4) -- (C4);
\draw (B5) -- (C5);
\draw (B6) -- (C6);

\coordinate (D1) at ([shift=(0:\y cm)]B1);
\coordinate (D2) at ([shift=(\x:\y cm)]B2);
\coordinate (D3) at ([shift=({2*\x}:\y cm)]B3);
\coordinate (D4) at ([shift=({3*\x}:\y cm)]B4);
\coordinate (D5) at ([shift=({4*\x}:\y cm)]B5);
\coordinate (D6) at ([shift=({5*\x}:\y cm)]B6);

\coordinate (E1) at ([shift=(0:\y cm)]C1);
\coordinate (E2) at ([shift=(\x:\y cm)]C2);
\coordinate (E3) at ([shift=({2*\x}:\y cm)]C3);
\coordinate (E4) at ([shift=({3*\x}:\y cm)]C4);
\coordinate (E5) at ([shift=({4*\x}:\y cm)]C5);
\coordinate (E6) at ([shift=({5*\x}:\y cm)]C6);

\draw (B1) -- (D1); \draw (C1) -- (E1);
\draw (B2) -- (D2); \draw (C2) -- (E2);
\draw (B3) -- (D3); \draw (C3) -- (E3);
\draw (B4) -- (D4); \draw (C4) -- (E4);
\draw (B5) -- (D5); \draw (C5) -- (E5);
\draw (B6) -- (D6); \draw (C6) -- (E6);

\end{tikzpicture}
\caption{When considering ribbon edges and ribbon vertices, the face on the left bounded by edges $e_1,\cdots,e_6$ becomes bounded by long edges on the right. All matrices $u_{e_iv_{i+1}}\,,i=1,\cdots,6$ are oriented counter-clockwise around the face.}
\label{fig:ribbon_face}
\end{figure}

Gauss constraints generate local $\SU(2)$ transformations through the Poisson brackets \cite{Bonzom:2014wva,Bonzom:2022bpv}. As usual in symplectic geometry, first-class constraints are not only imposed but one also needs to quotient out the phase space by the orbits they generate. This is called the symplectic quotient. Here, one obtains $\cP_{\kin}=\SL(2,\bC)^E//\SU(2)^V$ which is called the \emph{kinematical phase space}, where $E$ and $V$ denotes the number of edges and vertices in $\Gamma$. 

It was shown in \cite{Bonzom:2014wva} that the Gauss constraint for a trivalent vertex geometrically represents the hyperbolic cosine law, implying that the kinematical phase space describes hyperbolic geometry (at least hyperbolic triangles in that case). 

On the other hand, flatness constraints generate (deformed) translations \cite{Bonzom:2014wva}. The \emph{physical phase space} is then obtained via the symplectic quotient of the kinematical phase space by the flatness constraints, $\cP_{\phys}=\cP_{\kin}//\AN(2)^F$ with $F$ the number of the faces in $\Gamma$.

In the $\Lambda\to 0$ limit, one recovers the Poincar\'e phase space structure of first-order 3D gravity. In particular, the flatness constraints generate an $\R^3$ action, \ie translations. Geometrically, those translations simply move the vertices of the triangulation (dual to the faces of $\Gamma$) around (three directions for the three components of the constraints). The flatness constraints also enforce the dihedral angles to be functions of the angles within triangles as in flat, Euclidean geometry (recall that dihedral angles measure the extrinsic curvature at the discrete level) \cite{Bonzom:2011hm}. This geometric picture arises when the constraints are written on the basis determined by the fluxes themselves. It is also possible to describe them on a spinor basis \cite{Bonzom:2011nv}, which is what we will focus on in this paper.

%%%%%%%%%%%%%%%%%%%%%%%%
\section{Spinorial representation}
\label{sec:spinor_rep}

In this section, we rewrite the $q$-deformed loop gravity phase space structure described above in the spinor representation. In particular, we define the deformed spinors which can be naturally associated to the ribbon graph and reproduce the $\SU(2)$ holonomies and the $\AN(2)$ fluxes. We also define the scalar products of these deformed spinors, living at the corners of the ribbon graph, which are $\SU(2)$-invariant quantities hence live in the kinematical phase space. These scalar products are especially useful in constructing the Hamiltonian.  

\subsection{Deformed spinors}

Here we describe the $\kappa$-deformed spinors which can be used to describe the phase space and the constraints in place of the variables $u_{ev}, \ell_{ev}$. We will only give the main ingredients needed to construct the Hamiltonian constraint. The fully detailed construction appears in \cite{Bonzom:2022bpv}.

The building blocks are two {\it independent} pairs of $\ka$-deformed spinor variables $(\zeta^\ka_0,\zeta^\ka_1), (\tzeta^\ka_0,\tzeta^\ka_1)$ and their complex conjugates $(\bzeta^\ka_0,\bzeta^\ka_1),(\btzeta^\ka_0,\btzeta^\ka_1)$. The norms of these $\ka$-deformed spinor variables are $\bar{\zeta}_A^\kappa \zeta_A^\kappa = \f{2}{\kappa}\sinh(\f{\kappa N_A}{2})\,,\bar{\tzeta}_A^\kappa \tzeta_A^\kappa = \f{2}{\kappa}\sinh(\f{\kappa \Nt_A}{2})\,,A=0,1$ where $N_A$ and $\Nt_A$ are real functions of the $\ka$-deformed spinor variables 
\footnotemark{}. They satisfy the Poisson brackets
\be\ba{llll}
\{\zeta_A^\kappa, \bar{\zeta}^\kappa_B\}= -i \,\delta_{AB}\, %\red{\cancel{\kappa}}
 \cosh(\f{\kappa N_A}{2}),
&\quad
\{N_{A}, \zeta_B^\kappa\}=i \,\delta_{AB} \,\zeta^\kappa_A,
&\quad
\{ N_{A},\bar{\zeta}_B^\kappa\}=-i \,\delta_{AB}\, \bar{\zeta}^\kappa_A\,,
&
\\[0.3cm]
\{\tzeta_A^\kappa, \bar{\tzeta}^\kappa_B\}= -i \,\delta_{AB}\, 
 \cosh(\f{\kappa \Nt_A}{2}),
&\quad
\{\Nt_{A}, \tzeta_B^\kappa\}=i \,\delta_{AB} \,\tzeta^\kappa_A,
&\quad
\{ \Nt_{A},\bar{\tzeta}_B^\kappa\}=-i \,\delta_{AB}\, \bar{\tzeta}^\kappa_A\,,
&\quad
A,B=0,1\,,
\ea
\label{eq:Poisson_spinors}
\ee
and all other Poisson brackets vanish.
\footnotetext{
$N_A$ $(A=0,1)$ is the norm of the $\ka$-spinors at $\ka\rightarrow 0$ thus the norm of the standard spinor variables and likewise for $\Nt_A$. 
}

Let $\epsilon = \left(\begin{smallmatrix} 0&-1\\1&0\end{smallmatrix}\right)$. These $\ka$-deformed spinor variables can be used to define \emph{two} types of deformed spinors:
\begin{itemize}
\item The $\SU(2)$-covariant spinors, transforming under $\SU(2)$ gauge transformations in a covariant way. We denote them as $|t\ra$ and $|\tt\ra$, and their duals as $|t]$ and $|\tt]$,
\be\ba{ll}
|t\ra = \mat{c}{t_{-}\\ t_{+}} =\mat{c}{e^{\f{\ka N_1}{4}}\zeta^\ka_0 \\ e^{-\f{\ka N_0}{4}}\zeta^\ka_1}\,,
\quad &\qquad 
|t]=\epsilon |\overline{t}\ra = \mat{c}{-\tb_{+}\\ \tb_{-}}
=\mat{c}{-e^{-\f{\ka N_0}{4}}\bzeta^\ka_1 \\ e^{\f{\ka N_1}{4}}\bzeta^\ka_0}\,, \\
|\tt\ra=
\mat{c}{\tt_-\\ \tt_+} =
\mat{c}{e^{\f{\ka \Nt_1}{4}}\tzeta^\ka_0 \\ e^{-\f{\ka \Nt_0}{4}}\tzeta^\ka_1}
\,,\quad & \qquad
|\tt]=\epsilon |\btt\ra
=\mat{c}{-\btt_+\\\btt_-}
=\mat{c}{-e^{-\f{\ka \Nt_0}{4}}\btzeta^\ka_1 \\ e^{\f{\ka \Nt_1}{4}}\btzeta^\ka_0} \,.
\ea
\label{eq:def_deformed_spinors}
\ee
\item The braided-covariant spinors, transforming in a braided-covariant way \cite{Bonzom:2022bpv}. We denote them as $|\tau\ra$ and $|\ttau\ra$ and their duals as $|\tau]$ and $|\ttau]$,
\be\ba{ll}
|\tau\ra= \mat{c}{\tau_-\\\tau_+} 
= \mat{c}{e^{-\f{\ka N_1}{4}}\zeta_0^\ka \\e^{\f{\ka N_0}{4}}\zeta^\ka_1}\,,\quad &\qquad 
|\tau]
=\epsilon |\btau \ra = \mat{c}{-\btau_{+}\\  \btau_{-}}
=\mat{c}{-e^{\f{\ka N_0}{4}}\bzeta^\ka_1 \\ e^{-\f{\ka N_1}{4}}\bzeta_0^\ka}
\,,\\
|\ttau\ra=
\mat{c}{\ttau_- \\ \ttau_+}=
\mat{c}{e^{-\f{\ka \Nt_1}{4}}\tzeta^\ka_0 \\ e^{\f{\ka \Nt_0}{4}}\tzeta^\ka_1}
\,,\quad & \qquad
|\ttau]= \epsilon |\bttau\ra 
=\mat{c}{-\bttau_+ \\\bttau_-}
=\mat{c}{ -e^{\f{\ka \Nt_0}{4}}\btzeta^\ka_1 \\ e^{-\f{\ka \Nt_1}{4}}\btzeta^\ka_0 }\,.
\ea
\label{eq:def_deformed_spinors}
\ee
\end{itemize}
The norms are $\langle t|t\rangle = \langle \tau| \tau\rangle = \tfrac{2}{\kappa} \sinh \tfrac{\kappa}{2} (N_0+N_1)$, and similarly $\langle \tt |\tt\rangle = \langle \ttau| \ttau\rangle = \tfrac{2}{\kappa} \sinh \tfrac{\kappa}{2} (\Nt_0+\Nt_1)$ (the norm of a dual is the same since $\epsilon^\dagger = \epsilon^{-1}$). They match if the so-called norm matching condition holds, which is just $N_0+N_1\cong \Nt_0+\Nt_1$. 

\begin{figure}[h!]
	\centering
\begin{tikzpicture}[scale=1.3]
	\coordinate (O) at (0,0);
	\coordinate (A) at (3,0);
	\coordinate (B) at (3,1);
	\coordinate (C) at (0,1);
	
	\draw[thick,dashed,decoration={markings,mark=at position 0.55 with {\arrow[scale=1.3,>=stealth]{>}}},postaction={decorate}] (O) -- node[midway, left]{$\ell$}(C); 
	\draw[thick,dashed,decoration={markings,mark=at position 0.55 with {\arrow[scale=1.3,>=stealth]{>}}},postaction={decorate}] (A) -- node[midway, right]{$\lt$}(B);
	\draw[thick,decoration={markings,mark=at position 0.55 with {\arrow[scale=1.3,>=stealth]{>}}},postaction={decorate}] (B) -- node[midway, above]{$\ut$}(C);
	\draw[thick,decoration={markings,mark=at position 0.55 with {\arrow[scale=1.3,>=stealth]{>}}},postaction={decorate}] (A) -- node[midway, below]{$u$}(O);
	
	\draw (O) node{$\bullet$} node[below left]{$|\tau\ra$};
	\draw (A) node{$\bullet$} node[below right]{$|\tt]$};
	\draw (B) node{$\bullet$} node[above right]{$|\ttau]$};
	\draw (C) node{$\bullet$} node[above left]{$|t\ra$};
		
\end{tikzpicture}
	\caption{The ribbon edge with the holonomies on its long edges, fluxes on its short edges and spinors on its corners.}
	\label{fig:ribbon}
\end{figure}
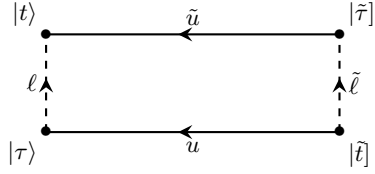

Holonomies and fluxes can be reconstructed as follows,
\be\ba{ll}
\ell=\mat{cc}{\exp(\f{\ka}{4}(N_1-N_0)) & 0 \\
-\ka\bzeta_0^\ka \zeta_1^\ka & \exp(\f{\ka}{4}(N_0-N_1))}\,,
& \qquad \lt=\mat{cc}{\exp(\f{\ka}{4}(\Nt_0-\Nt_1)) & 0 \\
\ka \btzeta^\ka_0 \tzeta^\ka_1 & \exp(\f{\ka}{4}(\Nt_1-\Nt_0))}\,,
\\[0.6cm]
  u
=\dfrac{|\tau \ra [ \tt|-|\tau]\la \tt|}{\sqrt{\la \tau|\tau \ra \la \tt|\tt\ra}}\,,
&\qquad \ut=
\dfrac{|t\ra [\ttau|-|t] \la \ttau|}{\sqrt{\la t|t\ra\la\ttau|\ttau\ra}}\,,
\ea
\label{eq:flux_holonomy_from_spinors}
\ee
with $N_0+N_1=\Nt_0+\Nt_1$. It is straightforward to check that the deformed spinors are related to one another by parallel transport via fluxes and holonomies. 
\be
|\tau\ra=e^{-\f{\kappa (N_0+N_1)}{4}}\ell^{-1}|t\ra\,,\quad
|\ttau]=e^{\f{\ka(\Nt_0+\Nt_1)}{4}}\lt |\tt]\,,\quad
u|\tt]= |\tau\ra\,,\quad
\ut|\ttau]=|t\ra\,.
\label{eq:paral_spinors}
\ee
Those relations have a natural graphical interpretation: the spinors can be assigned to the \emph{corners} of the ribbon edge. For instance, $|\tt]$ is at the source end of the long edge carrying $u$ and $|\tau\rangle$ is at its target end. This is depicted in Figure \ref{fig:ribbon} (we do not include the factors $e^{-\f{\kappa (N_0+N_1)}{4}}$ and $e^{\f{\ka(\Nt_0+\Nt_1)}{4}}$ in the graphical representation).

To avoid writing separate equations for $|t\rangle$ and $|t]$, we introduce the following notations,
\be\label{eq:t^pm}
\begin{aligned}
t_A^- &:= (-1)^{\f12 + A} t_A, \\\
t_A^+ &:=\bar{t}_{-A},
\end{aligned}
\qquad \text{and}\qquad 
\begin{aligned}
\tt_A^- &:= (-1)^{\f12 + A} \tt_A\,,\\
\tt_A^+ &:= \btt_{-A}\,.
\end{aligned}
\ee
for $A=\pm1/2$. Similar but exchanged notations are used for $\tau$ and $\ttau$,
\be \label{eq:tau^pm}
\begin{aligned}
\tau_A^- &:= \tau_A\,,\\
\tau_A^+ &:= (-1)^{\f12 - A} \btau_{-A}\,,
\end{aligned}
\qquad \text{and}\qquad 
\begin{aligned}
\ttau_A^- &:= \ttau_A\,, \\
\ttau_A^+ &:= (-1)^{\f12 - A}\bar{\ttau}_{-A}\,.
\end{aligned}
\ee
For reference, we explicitly write the spinors and dual spinors with those notations in a footnote\footnote{
\be\ba{llll}
|t\ra = \mat{c}{t_-^-\\ -t_+^-}\,, &
|t]=\mat{c}{-t_-^+\\ t_+^+}\,, &
|\tt\ra =\mat{c}{\tt_-^- \\ -\tt_+^-}\,, &
|\tt]=\mat{c}{-\tt_-^+\\\tt_+^+ }\,,\\[0.3cm]
\la t|=\mat{cc}{t_+^+,&t_-^+}\,, &
[t|=\mat{cc}{t_+^-,&t_-^-}\,, &
\la\tt|=\mat{cc}{\tt_+^+,&\tt_-^+}\,,&
[\tt|=\mat{cc}{\tt_+^-,&\tt_-^-}\,,\\[0.3cm]
|\tau\ra = \mat{c}{\tau_-^-\\ \tau_+^-}\,,&
|\tau]=\mat{c}{\tau_-^+ \\ \tau_+^+}\,, &
|\ttau\ra =\mat{c}{\ttau_-^-\\ \ttau_+^-}\,, &
|\ttau]=\mat{c}{\ttau_-^+\\ \ttau_+^+}\,, \\[0.3cm]
\la\tau|=\mat{cc}{\tau_+^+,&-\tau_-^+}\,,&
[\tau|=\mat{cc}{-\tau_+^-,& \tau_-^-}\,,&
\la\ttau|=\mat{cc}{\ttau_+^+,&-\ttau_-^+}\,, &
[\ttau|=\mat{cc}{-\ttau_+^-,& \ttau_-^-}\,,
\ea\ee
where the subscripts $A=\pm\f12$ have been notated as $A=\pm$ for simplicity.}. The norms read
\be\ba{ll}
\la t|t\ra = \f12 \sum_{\epsilon=\pm}\sum_{A=\pm \f12} \epsilon (-1)^{\f12+A}\, t_A^\epsilon t_{-A}^{-\epsilon}\,, &
\la \tt|\tt\ra=\f12 \sum_{\epsilon=\pm}\sum_{A=\pm \f12}  \epsilon (-1)^{\f12+A}\, \tt_A^\epsilon \tt_{-A}^{-\epsilon}\,, \\[0.15cm]
\la \tau|\tau \ra =\f12 \sum_{\epsilon=\pm}\sum_{A=\pm \f12}  \epsilon (-1)^{\f12+A}\, \tau_A^\epsilon \tau_{-A}^{-\epsilon}\,, &
\la \ttau|\ttau \ra = \f12 \sum_{\epsilon=\pm}\sum_{A=\pm \f12}  \epsilon (-1)^{\f12+A} \,\ttau_A^\epsilon \ttau_{-A}^{-\epsilon} \,.
\label{eq:spinor_norm}
\ea\ee
and the holonomies 
\begin{equation}
u_{AB} = -\frac{1}{\sqrt{\langle \tau|\tau\rangle \langle \tt|\tt\rangle}} \sum_{\epsilon=\pm} \epsilon
\tau^\epsilon_A \tt^\epsilon_{-B} \qquad 
\tilde{u}^{-1}_{AB} = \frac{1}{\sqrt{\langle \ttau|\ttau\rangle \langle t|t\rangle}} \sum_{\epsilon=\pm} \epsilon\ \ttau^\epsilon_A t^\epsilon_{-B}\,.
\end{equation}

%%%%%%%%%%
\subsection{Gauge invariant quantities}
%%%%%%%%%%

\begin{figure}
\includegraphics[scale=.75]{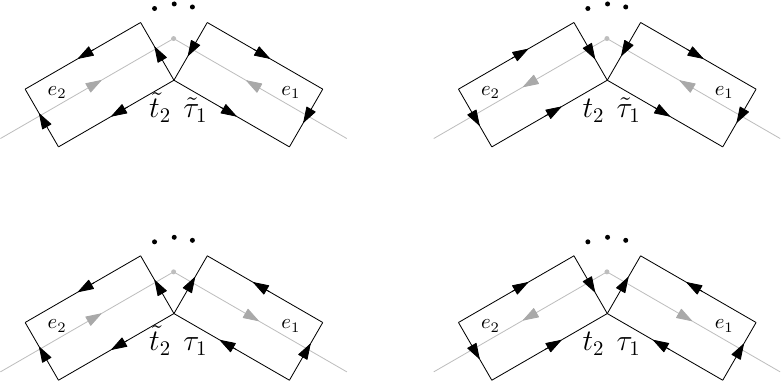}
\caption{\label{fig:TwoEdges} Two edges meet at a vertex and share a corner. There are four configurations of orientations and we indicate the spinors at the common corner.}
\end{figure}

We have described the spinors associated to an edge. Consider now two edges $e_1, e_2$ meeting at a vertex $v$ and incident to the same corner of $\Gamma$. The ribbon edges $R(e_1)$ and $R(e_2)$ share a corner where we have a spinor of $e_1$ and a spinor of $e_2$. Their scalar product is gauge-invariant. Given fixed orientations of the edges, there are four possible products (each spinor and its dual). There are moreover four configurations of orientations, shown in Figure \ref{fig:TwoEdges}. For instance, the four scalar products for the bottom right configuration are $\langle t_2| \tau_1\rangle$, $\langle t_2| \tau_1]$, $[t_2|\tau_1 \rangle$ and $[t_2|\tau_1]$.

Using the notations \eqref{eq:t^pm} and \eqref{eq:tau^pm}, we can give a uniform expression to the four scalar products at fixed orientations. For example, when both edges are outgoing,
\begin{equation}
E^{\epsilon_2, \epsilon_1}_{e_2 e_1} := \sum_{A=\pm1/2} \tau^{\epsilon_2}_{2, -A} t^{\epsilon_1}_{1,A} =  \begin{cases} \langle t_{2}| \tau_{1}] & \text{for $\epsilon_1=+, \epsilon_2=+$}\\
\langle t_{2}|\tau_{1}\rangle & \text{for $\epsilon_1=-, \epsilon_2=+$}\\
[t_{2}|\tau_{1}] & \text{for $\epsilon_1=+, \epsilon_2=-$}\\
[t_{2}|\tau_{1}\rangle & \text{for $\epsilon_1=-, \epsilon_2=-$}
\end{cases}\,.
\end{equation}
The other orientations are obtained by changing $\tau_{1}$ to $\tilde{\tau}_{1}$ and $t_{2}$ to $\tilde{t}_{2}$ and the invariant is still denoted $E^{\epsilon_2, \epsilon_1}_{e_2, e_1}$.

It will be convenient to encode all the orientations and have a fully uniform way of writing the invariant. We orient the corner between $e_1$ and $e_2$ counter-clockwise. We say that the orientation $o_i$ of $e_i$ for $i=1,2$ is positive if it matches that of the corner, and negative otherwise. We denote the spinors meeting there as $t_{e_1 v}$ and $t_{e_2 v}$ according to 
\begin{equation}
\begin{array}{|l|l|}
\hline
 & t_{e_1 v}\\
\hline
o_1=1 & \tilde{\tau}_1\\
\hline
o_1=-1 & \tau_1\\
\hline
\end{array}
\hspace{2cm}
\begin{array}{|l|l|}
\hline
 & t_{e_2 v}\\
\hline
o_2=1 & t_2\\
\hline
o_2=-1 & \tt_2\\
\hline
\end{array}
\end{equation}
so that
\begin{equation} \label{eq:scalar_product}
E^{\epsilon_2, \epsilon_1}_{e_2 e_1} = \begin{cases} \langle t_{e_2v}|t_{e_1v}] & \text{for $\epsilon_1=+, \epsilon_2=+$}\\
\langle t_{e_2v}|t_{e_1v}\rangle & \text{for $\epsilon_1=-, \epsilon_2=+$}\\
[t_{e_2v}|t_{e_1v}] & \text{for $\epsilon_1=+, \epsilon_2=-$}\\
[t_{e_2v}|t_{e_1v}\rangle & \text{for $\epsilon_1=-, \epsilon_2=-$}
\end{cases}\,.
\end{equation}

%%%%%%%%%%
\subsection{Hamiltonian constraint}
%%%%%%%%%%

By plugging $u$ and $\tilde{u}$ from \eqref{eq:flux_holonomy_from_spinors} into the flatness constraint \eqref{eq:flatness}, one obtains a spinorial expression of the constraint. By then taking the matrix elements of the constraints between different spinors, we get some scalar constraints which we call the Hamiltonian constraints. They are the $\kappa$-deformed versions of \cite{Bonzom:2011nv}.

We first write the Hamiltonian constraints generally, \ie on faces of arbitrary lengths, then specialize them to the case of faces of length 3.

\subsubsection{The Hamiltonian on a face of arbitrary degree}
%%%%%%%%%%%

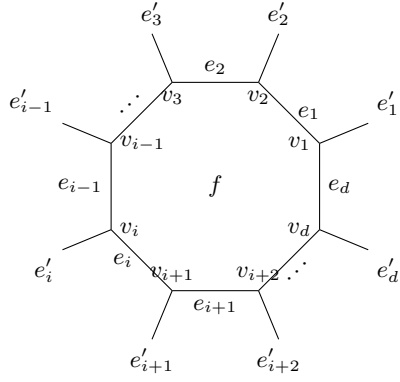
\begin{figure}[h!]
\centering
\begin{tikzpicture}
\coordinate (O) at (0,0);
\def\x{45};
\coordinate (A1) at ([shift=(22.5:1.5)]O);
\coordinate (A2) at ([shift=({22.5+\x}:1.5)]O);
\coordinate (A3) at ([shift=({22.5+2*\x}:1.5)]O);
\coordinate (A4) at ([shift=({22.5+3*\x}:1.5)]O);
\coordinate (A5) at ([shift=({22.5+4*\x}:1.5)]O);
\coordinate (A6) at ([shift=({22.5+5*\x}:1.5)]O);
\coordinate (A7) at ([shift=({22.5+6*\x}:1.5)]O);
\coordinate (A8) at ([shift=({22.5+7*\x}:1.5)]O);

\coordinate (B1) at ([shift=(22.5:0.7)]A1);
\coordinate (B2) at ([shift=(22.5+\x:0.7)]A2);
\coordinate (B3) at ([shift=(22.5+2*\x:0.7)]A3);
\coordinate (B4) at ([shift=(22.5+3*\x:0.7)]A4);
\coordinate (B5) at ([shift=(22.5+4*\x:0.7)]A5);
\coordinate (B6) at ([shift=(22.5+5*\x:0.7)]A6);
\coordinate (B7) at ([shift=(22.5+6*\x:0.7)]A7);
\coordinate (B8) at ([shift=(22.5+7*\x:0.7)]A8);

\draw (A1) -- node[midway, right]{$e_1$}(A2); 
\draw (A2) -- node[midway, above]{$e_2$}(A3);
\draw (A3) -- node[dashed,midway, sloped,above]{$\cdots$}(A4);
\draw (A4) -- node[midway, left]{$e_{i-1}$}(A5);
\draw (A5) -- node[midway, left]{$e_i$}(A6);
\draw (A6) -- node[dashed,midway, below]{$e_{i+1}$}(A7);
\draw (A7) -- node[midway, sloped,below]{$\cdots$}(A8);
\draw (A8) -- node[midway, right]{$e_d$}(A1);

\draw (A1) -- node[at end, above right]{$e'_1$}(B1); 
\draw (A2) -- node[at end, above]{$e'_2$}(B2); 
\draw (A3) -- node[at end, above]{$e'_3$}(B3); 
\draw (A4) -- node[at end, above left]{$e'_{i-1}$}(B4); 
\draw (A5) -- node[at end, below left]{$e'_i$}(B5); 
\draw (A6) -- node[at end, below]{$e'_{i+1}$}(B6); 
\draw (A7) -- node[at end, below]{$e'_{i+2}$}(B7); 
\draw (A8) -- node[at end, below right]{$e'_d$}(B8); 

\draw (A1) node[left]{$v_1$};
\draw (A2) node[below]{$v_2$};
\draw (A3) node[below]{$v_3$};
\draw (A4) node[right]{$v_{i-1}$};
\draw (A5) node[right]{$v_i$};
\draw (A6) node[above]{$v_{i+1}$};
\draw (A7) node[above]{$v_{i+2}$};
\draw (A8) node[left]{$v_d$};
\draw (O) node{$f$};

\end{tikzpicture}
\caption{A sunny graph with edges $e_1,\cdots,e_d$ counterclockwise oriented around the face $f$. Each triple of edges $(e_i,e_{i-1},e'_i)$ are incident to a vertex $v_i$.}
\label{fig:sunny}
\end{figure}

Let $f$ be a face of length $d$. We will introduce a constraint, derived from the flatness constraint, for every pair of edges $(e, e')$ around $f$. Label the edges counter-clockwise around $f$ as $e_1, \dotsc, e_d$. Without loss of generality, we set the pair $(e,e')$ which labels our function to $(e_1, e_k)$ for $k\in\{ 2, \dotsc, d\}$. Label the vertices around $f$ as $v_1,\dotsc, v_d$ counter-clockwise, such that $e_i$ is incident to $v_i$ and $v_{i+1}$, for $i=1,\dotsc, d \mod d$, as shown in Figure \ref{fig:sunny}. We assume that $f$ visits each vertex and edge exactly once (as when $\Gamma$ is dual to a simplicial complex), so that all $e_i$s and $v_i$s are distinct.

By convention, we denote the orientation $o_i = 1$ if $e_i$ is counter-clockwise and $o_i=-1$ elsewise (this is the relative orientation of the edge with respect to the counter-clockwise orientation of $f$). With the notation $u_{ev}$ introduced earlier, the flatness constraint reads $u_{e_d v_1} \dotsm u_{e_2 v_3} u_{e_1 v_2} = \mathbbm{1}$ in $\SU(2)$. In order to simplify the notations a bit, we will use 
\be
u_{e_i f} := u_{e_i v_{i+1}} =
\begin{cases}
\tu_{e_i}^{-1}\,, \qquad \text{if $o_{1}=1$}\,, \\
u_{e_i}\,, \qquad \text{if $o_{1}=-1$}\,.
\end{cases}
\ee
Furthermore we denote $t_{e_i v_i}$ the spinor along the long edge of $R(e_i)$ which is incident to both $f$ and $v_i$. It is determined by the orientation of $e_i$,
\begin{equation} \label{RuleSpinorsOrientations}
\begin{aligned}
o_{i} = 1&\quad \Rightarrow\quad t_{e_i v_i} = t_{e_i}\quad \text{and}\quad t_{e_i v_{i+1}} = \ttau_{e_i}\,,\\
o_{i} = -1&\quad \Rightarrow\quad t_{e_i v_i} = \tt_{e_i}\quad \text{and}\quad t_{e_i v_{i+1}} = \tau_{e_i}\,.
\end{aligned}
\end{equation}
Notice that we can combine the parallel transport relations \eqref{eq:paral_spinors} with the notations \eqref{eq:t^pm}, \eqref{eq:tau^pm} to relate the spinors which are on both ends of the long edge of $e_i$ incident to $f$,
\be
t^{\epsilon_i}_{e_i v_i,-A} 
= -o_i\sum_{B=\pm 1/2} t_{e_i v_{i+1},-B}^{-\epsilon_i}(-1)^{\f12+B}\, u_{e_i f,BA} \,,\qquad
t_{e_iv_{i+1},A}^{\epsilon_i}
= o_i\sum_{B=\pm 1/2}u_{e_if,AB} \,(-1)^{\f12+B}t_{e_iv_i,B}^{-\epsilon_i}\,. 
\label{eq:ParallelTAroundF}
\ee

The flatness constraint on $f$ is thus $u_{e_d f} \dotsm u_{e_1 f} = \mathbbm{1}$. Assume momentarily that all edges are counter-clockwise. Then, $\tu_{e_d}^{-1} \dotsm \tu_{e_1}^{-1} = \mathbbm{1}$ implies for all $k$
\be
[t_{e_k} | \tu_{e_{k-1}}^{-1} \dotsm \tu_{e_2}^{-1} | \ttau_{e_1} \ra 
= [t_{e_k} | \tu_{e_{k}} \tu_{e_{k+1}} \dotsm \tu_{e_d} \tu_{e_1}| \ttau_{e_1} \ra 
= \la \ttau_{e_k} | \tu_{e_{k+1}} \dotsm \tu_{e_d} | t_{e_1} ]\,.
\label{eq:ToyConstraint}
\ee 
In the first equality, we have used the constraint itself, while in the second equality we have used the parallel transport relations on the edges $e_1$ and $e_k$. Then by rewriting $\tu_{1_1}, \dotsc, \tu_{e_d}$ with \eqref{eq:flux_holonomy_from_spinors}, one obtains the following result: a constraint written as a sum of products of scalar invariants living on the corners around the face. Obviously, one can change $[t_{e_k}|$ to $\langle t_{e_k}|$ and $|\ttau_{e_1}\rangle$ to $|\ttau_{e_1}]$ without changing that result (qualitatively). Similarly, one should be able to write this function for arbitrarily chosen edge orientations. The notations we have introduced will help us write it in the most generic way.

Going back to arbitrary edge orientations around $f$, we consider
\be
E_{e_1\to e_k}^{\epsilon_1, \epsilon_k} 
= \sum_{A,B=\pm 1/2} t_{e_kv_{k},-A}^{\epsilon_k}
\lb u_{e_{k-1}f}\cdots u_{e_2f} \rb_{AB}
t_{e_1v_2,B}^{\epsilon_1}
\ee
as the generalization of the left-hand side of \eqref{eq:ToyConstraint}. Using the parallel transport relations \eqref{eq:ParallelTAroundF}, it reads
\be
E_{e_1\to e_k}^{\epsilon_1, \epsilon_k} 
= -o_1o_k\sum_{C,D=\pm 1/2} t_{e_k v_{k+1}, -C}^{-\epsilon_k} (-1)^{\f12-C}
\bigl(u_{e_kf} u_{e_{k-1} f} \dotsm u_{e_2 f} u_{e_1 f}\bigr)_{CD} 
(-1)^{\f12-D} t_{e_1 v_1, D}^{-\epsilon_1}\,.
\label{eq:E1k_2}
\ee
If the flatness constraint holds, the holonomy going counter-clockwise from $e_1$ to $e_k$ can then be replaced with the holonomy the other way around $f$, \ie clockwise. We thus define
\begin{equation}
E_{e_1 \leftarrow e_k}^{\epsilon_1, \epsilon_k} = \sum_{A,B=\pm 1/2} t_{e_1 v_1,- A}^{-\epsilon_1}  
\bigl(u_{e_d f} \dotsm u_{e_{k+1} f} \bigr)_{AB} 
 t_{e_k v_{k+1}, B}^{-\epsilon_k}\,.
\end{equation}
So if the flatness constraint holds, then
\begin{equation}
E_{e_1 \to e_k}^{\epsilon_1, \epsilon_k} + o_1o_k E_{e_1 \leftarrow e_k}^{\epsilon_1, \epsilon_k} = 0.
\end{equation}
Indeed, using the flatness constraint in \eqref{eq:E1k_2} we get
\begin{equation}
E_{e_1\to e_k}^{\epsilon_1, \epsilon_k} 
= -o_1o_k\sum_{C,D=\pm 1/2} t_{e_k v_{k+1}, -C}^{-\epsilon_k} (-1)^{\f12-C} \bigl(u_{e_d f} \dotsm u_{e_{k+1} f}\bigr)^{-1}_{CD} (-1)^{\f12-D} t_{e_1 v_1, D}^{-\epsilon_1}\,.
\end{equation}
For any matrix $g\in \SU(2)$, the matrix elements of the inverse can be written $g^{-1}_{CD} = (-1)^{\f12 - D} g_{-D -C} (-1)^{\f12 - C}$. This can be used to transform the above expression into $o_1o_k E_{e_1 \leftarrow e_k}^{\epsilon_1, \epsilon_k}$.
The last step to define our Hamiltonian constraints is to rewrite $E_{e_1 \to e_k}^{\epsilon_1, \epsilon_k}$ and $E_{e_1 \leftarrow e_k}^{\epsilon_1, \epsilon_k}$ in terms of scalars like \eqref{eq:scalar_product}. The matrix elements of the holonomies are indeed
\begin{align}
u_{e_i f, A_{i+1} A_i} &= o_{i}\f{1}{N_{e_i}} \sum_{\epsilon_i=\pm} \epsilon_i\ t_{e_i v_{i+1}, A_{i+1}}^{\epsilon_i}\ t_{e_i v_i, -A_i}^{\epsilon_i}\,,\\
\text{with}\quad N_{e_i}
&=\f12 \sqrt{\sum_{\epsilon_i,\epsilon'_i=\pm}\sum_{A,B=\pm 1/2} 
\epsilon \epsilon' (-1)^{\f12-A}(-1)^{\f12-B}\,
t_{e_i,v_{i+1},A}^{\epsilon_i} t_{e_iv_{i+1},-A}^{-\epsilon_i}\, t_{e_i,v_i,B}^{\epsilon'_i} t_{e_iv_i,-B}^{-\epsilon'_i}}\,,
\end{align}
so that one can re-organize the products over the vertices instead of edges,
\begin{align}
E_{e_1\to e_k}^{\epsilon_1, \epsilon_k} &= \sum_{\epsilon_2, \dotsc, \epsilon_{k-1}=\pm \atop A_2, \dotsc, A_k=\pm1/2} \biggl(\prod_{i=2}^{k-1} \frac{o_i \epsilon_i}{N_{e_i}}\biggr) 
\biggl(\prod_{i=2}^{k} t^{\epsilon_i}_{e_i v_i, -A_i} t^{\epsilon_{i-1}}_{e_{i-1} v_i, A_i} \biggr)\,,\\
E_{e_1\leftarrow e_k}^{\epsilon_1, \epsilon_k} &= (-1)^{d-k} \sum_{\epsilon_{k+1}, \dotsc, \epsilon_d=\pm \atop A_{k+1}, \dotsc, A_{d+1}=\pm1/2} \biggl(\prod_{i=k+1}^{d} \frac{o_i \epsilon_i}{N_{e_i}}\biggr) \biggl(\prod_{i=k+1}^{d+1} t^{-\epsilon_i}_{e_i v_i, -A_i} t^{-\epsilon_{i-1}}_{e_{i-1} v_i, A_i} \biggr)\,.
\end{align}

We can now use the quadratic invariants defined in \eqref{eq:scalar_product}, $E_{e_i e_{i-1}}^{\epsilon_i, \epsilon_{i-1}} 
= \sum_{A=\pm 1/2} t_{e_i v_i, -A}^{\epsilon_i} t_{e_{i-1} v_i, A}^{\epsilon_{i-1}}$, which encodes all four scalar products of the two spinors meeting at $v_i$, \ie
\begin{equation}
E_{e_i e_{i-1}}^{\epsilon_i, \epsilon_{i-1}}  =  \begin{cases} \langle t_{e_i v_i}|t_{e_{i-1} v_i}] & \qquad \text{for $\epsilon_i=+, \epsilon_{i-1}=+$}\\
\langle t_{e_i v_i}|t_{e_{i-1} v_i}\rangle & \qquad \text{for $\epsilon_i=+, \epsilon_{i-1}=-$}\\
[t_{e_i v_i}|t_{e_{i-1} v_i}] & \qquad \text{for $\epsilon_i=-, \epsilon_{i-1}=+$}\\
[t_{e_i v_i}|t_{e_{i-1} v_i}\rangle & \qquad \text{for $\epsilon_i=-, \epsilon_{i-1}=-$},
\end{cases}
\end{equation}
where the spinors $t_{e_i v_i}$ and $t_{e_{i-1} v_{i-1}}$ are given by the rule \eqref{RuleSpinorsOrientations} according to the orientations. This leads us to the following definition of the Hamiltonian constraints.
\begin{definition}
Let $f$ be a face of length $d$, with edges labelled by $e_1, \dotsc, e_d$ counter-clockwise around $f$. A Hamiltonian is associated to $f$ and a pair of edges along $f$ with a sign attached to each of them. Without loss of generality, the pair can be chosen to be $(e_1, e_k)$ with signs $(\epsilon_1, \epsilon_k)\in\{+,-\}^2$, for $k\in\{2, \dotsc, d\}$, and the Hamiltonian is
\begin{equation}
h_{f, e_1, e_k}^{\epsilon_1, \epsilon_k} 
= \sum_{\epsilon_2, \dotsc, \epsilon_{k-1} = \pm} 
\biggl(\prod_{i=2}^{k}  \f{o_{i} \epsilon_i}{N_{e_i}} 
E_{e_i e_{i-1}}^{\epsilon_i, \epsilon_{i-1}} \biggr)
+(-1)^{d-k}  \epsilon_1 \epsilon_k \f{N_{e_1}}{N_{e_k}}
\sum_{\epsilon_{k+1}, \dotsc, \epsilon_d = \pm} 
\biggl(\prod_{i=k+1}^{d+1} \f{o_{i} \epsilon_i}{N_{e_i}} E_{e_i e_{i-1}}^{-\epsilon_i, -\epsilon_{i-1}}\biggr)\,.
\label{eq:ClassicalHamiltonian}
\end{equation}
\end{definition}

The Hamiltonian constraint \eqref{eq:ClassicalHamiltonian} captures the flatness constraint completely with all choices of pairs $(e_1,e_k)$ and of signs $(\epsilon_1,\epsilon_k)$. The proof is the same as in the vector case at $\kappa=0$, see \cite{Bonzom:2011hm}.

%%%%%%%%%%%
\subsubsection{Application to faces of degree three}
%%%%%%%%%%%

\begin{figure}[h!]
\centering
\begin{tikzpicture}[scale=1.3]
	\coordinate (A) at (0,0);
	\coordinate (B) at (2,0);
	\coordinate (C) at ([shift=(-60:2cm)]A);
	\coordinate (A1) at ([shift=(150:0.8cm)]A);
	\coordinate (B1) at ([shift=(30:0.8cm)]B);
	\coordinate (C1) at ([shift=(-90:0.8cm)]C);
	
	\draw[thick,decoration={markings,mark=at position 0.55 with {\arrow[scale=1.5,>=stealth]{<}}},postaction={decorate}] (A) -- node[above,pos=.5]{$e_2$}(B);
	\draw[thick,decoration={markings,mark=at position 0.55 with {\arrow[scale=1.5,>=stealth]{<}}},postaction={decorate}] (C) -- node[right,pos=.5]{$e_6$}(B);
	\draw[thick,decoration={markings,mark=at position 0.55 with {\arrow[scale=1.5,>=stealth]{<}}},postaction={decorate}] (A) -- node[left,pos=.5]{$e_1$}(C);
	\draw[thick,decoration={markings,mark=at position 0.55 with {\arrow[scale=1.5,>=stealth]{<}}},postaction={decorate}] (A) -- node[left,pos=.5]{$e_3$}(A1);
	\draw[thick,decoration={markings,mark=at position 0.55 with {\arrow[scale=1.5,>=stealth]{<}}},postaction={decorate}] (B) -- node[below,pos=.5]{$e_4$}(B1);
	\draw[thick,decoration={markings,mark=at position 0.55 with {\arrow[scale=1.5,>=stealth]{<}}},postaction={decorate}] (C1) -- node[left,pos=.5]{$e_5$}(C);
	
	\draw[->] (3,-1) -- (5,-1);
	
	\coordinate (a1) at (6,0);
	\coordinate (a2) at ([shift=(90:0.5cm)]a1);
	\coordinate (a3) at ([shift=(0:2cm)]a2);
	\coordinate (a4) at ([shift=(-90:0.5cm)]a3);
	\coordinate (b1) at (a1);
	\coordinate (b2) at ([shift=(-150:0.5cm)]b1);
	\coordinate (b3) at ([shift=(-60:2cm)]b2);
	\coordinate (b4) at ([shift=(30:0.5cm)]b3);
	\coordinate (c1) at (b4);
	\coordinate (c2) at ([shift=(-30:0.5cm)]c1);
	\coordinate (c3) at ([shift=(60:2cm)]c2);
	\coordinate (c4) at ([shift=(150:0.5cm)]c3);
	
	\draw[thick,dashed,decoration={markings,mark=at position 0.65 with {\arrow[scale=1.5,>=stealth]{<}}},postaction={decorate}] (a1) -- node[right,pos=.5]{$\lt_2$}(a2);
	\draw[thick,decoration={markings,mark=at position 0.55 with {\arrow[scale=1.5,>=stealth]{>}}},postaction={decorate}] (a2) -- node[above,pos=.5]{$u_2$}(a3);
	\draw[thick,dashed,decoration={markings,mark=at position 0.65 with {\arrow[scale=1.5,>=stealth]{>}}},postaction={decorate}] (a3) -- node[left,pos=.5]{$\ell_2$}(a4);
	\draw[thick,decoration={markings,mark=at position 0.55 with {\arrow[scale=1.5,>=stealth]{<}}},postaction={decorate}] (a4) -- node[below,pos=.5]{$\ut_2$}(a1);
	
	\draw[thick,dashed,decoration={markings,mark=at position 0.65 with {\arrow[scale=1.5,>=stealth]{>}}},postaction={decorate}] (b1) -- node[below right,pos=.5]{$\lt_1$}(b2);
	\draw[thick,decoration={markings,mark=at position 0.55 with {\arrow[scale=1.5,>=stealth]{>}}},postaction={decorate}] (b2) -- node[left,pos=.5]{$\ut_1$}(b3);
	\draw[thick,dashed,decoration={markings,mark=at position 0.65 with {\arrow[scale=1.5,>=stealth]{<}}},postaction={decorate}] (b3) -- node[above left,pos=.5]{$\ell_1$}(b4);
	\draw[thick,decoration={markings,mark=at position 0.55 with {\arrow[scale=1.5,>=stealth]{<}}},postaction={decorate}] (b4) -- node[right,pos=.5]{$u_1$}(b1);

	\draw[thick,dashed,decoration={markings,mark=at position 0.65 with {\arrow[scale=1.5,>=stealth]{>}}},postaction={decorate}] (c1) -- node[above right,pos=.5]{$\lt_6$}(c2);
	\draw[thick,decoration={markings,mark=at position 0.55 with {\arrow[scale=1.5,>=stealth]{>}}},postaction={decorate}] (c2) -- node[right,pos=.5]{$\ut_6$}(c3);
	\draw[thick,dashed,decoration={markings,mark=at position 0.65 with {\arrow[scale=1.5,>=stealth]{<}}},postaction={decorate}] (c3) -- node[below left,pos=.5]{$\ell_6$}(c4);
	\draw[thick,decoration={markings,mark=at position 0.55 with {\arrow[scale=1.5,>=stealth]{<}}},postaction={decorate}] (c4) -- node[left,pos=.5]{$u_6$}(c1);
	
	\draw[thick,dashed,decoration={markings,mark=at position 0.65 with {\arrow[scale=1.5,>=stealth]{>}}},postaction={decorate}] (b2) -- node[left,pos=.5]{$\lt_3$}(a2);
	\draw[thick,dashed,decoration={markings,mark=at position 0.65 with {\arrow[scale=1.5,>=stealth]{>}}},postaction={decorate}] (c3) -- node[right,pos=.5]{$\lt_4$}(a3);
	\draw[thick,dashed,decoration={markings,mark=at position 0.65 with {\arrow[scale=1.5,>=stealth]{>}}},postaction={decorate}] (b3) -- node[below,pos=.5]{$\ell_5$}(c2);

\end{tikzpicture}

\caption{ On the left, a triangular face with its adjacent edges. On the right, the ribbon graph it gives rise to.}
\label{fig:Face126}
\end{figure}
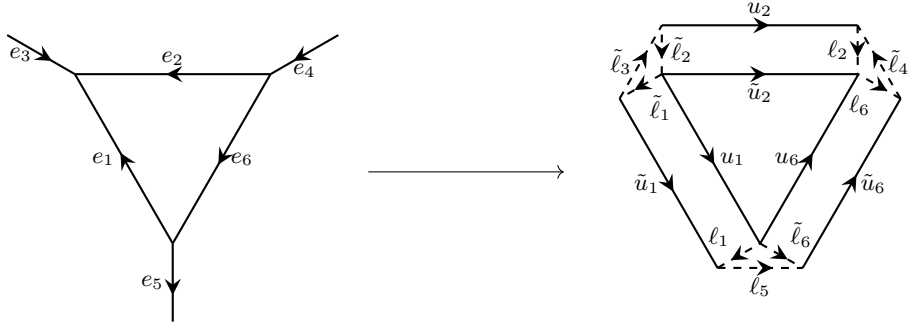

Let us discuss more explicitly the case of triangular faces. We use the notations and orientations of Figure \ref{fig:Face126} as an example. In particular $o_1=o_6=-1$ and $o_2=1$. Here there are three choices of pairs of edges (which label the Hamiltonians \eqref{eq:ClassicalHamiltonian}), which correspond to the three corners of the face.

On a corner, say between the edges $e_2$ and $e_6$, there are four invariant quantities quadratic in the spinors, $\langle t_2|\tau_6\rangle, \langle t_2| \tau_6], [t_2| \tau_6\rangle, [t_2| \tau_6]$ which are encoded in the scalar product \eqref{eq:scalar_product},
\begin{equation}
E_{26}^{\epsilon_2, \epsilon_6} =  \sum_{A=\pm 1/2} t_{2, -A}^{\epsilon_2} \tau_{6, A}^{\epsilon_6} = 
\begin{cases} 
\la t_2|t_6] & \text{for $\epsilon_2= \epsilon_6 = +$},\\
\la t_2|t_6\ra & \text{for $\epsilon_2 = -\epsilon_6 = +$},\\
[t_2|t_6]& \text{for $\epsilon_2 =- \epsilon_6 = -$},\\
[t_2|t_6\ra & \text{for $\epsilon_2 = \epsilon_6 = -$}\,.
\end{cases}
\end{equation}
Similarly at the corners between $e_1, e_2$ and $e_6, e_1$, 
\be
E_{12}^{\epsilon_1, \epsilon_2} 
= \sum_{A=\pm 1/2} 
 \tt_{1, -A}^{\epsilon_1}\ttau_{2, A}^{\epsilon_2},\qquad
E_{61}^{\epsilon_6, \epsilon_1} 
=   \sum_{A=\pm 1/2} 
\tt_{6, -A}^{\epsilon_6}\tau_{1, A}^{\epsilon_1} \,.
\ee

The flatness constraint  
$\tu_2u_1^{-1} u_6^{-1}  = \id$ 
implies that if $|\tau_6\rangle$ and $|\tau_6]$ are transported around the face via $\tu_2u_1^{-1} u_6^{-1}  $, the above quadratic quantities are left unchanged, that is
\be \label{eq:Constraints16}
\begin{aligned}
&\la t_2|\tu_2u_1^{-1}  u_6^{-1} |\tau_6\ra = \la t_2|\tau_6\ra\,,
&\qquad &
\la t_2|\tu_2 u_1^{-1} u_6^{-1} |\tau_6] = \la t_2|\tau_6]\,,
\\
&[ t_2|\tu_2 u_1^{-1} u_6^{-1} |\tau_6\ra = [ t_2|\tau_6\ra\,,
&\qquad &
[ t_2|\tu_2 u_1^{-1} u_6^{-1} |\tau_6] =[t_2|\tau_6]\,.
\end{aligned}
\ee
Similarly at the corners where $e_1, e_2$ and $e_6, e_1$ meet respectively, 
\be \label{Constraints12-16}
\begin{aligned}
&\la \tt_1| u_1^{-1} u_6^{-1} \tu_2 |\ttau_2\ra = \la \tt_1|\ttau_2\ra\,, &\qquad & 
\la \tt_6| u_6^{-1} \tu_2 u_1^{-1}| \tau_1\ra = \la \tt_6|\tau_1\ra\,, \\
&\la \tt_1|u_1^{-1} u_6^{-1} \tu_2  | \ttau_2] = \la \tt_1|\ttau_2]\,, &\qquad & 
\la \tt_6| u_6^{-1} \tu_2 u_1^{-1}| \tau_1] = \la \tt_6|\tau_1]\,, \\
&[ \tt_1| u_1^{-1} u_6^{-1} \tu_2  | \ttau_2\ra = [ \tt_1|\ttau_2\ra \,,&\qquad & 
[ \tt_6| u_6^{-1} \tu_2 u_1^{-1}| \tau_1\ra = [ \tt_6|\tau_1\ra\,, \\
&[ \tt_1| u_1^{-1} u_6^{-1} \tu_2  | \ttau_2] = [ \tt_1|\ttau_2]\,, &\qquad & 
[ \tt_6| u_6^{-1} \tu_2 u_1^{-1}| \tau_1] = [ \tt_6|\tau_1]\,.
\end{aligned}
\ee
In fact, this set of constraints simply amounts to rewriting the constraint $\tu_2u_1^{-1} u_6^{-1} = \id$ in the coherent state basis. Therefore, as long as those vectors are generic (hence linearly independent), this whole set is equivalent to $\tu_2u_1^{-1} u_6^{-1}=\id$.

Let us consider the constraint $\la t_2|\tu_2u_1^{-1} u_6^{-1} |\tau_6]- \la t_2 |\tau_6]$ and rewrite it like in \eqref{eq:ClassicalHamiltonian}. 
Use the parallel transport relations $u_6^{-1} |\tau_6] = -|\tt_6\ra$ and $\la t_2| \tu_2= [ \ttau_2|$ which gives $\la t_2|\tu_2u_1^{-1} u_6^{-1} |\tau_6] = - [\ttau_2|u_1^{-1}|\tt_6\ra = [ \tt_6|u_1|\ttau_2\ra$. Then use $u_1 = \frac{1}{N_{e_1}}(|\tau_1\rangle [\tt_1| - |\tau_1]\langle \tt_1|)$ so that the constraint becomes
\begin{equation}
\begin{aligned}
\la t_2|\tu_2u_1^{-1} u_6^{-1} |\tau_6]- \la t_2 |\tau_6] 
&= \frac{1}{N_{e_1}} \bigl([ \tt_6|\tau_1\ra [\tt_1|\ttau_2\ra - [ \tt_6|\tau_1]\la \tt_1|\ttau_2\ra\bigr) - \la t_2|\tau_6]\\
&= - \frac{1}{N_{e_1}} \sum_{\epsilon=\pm} \epsilon\, E^{-,\epsilon}_{61} E^{\epsilon,-}_{12} - E_{26}^{+,+}
\end{aligned}
\end{equation}
which is exactly the specialization of \eqref{eq:ClassicalHamiltonian} to $d=3$, $k=2$, $o_1=o_6=-o_2=-1$ and $\epsilon_2=\epsilon_6=+$,
\be 
h_{e_2 e_6}^{\epsilon_2, \epsilon_6} 
= E_{26}^{\epsilon_2,\epsilon_6}
+
\f{1}{N_{e_1}}
\sum_{\epsilon_1=\pm} \epsilon_1 E_{12}^{\epsilon_1,-\epsilon_2} E_{61}^{-\epsilon_6,\epsilon_1}\,.
\label{eq:ClassicalTriangularHamiltonian}
\ee
where we recall that $N_{e_1} = \sqrt{\la \tau_1|\tau_1\ra \la \tt_1|\tt_1\ra}$.

This way, the Hamiltonian constraint does not involve holonomy variables anymore like in \eqref{eq:Constraints16}, but only the quadratic invariants of spinors.

%%%%%%%%%%%%%%%%%%%%%%%%%%%%%%%%%%%%%%%%%%%%%%%%%%%%%%%
\section{Quantum hamiltonian constraint}
\label{sec:quantum_hamiltonian}
%%%%%%%%%%%%%%%%%%%%%%%%%%%%%%%%%%%%%%%%%%%%%%%%%%%%%%%

We now proceed to the quantization of the system. The aim is to quantize the Hamiltonian constraints \eqref{eq:ClassicalHamiltonian} and solve them at the quantum level. This requires quantizing the quadratic invariant $E^{\epsilon_i,\epsilon_{i-1}}_{e_ie_{i-1}}$. It has been constructed consistently with the quantization of holonomies and fluxes in the companion paper \cite{Bonzom:2022bpv}. Therefore, we start with recalling the main ingredients before proceeding to the construction of the quantum Hamiltonian.

%%%%%%%%%%%%%%%%%%%%%%%%%%%%%%%%%%%%%%%%%%%%%%%%%%%%%%
\subsection{Quantum deformed spinors}

Let $q:=e^{\hbar\kappa}$ and denote $[n]:=\f{q^{\f{n}{2}}-q^{-\f{n}{2}}}{q^{\f12}-q^{-\f12}}$ the $q$-numbers. The $\ka$-deformed spinors are quantized as $q$-bosons (in the same way they are at $\kappa=0$  \cite{Livine:2012}). Consider two independent pairs of $q$-boson operators $(a,a^\dagger)$ and $(b,b^\dagger)$ satisfying the relations
\be \ba{llll}
a a^\dagger - q^{\mp\demi}a^\dagger a = q^{\pm \f{N_a}{2}}\,,\quad &
a^{\dagger}a-q^{\pm \f12}aa^{\dagger} =-q^{\pm\f{N_a+1}{2}}\,,\quad &
[N_a,a^\dagger]=a^\dagger\,,\quad &
[N_a,a]=-a\,, \\
b b^\dagger - q^{\mp\demi}b^\dagger b = q^{\pm \f{N_b}{2}}\,,\quad &
b^{\dagger}b-q^{\pm \f12}bb^{\dagger} =-q^{\pm\f{N_b+1}{2}}\,,\quad &
[N_b,b^\dagger]=b^\dagger\,,\quad &
[N_b,b]=-b \,.
\ea
\label{eq:a_adagger}
\ee
with $a^\dagger a = [N_a]$, $aa^\dagger= [N_{a}+1]$, $b^\dagger b = [N_b]$, $bb^\dagger= [N_{b}+1]$. We furthermore introduce two other independent pairs of $q$-bosons denoted $(\at,\at^\dagger)$ and $(\bt,\bt^\dagger)$, and satisfying the same relations as above. The tilde and non-tilde operators are chosen to commute with each other. 

We will then use the following quantization map
\be \ba{lll}
(\zeta^\kappa_0, \zeta^\kappa_1)\dr (a,b)\,, \quad &
(\bzeta^\kappa_0, \bzeta^\kappa_1)\dr   (a^\dagger, b^\dagger)\,, \quad & 
(N_0, N_1) \dr (N_a, N_b)\,,\\
(\tzeta^\ka_0,\tzeta^\ka_1 )\dr (\at,\bt)\,,\quad &
(\btzeta^\ka_0,\btzeta^\ka_1 )\dr (\at^\dagger ,\bt^\dagger)\,,\quad &
(\Nt_0,\Nt_1) \dr (\Nt_a,\Nt_b)\,.
\ea
\label{eq:quantization_spinors}
\ee

In previous works by the first and second authors, the fluxes $\ell$ and $\tell$ had been quantized in terms of the quantum algebra $\UQ$. This can be replicated in a manner which is consistent with the $q$-bosons. Indeed, the Jordan map builds generators of $\UQ$ out of the above $q$-bosons,
\be\ba{lll}
J_+=a^\dagger b\,,\quad &
J_-=ab^\dagger\,,\quad &
K=q^{\f{J_z}{2}}= q^{\f{N_a-N_b}{4}}\,,\\
\Jt_+=\at^\dagger \bt\,,\quad &
\Jt_-=\at\bt^\dagger\,,\quad &
\Kt=q^{\f{\Jt_z}{2}}= q^{\f{\Nt_a-\Nt_b}{4}}\,,
\ea
\label{eq:Jordan_map}
\ee
where $J_\pm\,,K\equiv q^{\f{J_z}{2}}$ and $\Jt_\pm\,,\Kt\equiv q^{\f{\Jt_z}{2}}$ are two {\it independent} copies of the $\UQ$ generators satisfying the relations
\be
KJ_\pm K\mone = q^{\pm \demi} J_\pm\,,\quad
[J_+,J_-]=[2J_z]\,,\quad 
\Kt\Jt_\pm \Kt\mone = q^{\pm \demi} \Jt_\pm\,,\quad
[\Jt_+,\Jt_-]=[2\Jt_z]\,,
\label{eq:UQ_generators_2}
\ee 
and others vanish. Performing the quantization as follows,
\be\ba{lcl}
\ell=\mat{cc}{\exp(\f{\ka}{4}(N_1-N_0)) & 0 \\
-\ka\bzeta_0^\ka \zeta_1^\ka & \exp(\f{\ka}{4}(N_0-N_1))}
&\quad\longrightarrow\quad &
\lh=\mat{cc}{K^{-1} & 0 \\ -q^{\f14}(q^{\f12}-q^{-\f12})J_+ & K}\,, \\
\lt=\mat{cc}{\exp(\f{\ka}{4}(\Nt_0-\Nt_1)) & 0 \\
\ka \btzeta^\ka_0 \tzeta^\ka_1 & \exp(\f{\ka}{4}(\Nt_1-\Nt_0))}
&\quad\longrightarrow\quad &
\lth=\mat{cc}{\Kt & 0 \\ q^{-\f14}(q^{\f12}-q^{-\f12})\Jt_+ & \Kt^{-1}}\,,
\ea
\label{eq:quantum_fluxes}
\ee
one finds, as in \cite{Bonzom:2022bpv}, that $\lh\in F_{q^{-1}}(\AN(2))\cong\UQI$ and $\lth\in F_{q}(\AN(2))\cong\UQ$ \cite{Semenov:1994po}. 

The map \eqref{eq:quantization_spinors} quantizes the $\kappa$-deformed spinors \eqref{eq:def_deformed_spinors} as follows
\be\ba{lllllll}
|t\ra=\mat{c}{e^{\f{\ka N_1}{4}}\zeta^\ka_0 \\ e^{-\f{\ka N_0}{4}}\zeta^\ka_1}
&\longrightarrow &
\bft^-=\mat{c}{\bft^-_+\\\bft^-_-}=\mat{c}{q^{\f{N_b}{4}}a \\q^{-\f{N_a}{4}}b}\,,
&\quad &
|t] =\mat{c}{-e^{-\f{\ka N_0}{4}}\bzeta^\ka_1 \\ e^{\f{\ka N_1}{4}}\bzeta^\ka_0}
&\longrightarrow &
\bft^+=\mat{c}{\bft^+_+\\ \bft^+_-} =\mat{c}{-b^\dagger q^{-\f{N_a+1}{4}}\\ a^\dagger q^{\f{N_b+1}{4}}}\,,
\\[0.4cm]
|\tau\ra =\mat{c}{e^{-\f{\ka N_1}{4}}\zeta_0^\ka \\ e^{\f{\ka N_0}{4}}\zeta^\ka_1}
&\longrightarrow &
\mt^-=\mat{c}{\mt^-_+\\ \mt^-_-}=\mat{c}{q^{-\f{N_b}{4}}a\\ q^{\f{N_a}{4}}b }\,,
 &\quad &
|\tau] = \mat{c}{- e^{\f{\ka N_0}{4}}\bzeta^\ka_1 \\ e^{-\f{\ka N_1}{4}}\bzeta_0^\ka}
&\longrightarrow &
\mt^+ =\mat{c}{\mt^+_+\\ \mt^+_-} =\mat{c}{-b^\dagger q^{\f{N_a+1}{4}} \\ a^\dagger q^{-\f{N_b+1}{4}}} \,,\\[0.4cm]
|\tt\ra =\mat{c}{e^{\f{\ka \Nt_1}{4}}\tzeta^\ka_0 \\ e^{-\f{\ka \Nt_0}{4}}\tzeta^\ka_1}
&\longrightarrow &
\bftt^-=\mat{c}{\bftt^-_+\\ \bftt^-_-}=\mat{c}{q^{\f{\Nt_b}{4}}\at \\ q^{-\f{\Nt_a}{4}}\bt}\,,
 &\quad &
|\tt]=\mat{c}{-e^{-\f{\ka \Nt_0}{4}}\btzeta^\ka_1 \\ e^{\f{\ka \Nt_1}{4}}\btzeta^\ka_0}
&\longrightarrow &
\bftt^+=\mat{c}{\bftt^+_+\\ \bftt^+_-}=\mat{c}{-\bt^\dagger q^{-\f{\Nt_a+1}{4}}\\ \at^\dagger q^{\f{\Nt_b+1}{4}}}\,,\\[0.4cm]
|\ttau\ra =\mat{c}{e^{-\f{\ka \Nt_1}{4}}\tzeta^\ka_0 \\ e^{\f{\ka \Nt_0}{4}}\tzeta^\ka_1}
&\longrightarrow &
\mtt^-=\mat{c}{\mtt^-_+\\\mtt^-_-}=\mat{c}{q^{-\f{\Nt_b}{4}}\at\\ q^{\f{\Nt_a}{4}}\bt }\,,
 &\quad &
|\ttau]=\mat{c}{ -e^{\f{\ka \Nt_0}{4}}\btzeta^\ka_1 \\ e^{-\f{\ka \Nt_1}{4}}\btzeta^\ka_0 }
&\longrightarrow &
\mtt^+ =\mat{c}{\mtt^+_+\\ \mtt^+_-}=\mat{c}{-\bt^\dagger q^{\f{\Nt_a+1}{4}} \\ \at^\dagger q^{-\f{\Nt_b+1}{4}}}\,.
\ea
\label{eq:def_quantum_spinor}
\ee
These objects are in fact really spinors for some $\UQ$ actions: $\bft^\epsilon$ and $\bftt^\epsilon$ are spinors under the $\UQ$ adjoint right action, while $\mt^\epsilon$ and $\mtt^\epsilon$ are spinors under the $\UQI$ adjoint right action \cite{Bonzom:2022bpv}.

Notice that the map $q\to q^{-1}$ exchanges $\bft^\epsilon$ with $\mt^\epsilon$ and $\bftt^\epsilon$ with $\mtt^\epsilon$ (the operators $a, b, \at, \bt$ are invariant under $q\to q^{-1}$, \eqref{eq:a_adagger}). 

%%%%%%%%%%%%%%%%%%%%%%%%%%%%%%%%%%%%%%%%%%%%%%%%%%%%%%
\subsection{Kinematical Hilbert space}

The kinematical state space was defined in \cite{Bonzom:2022bpv}. We here describe the corresponding spin network basis. Each edge $e$ of $\Gamma$ carries an irreducible representation $V_{j_e}$ of $\UQ$, characterized by its spin $j_e\in\N/2$. The Gauss constraint then projects the tensor product of the incident representations at each vertex onto the invariant subspace.

We consider the usual magnetic basis $\{|j,m\rangle\}_{m=-j,\dotsc, j}$ on each $V_j$, which diagonalizes $K$, i.e. $K|j,m\rangle = q^{\frac{m}{2}}|j, m\rangle$. The $q$-bosons act on $V_j$ as
\be\ba{ll}
a^{\dagger} \jm =\sqrt{[j+m+1]}\,|j+\f12,m+\f12\ra\,, &
a \jm =\sqrt{[j+m]}\,|j-\f12,m-\f12\ra\,, \\[0.15cm]
b^{\dagger} \jm = \sqrt{[j-m+1]}\, |j+\f12,m-\f12\ra\,, &
b \jm = \sqrt{[j-m]}\, |j-\f12,m+\f12\ra\,, \\[0.15cm]
N_a \jm = (j+m)\,\jm\,, &
N_b \jm =(j-m) \,\jm\,.
\ea
\label{eq:boson_on_basis}
\ee
In particular $a^\dagger, b^\dagger$ ($a, b$) map $V_j$ to $V_{j+1/2}$ (to $V_{j-1/2}$). The tilde $q$-bosons $\at,\at^{\dagger},\bt,\bt^\dagger$ act on these basis in the same way by definition. It leads to the Wigner-Eckart theorem for the quantum spinors \eqref{eq:def_quantum_spinor}:
\begin{subequations} \label{WignerEckart}
\begin{alignat}{4}
&\jml \bft^\epsilon_m \jmr =  
\delta_{j_{1}, j_2+\epsilon/2}\, \sqrt{[d_{j_1}]}\,
_{q}C^{j_1\,\, \,\,\,\f12\, \,\,\,\,j_2}_{m_1 \,-m \,m_2}
\,, 
\label{eq:matrix_t}
\\
&\jml  \mt^\epsilon_m \jmr =  
\delta_{j_{1}, j_2+\epsilon/2} \,\sqrt{[d_{j_1}]}\, 
_{q^{-1}}C^{j_1\,\, \,\,\,\f12\, \,\,\,\,j_2}_{m_1 \,-m \,m_2}
\,,   
\label{eq:matrix_tau}
\\
&\jml  \bftt^\epsilon_m \jmr =   
\delta_{j_{1}, j_2+\epsilon/2}\, \sqrt{[d_{j_1}]}\,
_{q}C^{j_1\,\, \,\,\,\f12\, \,\,\,\,j_2}_{m_1 \,-m \,m_2}
\,,   
\label{eq:matrix_tt}\\
&\jml  \mtt^\epsilon_m \jmr =   
\delta_{j_{1}, j_2+\epsilon/2} \,\sqrt{[d_{j_1}]}\, 
_{q^{-1}}C^{j_1\,\, \,\,\,\f12\, \,\,\,\,j_2}_{m_1 \,-m \,m_2}
\,,
\label{eq:matrix_ttau}
\end{alignat}
\end{subequations}
where ${}_qC^{j_1}_{m_1}{}^{j_2}_{m_2}{}^{j_3}_{m_3}$ is the Clebsch-Gordan coefficient for $\UQ$. 

Before enforcing the Gauss constraints, the space of states is $\bigoplus_{\{j_e\}} \bigotimes_e V_{j_e}\otimes V_{j_e}^*$, where $V_{j_e}$ is associated to the target end of $e$ and $V^*_{j_e}$ to its source. At each vertex $v$, the Gauss constraint enforces a projection of the tensor product of the vectors meeting at $v$ onto the trivial representation. If the edges meeting at $v$ are denoted $e_1, \dotsc, e_n$, we further denote $\operatorname{Inv}(j_{e_1 v} \otimes \dotsb \otimes j_{e_n v})$ the space of \emph{intertwiners}, \ie the invariant subspace of the tensor product $V_{j_{e_1}}\otimes \dotsb \otimes V_{j_{e_n}}$ if all $e_i$s are incoming at $v$, and we dualize to $V_{j_{e_i}}^*$ if $e_i$ is outgoing at $v$. Therefore, the kinematical Hilbert space is given by 
\be 
\mathcal{H}_{\kin} = \bigoplus_{\{j_e\in \N/2\}} \bigotimes_v \operatorname{Inv}(j_{e_1 v} \otimes \dotsb \otimes j_{e_n v})\,.
\label{eq:Hkin}
\ee 
A basis is obtained at fixed spins $\{j_e\}$ by specifying a basis of $\operatorname{Inv}(j_{e_1 v} \otimes \dotsb \otimes j_{e_n v})$ for all $v$. We denote an element of this space as $i_{j_{e_1 v} \dotsb  j_{e_n(v) v}}$ (the letter $i$ referring to intertwiner). A kinematical state $|\psi\rangle$ thus admits the expansion
\begin{equation}
|\psi\rangle = \sum_{\{j_e\}} \sum_{\{i_v\}} \psi(\{j_e, i_v\}) \ |\{j_e, i_v\}\rangle,
\end{equation}
with 
\begin{equation}
|\{j_e, i_v\}\rangle = \bigotimes_v i_{j_{e_1 v} \dotsb  j_{e_n(v) v}} 
\end{equation}
where the sum over each $i_v$ runs over a basis of the invariant space at $v$. The state $|\{j_e, i_v\}\rangle$ is called a \emph{spin network state}. They form a basis of $\mathcal{H}_{\kin}$. In the case of trivalent vertices, the invariant space $\operatorname{Inv}(j_{e_1 v} \otimes j_{e_2 v} \otimes j_{e_3 v})$ is one-dimensional.  
This is the case we are most interested in and will be considered when constructing the quantum Hamiltonian. Let us now give more details on the intertwiners in this case.

{\bf The $q$-intertwiner for a three-valent vertex. }
As shown in \cite{Bonzom:2014bua}, the order of the $\UQ$-invariant spaces on different nodes is irrelevant (only the linear order of the links incident to each node matters). 
In the basis $|j_1, m_1\rangle \otimes |j_2, m_2\rangle \otimes |j_3, m_3\rangle$ which diagonalizes the $\UQ$ generator $K$ for each particle, the components of the state (up to normalization) are the $q$-deformed Clebsch-Gordan coefficients \cite{Bonzom:2014bua,Bonzom:2022bpv}. Explicitly, the intertwiner for a vertex to which three incoming edges incident to reads
\be
i_{j_1j_2j_3} = \sum_{m_i} \f{(-1)^{j_3+m_3}}{\sqrt{[d_{j_3}]}} q^{-\f{m_3}{2}} 
 {}_qC^{\,j_1 \,\,\,\,j_2\,\,\, \,\,j_3}_{m_1\, m_2 \,-m_3} 
|j_1, m_1\ra \otimes |j_2, m_2\ra \otimes |j_3, m_3\ra\,.
\label{eq:intertwiner_iii}
\ee
It solves the quantum Gauss constraint equation $\hat{\cG}i_{j_1j_2j_3} := \cop^{(2)}\lth \,i_{j_1j_2j_3}\equiv \lth\otimes\lth\otimes\lth i_{j_1j_2j_3}=i_{j_1j_2j_3}$. Changing the orientation of each edge, say $e_i$, leads to the flipping of the vector space $V_{j_i}$ to the dual vector space $V^*_{j_i}$. To write down the expression of the corresponding intertwiner, we make use of the $\UQ$-invariant bilinear form, $\cB_q:V^j\otimes V^j\rightarrow \bC$, which is defined with the $q$-WCG coefficient projected on the trivial representation \cite{Biedenharn:1996vv}. Explicitly, for two given vectors $w=\sum_m w_m |j,m\ra, r=\sum_n r_n | j,n\ra \in V^j$,

\be
\cB_q(w,r) = \sum_{m} \,_qC^{\,\,\,j\,\,\,\,\,j\,\,\,0}_{-m\,m\,0}\, w_{-m} r_{m} = \sum_m \, (-1)^{j+m}q^{-\f{m}{2}} w_{-m} r_{m}\,.
\label{eq:bilinear}
\ee 
One can thus define the dual vector $w^*$ of $w$ as 
\be
w^* \equiv\sum_m\la j,m| w^*_m := \sum_m \la j,m | q^{-\f{m}{2}} (-1)^{j+m} w_{-m}\quad
\Longrightarrow\quad
w^*_m=q^{-\f{m}{2}} (-1)^{j+m} w_{-m}\,.
\label{eq:u_u*}
\ee
Apparently, this dual operation is not an involution.
\footnote{One can also define the dual vector with the $\UQI$-invariant bilinear form $\Bqi$, that is to replace $q$ with $q^{-1}$ in \eqref{eq:u_u*}. }

When edge $e_2$ is outgoing and $e_1,e_3$ incoming, one needs to dualize the vector on $e_2$, that is to change ${}_qC^{\,j_1 \,\,\,\,j_2\,\,\, \,\,j_3}_{m_1\, m_2 \,-m_3} |j_2,m_2\ra \rightarrow \,_qC^{\,j_1 \,\,\,\,\,\,j_2\,\,\,\, \,\,\,j_3}_{m_1\, -m_2 \,-m_3}\la j_2,m_2|$ and add $(-1)^{j_2+m_2}q^{-\f{m_2}{2}}$ according to \eqref{eq:u_u*}. Thus the correspondent intertwiner is
\be
i_{j_1j_2^*j_3} = \sum_{m_i} \f{(-1)^{j_3+m_3}}{\sqrt{[d_{j_3}]}} q^{-\f{m_3+m_2}{2}} (-1)^{j_2+m_2}
{}_qC^{\,j_1 \,\,\,\,\,\,j_2\,\,\,\,\,\,\,j_3}_{m_1\, -m_2 \,-m_3} 
|j_1, m_1\ra \otimes \la j_2, m_2| \otimes |j_3, m_3\ra\,,
\label{eq:intertwiner_ioi}
\ee
which can be checked to be the eigenstate for the quantum Gauss constraint $\hat{\cG}=\lth\otimes \lh\otimes \lth$.

When edge $e_1$ is outgoing and $e_2,e_3$ incoming, the intertwiner is obtained using the same dualization as in \eqref{eq:intertwiner_ioi} but for $j_1$ and $m_1$. 
The last case of keeping the orientation of $e_3$ unchanged is to switch both $e_1$ and $e_2$ to be outgoing, then the same dualization should be applied to both $(j_1,m_1)$ and $(j_2,m_2)$.

What needs special care is when one switches the orientation of $e_3$, $\ie$when $e_3$ is outgoing and $e_1,e_2$ incoming. In this case, one needs to dualize the vector on $e_3$ with a different rule.
This is because the $q$-WCG coefficient $\,_qC^{j_1\,\,\,j_2\,\,\,\,j_3}_{m_1\,m_2\,m_3}=\la j_1,m_1;j_2,m_2|(j_1j_2)j_3,m_3\ra$ can be viewed as the coefficient $w_{m_1}$ (\resp $w_{m_2}$) of a vector in $V^{j_1}$ (\resp $V^{j_2}$) or the coefficient $w^*_{m_3}$ of a dual vector in $V^{j_3\,*}$ in the sense of the decomposition \eqref{eq:u_u*}. 

Note that the factor $(-1)^{j_3+m_3}q^{-\f{m_3}{2}}$ in $i_{j_1j_2j_3}$ is 
the transformation factor from the coefficient $w_{m}$ of a vector $w$ to the coefficient $w^*_m$ of a dual vector $w^*$ as shown in \eqref{eq:u_u*}, thus one needs to change $(-1)^{j_3+m_3}q^{-\f{m_3}{2}}\,_qC^{\,j_1 \,\,\,\,j_2\,\,\, \,\,j_3}_{m_1\, m_2 \,-m_3} |j_3,m_3\ra \rightarrow(-1)^{j_3-m_3}q^{\f{m_3}{2}}\,_qC^{\,j_1 \,\,\,\,j_2\,\, \,\,j_3}_{m_1\, m_2 \,\,m_3} \la j_3,m_3|$ and add $(-1)^{j_3-m_3}q^{-\f{m_3}{2}}$ which is the factor of the inverse transformation of $w^*$. This leads to the intertwiner 
\be
i_{j_1j_2j_3^*} = \sum_{m_i} \f{1}{\sqrt{[d_{j_3}]}}  
{}_qC^{\,j_1 \,\,\,\,j_2\,\,\,j_3}_{m_1\, m_2 \,m_3} 
|j_1, m_1\ra \otimes | j_2, m_2\ra \otimes \la j_3, m_3|\,,
\label{eq:interetwiner_iio}
\ee
which is exactly the eigenstate for the quantum Gauss constraint $\hat{\cG}=\lth\otimes \lth \otimes \lh$.
\eqref{eq:interetwiner_iio} can also be used to define the $q$-WCG coefficient
\be
_qC^{j_1\,\,\,j_2\,\,\,\,j_3}_{m_1\,m_2\,m_3}:=\la j_1,m_1|\otimes\la j_2,m_2|i_{j_1j_2j_3^*}|j_3,m_3\ra\,.
\ee
Indeed, when we change the orientation of $e_3$ again, we recovers the original intertwiner $i_{j_1j_2j_3}$ by adding the regular factor $(-1)^{j_3+m_3}q^{-\f{m_3}{2}}$ as in obtaining $i_{j_1j_2^*j_3}$ from $i_{j_1j_2j_3}$.

Given the explicit expressions of the intertwiners for a three-valent vertex, our goal now is to construct the scalar operators in terms of the quantum spinors \eqref{eq:def_quantum_spinor} which act on the intertwiner in a uniform way regardless of the orientations of all the incident edges. This will largely simplify the construction of the quantum Hamiltonian constraint since we do not need to consider different orientations of relevant edges separately\footnote{However, in \cite{Bonzom:2022bpv}, we define the scalar operators differently so that the algebras they form have the same expression. The different forms of the scalar operators in this paper and in \cite{Bonzom:2022bpv} should be viewed as the same object represented in different bases. }.

%%%%%%%%%%%%%%%%%%%%%%%%%%%%%%%%%%%%%%%%%%%%%%%%%%%%%%
\subsection{Scalar operators}

We proceed to the quantization of the quadratic invariant \eqref{eq:scalar_product}, $E^{\epsilon_2, \epsilon_1}_{e_2 e_1}$. The quantization of the spinors $t_{e_1 v}, t_{e_2 v}$ themselves is given by \eqref{eq:def_quantum_spinor}. As can be seen from \eqref{WignerEckart}, the operators $\bft^\epsilon, \mt^\epsilon, \bftt^\epsilon, \mtt^\epsilon$ transform as spinors under $\UQ$ or $\UQI$. Therefore to ensure that $E^{\epsilon_2, \epsilon_1}_{e_2 e_1}$ is quantized as a quantum group invariant, one needs to contract the two spinor operators via some $q$-Clebsh-Gordan coefficients,
$_qC^{\f12\,\f12\,0}_{A\,B\,0} = (-1)^{\f12-A}q^{\f{A}{2}} \delta_{B, -A}$ or $_{q^{-1}}C^{\f12\,\f12\,0}_{A\,B\,0} = (-1)^{\f12-A}q^{-\f{A}{2}} \delta_{B, -A}$.
 
Since changing the orientation of an edge exchanges $\bft^\epsilon$ with $\bftt^\epsilon$, and $\mt^\epsilon$ with $\mtt^\epsilon$, and since $\bft^\epsilon$ and $\bftt^\epsilon$ are in fact the same operator (and also $\mt^\epsilon$ and $\mtt^\epsilon$), one would expect the quantum operator for $E^{\epsilon_2, \epsilon_1}_{e_2 e_1}$ to be independent of the orientations of $e_1$ and $e_2$. This is entirely possible to proceed this way. 

We will however not do so. Our motivation is that while $E^{\epsilon_2, \epsilon_1}_{e_2 e_1}$ would be independent of orientations, the vector space on which it acts does depend on orientations ($V_j^*$ versus $V_j$). Therefore the action on an intertwiner would in fact depend explicitly on the orientations. Instead, we decide to perform the quantization so that its action on intertwiners is independent of orientations.

This requires changing the spinor operator to its $q^{-1}$ version when flipping the orientation. Obviously, this exchanges the $\mathbf{t}$s with the $
\boldsymbol{\uptau}$s. However, we prefer to keep the same letter for the spinor operator because we think exchanging $\mathbf{t}$s with $
\boldsymbol{\uptau}$s could be confusing in the ribbon picture. We therefore define $\overline{{\bft}^{\epsilon}} := \mt^{\epsilon}$ and same with the tildes, and eventually\footnote{
Note that the definition of $o_1$ is opposite to that in \cite{Bonzom:2022bpv} which leads to a slight difference for the definition of the scalar operator \eqref{eq:def_E12} compared to that in \cite{Bonzom:2022bpv}. This is because, in \cite{Bonzom:2022bpv}, $o_1$ and $o_2$ are considered to be the orientation of edges relative to vertex $v$ and $+1$ (\resp $-1$) denotes outgoing (\resp incoming). Here, in contrast, $o_1$ and $o_2$ are considered to be the orientation of edges relative to the orientation of the face $f$. As an example, when edges $e_2$ and $e_1$ are both outgoing to $v$, $e_2$ is counter-clockwise while $e_1$ is clockwise relative to $f$.
}
\begin{multline}
\E^{\epsilon_2,\epsilon_1}_{e_2e_1} 
= -o_1\sqrt{[2]} (-1)^{\f{1-o_1}{2}\f{1+\epsilon_1}{2}} (-1)^{\f{1-o_2}{2}\f{1+\epsilon_2}{2}} 
\sum_{A=\pm\f12}
\,_{q^{o_1}}C^{\f12\,\,\,\,\f12\,\,0}_{A\,-A\,0}\,
{\bf T}_{e_1v,A}^{o_1\epsilon_1} \otimes {\bf T}_{e_2v,-A}^{-o_2\epsilon_2} \\
=
\begin{cases}
\sum_{A=\pm\f12}(-1)^{\f12-A}q^{\f{A}{2}} \epsilon_2\,
\bftt_A^{\epsilon_{1}}\otimes \mtt_{-A}^{\epsilon_{2}}
&\text{ for }\, -o_{1}=o_{2}=-1 \\[0.2cm]
\sum_{A=\pm\f12}(-1)^{\f12+A}q^{\f{A}{2}} 
\bftt_A^{\epsilon_{1}}\otimes \overline{\mt_{-A}^{-\epsilon_2}} 
& \text{ for }\,o_{1}=o_{2}=1 \\[0.2cm]
\sum_{A=\pm\f12}(-1)^{\f12-A}q^{-\f{A}{2}} \epsilon_{1}\epsilon_2\,
\overline{\bft_{A}^{-\epsilon_{1}}} \otimes \mtt_{-A}^{\epsilon_2} 
& \text{ for }\,-o_{1}=-o_{2}=1 \\[0.2cm]
\sum_{A=\pm\f12}(-1)^{\f12+A}q^{-\f{A}{2}}\epsilon_{1}\,
\overline{ \bft_{A}^{-\epsilon_{1}}} \otimes \overline{ \mt_{-A}^{-\epsilon_2} } 
& \text{ for }\,-o_{1}=o_{2}=1
\end{cases} \,,
\label{eq:def_E12}
\end{multline}
where ${\bf T}_{e_2v,A}^{-o_2\epsilon_2}={\bft}_{e_2v,A}^{\epsilon_2}$ if $o_2=-1$ while ${\bf T}_{e_2v,A}^{-o_2\epsilon_2}=\overline{{\bft}_{e_2v,A}^{-\epsilon_2}}$ if $o_2=1$, and similarly for ${\bf T}_{e_{1}v,A}^{o_{1}\epsilon_{1}}$. We then extend this definition to the space $\operatorname{Inv}(j_{e_1 v} \otimes \dotsb \otimes j_{e_n v})$ of invariant vectors at $v$ by tensoring with the identity as necessary. It comes
\begin{equation}
\E^{\epsilon_2,\epsilon_1}_{e_2e_1} \, i_{j_1 j_2 k} 
= \sqrt{[d_{j_1}][d_{j_2}][d_{l_1}][d_{l_2}]} \delta_{l_1,j_1+\f{\epsilon_1}{2}}\delta_{l_2,j_2+\f{\epsilon_2}{2}} 
\Mat{ccc}{l_1 & j_1 &\f12 \\ j_2 & l_2 & k}_q (-1)^{l_1+l_2+k} i_{l_1l_2k}\,.
\label{eq:E12_on_intertwiner}
\end{equation}
It thus maps the intertwiner space $\operatorname{Inv}(j_1\otimes j_2\otimes k)$ to $\operatorname{Inv}(l_1\otimes l_2\otimes k)$.

This definition also works for two edges $e_i, e_{i+1}$ sharing a corner in $\operatorname{Inv}(j_{e_1 v} \otimes \dotsb \otimes j_{e_n v})$, for $i=1, \dotsc, n-1$. In the trivalent case, this gives $E_{e_3 e_2}^{\epsilon_3, \epsilon_2} i_{j_1 j_2 j_3}$ exactly as in \eqref{eq:E12_on_intertwiner} with $e_1\to e_2, e_2\to e_3, e_3\to e_1$
\begin{equation}
\E^{\epsilon_3,\epsilon_2}_{e_3e_2} \, i_{j_1 j_2 j_3} 
= \sqrt{[d_{j_2}][d_{j_3}][d_{l_2}][d_{l_3}]} \delta_{l_2,j_2+\f{\epsilon_2}{2}}\delta_{l_3,j_3+\f{\epsilon_3}{2}} 
\Mat{ccc}{l_2 & j_2 &\f12 \\ j_3 & l_3 & j_1}_q (-1)^{l_2+l_3+j_1} i_{j_1 l_2 l_3}.
\end{equation}
For the case $i=n$, \ie $E^{\epsilon_1, \epsilon_n}_{e_1 e_n}$, the definition has to be amended to obtain an invariant operator \cite{Bonzom:2022bpv} and eventually one finds the same expression for $E^{\epsilon_1, \epsilon_3}_{e_1 e_3} i_{j_1 j_2 j_3}$ as \eqref{eq:E12_on_intertwiner} with the appropriate permutation of the indices, \ie
\begin{equation}
\E^{\epsilon_1,\epsilon_3}_{e_1e_3} \, i_{j_1 j_2 j_3} 
= \sqrt{[d_{j_3}][d_{j_1}][d_{l_3}][d_{l_1}]} \delta_{l_3,j_3+\f{\epsilon_3}{2}}\delta_{l_1,j_1+\f{\epsilon_1}{2}} 
\Mat{ccc}{l_3 & j_3 &\f12 \\ j_1 & l_1 & j_2}_q (-1)^{l_3+l_1+j_2} i_{l_1 j_2 l_3}.
\end{equation}

%%%%%%%%%%%%% 
\subsection{Quantum Hamiltonian constraint}

%%%%%%%%%%%%%%%%%

We now need to quantize the classical Hamiltonian \eqref{eq:ClassicalHamiltonian} as a well-defined operator on $\mathcal{H}_{\kin}$ (defined in \eqref{eq:Hkin}). The first step is obviously to use the quantization map described in the previous section to turn the observables $E_{e_i e_{i-1}}^{\epsilon_i, \epsilon_{i-1}} $ into operators $\E_{e_i e_{i-1}}^{\epsilon_i, \epsilon_{i-1}}$. The second step is concerned with quantization ambiguities. Indeed, factors $N_{e_i}$ appear in \eqref{eq:ClassicalHamiltonian} and they are expected to be diagonal on the spin network basis, as a function of $j_{e_i}$ only in fact. Notice however that the operator $E^{\epsilon_2, \epsilon_1}_{e_2 e_1}$ changes the spins of the edges $e_1, e_2$ by $\epsilon_1/2$ and $\epsilon_2/2$. There are therefore ordering ambiguities. The results differ according to whether $N_{e_i}$ is before or after some operators $E$ which changes $j_{e_i}$. We found an ordering, see below, which ultimately leads to a topological model, which would presumably not be true for other orderings.

Let us introduce
\be
\mathbf{h}_{f, e_1, e_p}^{\epsilon_1, \epsilon_p} 
= \f{1}{\mathbf{N}_{e_1 v_2}}\,\lb \sum_{\epsilon_2, \dotsc, \epsilon_{p-1} = \pm}  
 \prod_{i=2}^{p}  \E_{e_i e_{i-1}}^{\epsilon_i, \epsilon_{i-1}} 
 \f{o_{i} \epsilon_i}{\mathbf{N}_{e_iv_i}} \rb
+(-1)^{d-p} \epsilon_1 \epsilon_p 
 \f{1}{\mathbf{N}_{e_p v_{p+1}}}\,\lb \sum_{\epsilon_{p+1}, \dotsc, \epsilon_d = \pm} 
\prod_{i=p+1}^{d+1} \E_{e_i e_{i-1}}^{-\epsilon_i, -\epsilon_{i-1}}  \f{o_{i} \epsilon_i}{\mathbf{N}_{e_iv_i}}\rb\,,
\label{eq:QuantumHamiltonian}
\ee
where $\mathbf{N}_{e_i v_i}$ is diagonal on $V_{j_{e_i}}$ (or its dual), $\mathbf{N}_{e_iv_i}|j_{e_i},m_{e_i}\ra = [d_{j_{e_i}}]|j_{e_i},m_{e_i}\ra$. We include the vertex $v_i$ in the notation because here $\mathbf{N}_{e_i v_i}$ only acts on the space of intertwiners at $v_i$, where $e_i$ and $e_{i-1}$ meet. As already discussed, the ordering is important because $[\mathbf{N}_{e_iv_i}, E^{\epsilon_{i-1}, \epsilon_i}_{e_{i-1}e_i}]\neq 0$. However $[\mathbf{N}_{e_iv_i}, E^{\epsilon_{i}, \epsilon_{i+1}}_{e_{i}e_{i+1}}]= 0$ by definition, so that the operators $\E_{e_i e_{i-1}}^{\epsilon_i, \epsilon_{i-1}} \f{o_{i} \epsilon_i}{\mathbf{N}_{e_iv_i}}$ which act on the space of intertwiners at $v_i$ commute with one another. Here $\mathbf{N}_{e_i v_i}$ is placed to the right of $\E_{e_i e_{i-1}}^{\epsilon_i,\epsilon_{i-1}}$, which is also the case if one reconstructs the quantum holonomies ($\ie$the quantization of $u$ and $\ut^{-1}$) from the quantum spinors \cite{Bonzom:2022bpv}.

However, the operator $\mathbf{h}_{f, e_1, e_p}^{\epsilon_1, \epsilon_p}$ as such is not defined on $\mathcal{H}_{\kin}$. Indeed, a state in $\mathcal{H}_{\kin}$ is a superposition of spin network states which assigns a spin to each edge along with the space $V_{j}$ to the target end and $V^*_j$ to the source end. Say the edge $e_1$ gets the spin $j_1$. Then the first term of the above operator acts on $e_1$ with $\E_{e_2 e_1}^{\epsilon_2,\epsilon_1}$ which shifts the spin $j_1$ to $j_1+\epsilon_1/2$, on the intertwiner which sits at the vertex where $e_1$ and $e_2$ meet. It thus maps $V_{j_1}$ to $V_{j_1+\epsilon_1/2}$, or $V_{j_1}^*$ to $V_{j_1+\epsilon_1/2}^*$ depending on orientations, but not both, i.e. it does not shift $j_1$ at the vertex where $e_d$ and $e_1$ meet. Therefore the operator brings the state out of $\mathcal{H}_{\kin}$. 

Similarly, the second term of $\mathbf{h}_{f,e_1,e_p}^{\epsilon_1,\epsilon_p}$ acts on $e_1$ through $\E_{e_1 e_d}^{-\epsilon_1,-\epsilon_d}$. This shifts $j_1$ to $j_1-\epsilon_1/2$ at the vertex where $e_d$ and $e_1$ meet. If $\E_{e_2 e_1}^{\epsilon_2,\epsilon_1}$ in the first term acted on $V_{j_1}$, then this operator acts on $V^*_{j_1}$ (or the other way around).

We thus turn $\mathbf{h}_{f,e_1,e_p}^{\epsilon_1,\epsilon_p}$ into a well-defined operator on $\mathcal{H}_{\kin}$ by multiplying it by a product of operators $\E_{e_i e_{i-1}}^{\epsilon_i, \epsilon_{i-1}}$ so that the intertwiners of both ends of the same edge have the same spin. 
Notice that the first term in \eqref{eq:QuantumHamiltonian} only contains the shift operators for $i=2,\cdots,p$, one can add $\E_{e_i e_{i-1}}^{\epsilon_i, \epsilon_{i-1}}$ for all the remaining vertices, \ie $i=p+1,\cdots,d+1$, so that the change of spins for both ends of each edge are the same. For the second term in \eqref{eq:QuantumHamiltonian}, adding these shift operators also shift all the spins $j_i-\epsilon_i/2$ to $j_i$ thus drags the state back in $\cH_{\kin}$. This is the method which was already used in \cite{Bonzom:2011nv} to construct the quantum Hamiltonian in the spinor representation in the flat case.

\begin{definition}
We define the quantum Hamiltonian on the face $f$, labelled by the pair of edges $(e_1, e_p)$, to be
\be
\H_{f,e_1,e_p}^{\epsilon_1,\epsilon_p,\epsilon_{p+1},\dotsc,\epsilon_d} = \biggl[\prod_{i=p+1}^{d+1} \E_{e_i e_{i-1}}^{\epsilon_i, \epsilon_{i-1}}\biggr] \ \mathbf{h}_{f,e_1,e_p}^{\epsilon_1,\epsilon_p}.
\label{eq:Hamiltonian}
\ee
\end{definition}
Compared to the operator \eqref{eq:QuantumHamiltonian}, the quantum Hamiltonian defined as such not only depends on $\epsilon_1$ and $\epsilon_p$, but also $\epsilon_{p+1},\cdots,\epsilon_{d+1}$. The physical Hilbert space is spanned by the physical states which are solutions to the quantum Hamiltonian. In the spin representation, the coefficients of these physical spin network states satisfy a set of difference equations, which is stated in the following theorem. 
 
\begin{theorem} \label{thm:QuantumConstraint}
The constraint
\begin{equation}
\forall k_e\qquad \langle \{k_e\}|\ \H_{f,e_1,e_p}^{\epsilon_1,\epsilon_p,\epsilon_{p+1},\dotsc,\epsilon_d}\ |\psi\rangle = 0,
\label{eq:quantum_constraint_formal}
\end{equation}
is equivalent to the following set of \emph{difference equations} on the spin network coefficients 
$\psi(k_1,k_2,\cdots,k_d,\{k_e\}_{e\notin \partial f})$
of $|\psi\rangle$,
\begin{multline} \label{QuantumConstraint}
\sum_{\tilde{\epsilon}_2, \dotsc, \tilde{\epsilon}_{p-1} = \pm} \biggl(\prod_{i=2}^{p} A^{\tilde{\epsilon}_i,\tilde{\epsilon}_{i-1}}_{o_{i}}(k_i, k_{i-1},l_i)\biggr)\ \psi(k_1-\frac{\epsilon_1}{2}, k_2-\frac{\tilde{\epsilon}_2}{2}, \dotsc, k_{p-1} - \frac{\tilde{\epsilon}_{p-1}}{2}, k_p-\frac{\epsilon_p}{2},\dotsc, k_d - \frac{\epsilon_d}{2}, \{k_e\}_{e\not\in\partial f}) \\
+ (-1)^{d-p}  \alpha^{\epsilon_1,\epsilon_p}(k_1,k_p) 
 \sum_{\tilde{\epsilon}_{p+1}, \dotsc, \tilde{\epsilon}_d = \pm} \biggl(\prod_{i=p+1}^{d+1} B^{\tilde{\epsilon}_i,\tilde{\epsilon}_{i-1}}_{o_{i}}(k_i - \frac{\epsilon_i}{2}, k_{i-1}-\frac{\epsilon_{i-1}}{2}, l_i)\biggr)\\
 \psi(k_1, \dotsc, k_p, k_{p+1}-\frac{\epsilon_{p+1}}{2}+\frac{\tilde{\epsilon}_{p+1}}{2}, \dotsc, k_d-\frac{\epsilon_d}{2}+\frac{\tilde{\epsilon}_d}{2}, \{k_e\}_{e\not\in\partial f}) = 0.
\end{multline}
Here
\begin{itemize}
\item $l_1, \dotsc, l_d$ are the spins carried by the edges $e'_1, \dotsc, e'_d$ incident to $f$, see Figure \ref{fig:sunny}.
\item By definition, $\tilde{\epsilon}_1=\epsilon_1, \tilde{\epsilon}_p = \epsilon_p$, while $\epsilon_{p+1}, \dotsc, \epsilon_{d}$ are fixed.
\item The coefficients are
\begin{align} \label{ACoeff}
A^{\tilde{\epsilon}_i,\tilde{\epsilon}_{i-1}}_{o_{i}}(k_i, k_{i-1},l_i) 
&=
o_{i} \tilde{\epsilon}_i [d_{k_i}]
 (-1)^{k_i + k_{i-1} + l_i} 
\begin{Bmatrix} 
k_i & k_i-\frac{\tilde{\epsilon}_i}{2} & \frac12 \\
 k_{i-1}-\frac{\tilde{\epsilon}_{i-1}}{2} &k_{i-1} &l_i
  \end{Bmatrix}_q \,,\\
\label{BCoeff}
B^{\tilde{\epsilon}_i,\tilde{\epsilon}_{i-1}}_{o_{i}}(k_i, k_{i-1},l_i) 
&= 
o_{i} \tilde{\epsilon}_i [d_{k_i}] 
 (-1)^{k_i + k_{i-1} + l_i} 
\begin{Bmatrix} 
k_i & k_i+\frac{\tilde{\epsilon}_i}{2}  & \frac12\\
 k_{i-1}+\frac{\tilde{\epsilon}_{i-1}}{2} & k_{i-1}  & l_i
 \end{Bmatrix}_q \,,\\
 \alpha^{\epsilon_1,\epsilon_p}(k_1,k_p) 
 & =  
  \epsilon_1\epsilon_p   \f{[d_{k_p}]}{[d_{k_1-\f{\epsilon_1}{2}}]} 
 \,.
\end{align}
\end{itemize}
\end{theorem}

Those constraints are recursions on the physical states. They generalize the one found in \cite{Bonzom:2011nv} for a triangular face. Improving on \cite{Bonzom:2011hm}, the differences are shifts of the spins by $1/2$ instead of $1$. Moreover, edge orientations are kept arbitrary.

\begin{figure}
\includegraphics[scale=.6]{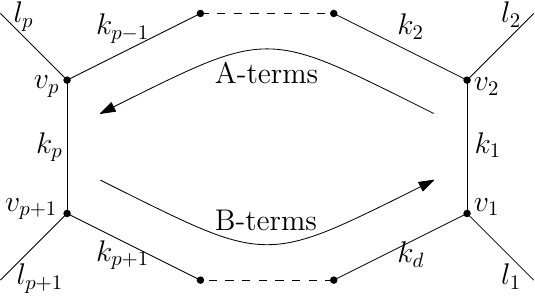}
\caption{\label{fig:QuantumConstraint} The schematic representation of the quantum constraint \eqref{QuantumConstraint} with its $A$-terms and $B$-terms associated to the corners around the face.}
\end{figure}

Those constraints have two types of contributions: the \emph{$A$-terms} and the \emph{$B$-terms}. Notice that $\mathbf{h}_{f, e_1, e_p}^{\epsilon_1, \epsilon_p}$ contains all the operators $\E_{e_i e_{i-1}}^{\tilde{\epsilon}_i, \tilde{\epsilon}_{i-1}}$ exactly once, for $i=1, \dotsc, d$. Whether an operator $\E_{e_i e_{i-1}}^{\tilde{\epsilon}_i, \tilde{\epsilon}_{i-1}}$ gives rise to an $A$-term or a $B$-term depends on the choice of the reference edges $e_1$ and $e_p$ around $f$. It is important that the coefficients $A^{\tilde{\epsilon}_i,\tilde{\epsilon}_{i-1}}_{o_{i}}(k_i, k_{i-1},l_i)$ and $B^{\tilde{\epsilon}_i,\tilde{\epsilon}_{i-1}}_{o_{i}}(k_i, k_{i-1},l_i)$ are \emph{local}: they only depend on the spins incident to the vertex and are determined by the choice of a corner on that vertex. As a consequence, for example, if one considers another constraint on the same face with $e_q$, $q<p$, choosing $e_1$ as reference edge, then the coefficients $A^{\tilde{\epsilon}_i,\tilde{\epsilon}_{i-1}}_{o_{i}}(k_i, k_{i-1},l_i)$ for $i=1, \dotsc, q$ would be the same as those appearing above, and similarly for the $B$-terms. The structure of the constraint is schematically pictured in Figure \ref{fig:QuantumConstraint}.

By exchanging the role of $e_1$ with $e_p$, the $A$-terms become the $B$-terms and  vice versa. The constraint obtained this way is equivalent to \eqref{QuantumConstraint}, as we now show. First, evaluate \eqref{QuantumConstraint} on $k_1 + \epsilon_1/2, \dotsc, k_d + \epsilon_d/2$, and then flip the signs of all the $\epsilon_i$ and $\tilde{\epsilon}_i$. That gives the constraint
\begin{multline}
\sum_{\tilde{\epsilon}_2, \dotsc, \tilde{\epsilon}_{p-1} = \pm} \biggl(\prod_{i=2}^{p} A^{-\tilde{\epsilon}_i, -\tilde{\epsilon}_{i-1}}_{o_{i}}(k_i-\frac{\epsilon_i}{2}, k_{i-1}-\frac{\epsilon_{i-1}}{2},l_i)\biggr)\ \psi\Bigl(k_1, k_2-\frac{\epsilon_2}{2}+\frac{\tilde{\epsilon}_2}{2}, \dotsc, k_{p-1} - \frac{\epsilon_{p-1}}{2} + \frac{\tilde{\epsilon}_{p-1}}{2}, k_p,\dotsc, k_d, \{k_e\}_{e\not\in\partial f}\Bigr)\\
 + (-1)^{d-p} 
 \alpha^{\epsilon_1,\epsilon_p}(k_1-\f{\epsilon_1}{2}, k_p-\f{\epsilon_p}{2}) 
 \sum_{\tilde{\epsilon}_{p+1}, \dotsc, \tilde{\epsilon}_d = \pm} \biggl(\prod_{i=p+1}^{d+1} B^{-\tilde{\epsilon}_i,-\tilde{\epsilon}_{i-1}}_{o_{i}}(k_i, k_{i-1}, l_i)\biggr)\\ \psi\Bigl(k_1-\frac{\epsilon_1}{2}, \dotsc, k_p-\frac{\epsilon_p}{2}, k_{p+1}-\frac{\tilde{\epsilon}_{p+1}}{2}, \dotsc, k_d-\frac{\tilde{\epsilon}_d}{2}, \{k_e\}_{e\not\in\partial f}\Bigr)
 = 0\,.
\end{multline}
We then use the key relation between the coefficients $A$ and $B$,
\be
B^{\tilde{\epsilon}_i,\tilde{\epsilon}_{i-1}}_{o_{i}}(k_i, k_{i-1},l_i) 
= - 
A_{o_i}^{-\tilde{\epsilon}_i, -\tilde{\epsilon}_{i-1}}(k_i, k_{i-1}, l_i)
\ee
to get 
\begin{multline}
(-1)^{p} \sum_{\tilde{\epsilon}_2, \dotsc, \tilde{\epsilon}_{p-1} = \pm} \biggl(\prod_{i=2}^{p} 
B^{\tilde{\epsilon}_i, \tilde{\epsilon}_{i-1}}_{o_{i}}(k_i-\frac{\epsilon_i}{2}, k_{i-1}-\frac{\epsilon_{i-1}}{2},l_i)\biggr)\ \psi\Bigl(k_1, k_2-\frac{\epsilon_2}{2}+\frac{\tilde{\epsilon}_2}{2}, \dotsc, k_{p-1} - \frac{\epsilon_{p-1}}{2} + \frac{\tilde{\epsilon}_{p-1}}{2}, k_p,\dotsc, k_d, \{k_e\}_{e\not\in\partial f}\Bigr)\\
+ \alpha^{\epsilon_1,\epsilon_p}(k_1-\f{\epsilon_1}{2}, k_p-\f{\epsilon_p}{2}) 
 \sum_{\tilde{\epsilon}_{p+1}, \dotsc, \tilde{\epsilon}_d = \pm} \biggl(\prod_{i=p+1}^{d+1} 
 A^{\tilde{\epsilon}_i,\tilde{\epsilon}_{i-1}}_{o_{i}}(k_i, k_{i-1}, l_i)\biggr)\\ \psi\Bigl(k_1-\frac{\epsilon_1}{2}, \dotsc, k_p-\frac{\epsilon_p}{2}, k_{p+1}-\frac{\tilde{\epsilon}_{p+1}}{2}, \dotsc, k_d-\frac{\tilde{\epsilon}_d}{2}, \{k_e\}_{e\not\in\partial f}\Bigr)
 = 0,
\end{multline}
where we recognize the matrix element $\langle \{k_e\}| \H_{f, e_p, e_1}^{\epsilon_1, \epsilon_2, \dotsc, \epsilon_p} |\psi\rangle$ and have shown the equivalence
\begin{equation}
\langle \{k_e\}| \H_{f, e_p, e_1}^{\epsilon_p, \epsilon_1, \epsilon_2, \dotsc, \epsilon_{p-1}} |\psi\rangle = 0 \quad \Leftrightarrow \quad \langle \{k_e\}|\ \H_{f,e_1,e_p}^{\epsilon_1,\epsilon_p,\epsilon_{p+1},\dotsc,\epsilon_d}\ |\psi\rangle = 0.
\end{equation}

{\bf Proof of Theorem \ref{thm:QuantumConstraint}.} 
There are two types of terms in \eqref{eq:Hamiltonian}, whose action on spin network states is now presented. First,
\begin{multline}
\prod_{i=p+1}^{d+1} \E_{e_i e_{i-1}}^{\epsilon_i, \epsilon_{i-1}}
\f{1} {\mathbf{N}_{e_1 v_2}}\lb \prod_{i=2}^{p} \E_{e_i e_{i-1}}^{\epsilon_i, \epsilon_{i-1}}\f{ o_{i} \epsilon_i}{\mathbf{N}_{e_i v_i}} \rb
 |\{j_e\}\ra \\
= \f{1}{[d_{k_1}]}\prod_{i=2}^{p} \f{ o_{i} \epsilon_i}{[d_{j_i}]}
\prod_{i=2}^{d+1} \delta_{k_i, j_i+\frac{\epsilon_i}{2}} 
[d_{k_i}][d_{j_i}]
(-1)^{k_i + k_{i-1} + l_i} 
\begin{Bmatrix} 
k_i & k_i-\frac{\epsilon_i}{2}& \frac12  \\
 k_{i-1}-\frac{\epsilon_{i-1}}{2} &k_{i-1} &l_i
\end{Bmatrix}_q 
|\{k_i\}_{i=1,\dotsc,d},\{j_e\}_{e\not\in \partial f}\ra\,,
\end{multline}
where we have applied the action \eqref{eq:E12_on_intertwiner} of $\E_{e_i e_{i-1}}^{\epsilon_i, \epsilon_{i-1}}$ on the intertwiner $i_{j_{i-1}j_il_i}$ at the vertex where $e_{i-1}\,,e_i$ and $e'_i$ meet for all $i=1,\cdots,d$. Each operator $1/\mathbf{N}_{e_i v_i}$ acts before the shift operator $\E_{e_i e_{i-1}}^{\epsilon_i, \epsilon_{i-1}}$ thus the result picks up a factor $1/[d_{j_i}]$. For $i=1, \dotsc, d$, the spin $j_i$ is shifted to $j_i+\f{\epsilon_i}{2}$ after the action of $\E_{e_i e_{i-1}}^{\epsilon_i, \epsilon_{i-1}}$. The spins $l_i$s of the edges $e'_i$s not on the boundary of the face $f$ remain unchanged. In addition, $\f{1}{\mathbf{N}_{e_1 v_2}}$ acts after $\E_{e_2e_1}^{\epsilon_2,\epsilon_1}$ thus the result picks up the factor $1/[d_{k_1}]$. As each edge is incident to two vertices, the assigned spin shows up in two intertwiners thus the term $\sqrt{[d_{k_i}][d_{j_i}]}$ appears twice in the result, which gives the factor $[d_{k_i}][d_{j_i}]$. The $q$-$6j$ symbols and the sign factors naturally follow from \eqref{eq:E12_on_intertwiner}. 

Secondly,
\begin{multline}
\prod_{i=p+1}^{d+1} 
\E_{e_i e_{i-1}}^{\epsilon_i, \epsilon_{i-1}} 
\f{1}{\mathbf{N}_{e_p v_{p+1}}} \lb \prod_{i=p+1}^{d+1}
\E_{e_i e_{i-1}}^{-\tilde{\epsilon}_i, -\tilde{\epsilon}_{i-1}}
\f{ o_{i} \epsilon_i}{\mathbf{N}_{e_i v_i}} 
\rb  |\{j_e\}\ra \\ 
= \f{1}{[d_{j_p-\f{\tilde{\epsilon}_p}{2}}]} 
\prod_{i=p+1}^{d+1} \f{ o_{i} \epsilon_i}{[d_{j_i}]}
\delta_{k_i, j_i - \frac{\tilde{\epsilon}_i}{2} + \frac{\epsilon_i}{2}} 
(-1)^{k_i - \frac{\epsilon_i}{2} + k_{i-1} - \frac{\epsilon_{i-1}}{2} + l_i} 
(-1)^{k_i + k_{i-1} + l_i}
[d_{j_i-\f{\tilde{\epsilon}_i}{2}}][d_{j_{i-1}-\f{\tilde{\epsilon}_{i-1}}{2}}]
\\
\sqrt{[d_{j_i}][d_{k_i}][d_{j_{i-1}}][d_{k_{i-1}}] } 
\begin{Bmatrix} 
k_i-\frac{\epsilon_i}{2} & j_i & \frac12 \\
 j_{i-1} & k_{i-1}-\frac{\epsilon_{i-1}}{2}  & l_i
\end{Bmatrix}_q
\begin{Bmatrix} 
k_i-\frac{\epsilon_i}{2} & k_i  & \frac12\\ 
k_{i-1} & k_{i-1}-\frac{\epsilon_{i-1}}{2}  & l_i
\end{Bmatrix}_q
\ |\{j_i\}_{i=1,\dotsc,p},\{k_i\}_{i=p+1,\dotsc,d},\{j_e\}_{e\not\in \partial f}\ra \\ 
=\f{[d_{k_p}]}{[d_{k_1}][d_{k_1-\f{\epsilon_1}{2}}]}
\prod_{i=p+1}^{d+1}  o_{i} \epsilon_i
\delta_{k_i, j_i - \frac{\tilde{\epsilon}_i}{2} + \frac{\epsilon_i}{2}} 
(-1)^{k_i - \frac{\epsilon_i}{2} + k_{i-1} - \frac{\epsilon_{i-1}}{2} + l_i} 
(-1)^{k_i + k_{i-1} + l_i}
[d_{k_{i}-\f{\epsilon_{i}}{2}}]^2[d_{k_i}]\\
\begin{Bmatrix} 
k_i-\frac{\epsilon_i}{2} & j_i & \frac12 \\
 j_{i-1} & k_{i-1}-\frac{\epsilon_{i-1}}{2}  & l_i
\end{Bmatrix}_q
\begin{Bmatrix} 
k_i-\frac{\epsilon_i}{2} & k_i  & \frac12\\ 
k_{i-1} & k_{i-1}-\frac{\epsilon_{i-1}}{2}  & l_i
\end{Bmatrix}_q
\ |\{j_i\}_{i=1,\dotsc,p},\{k_i\}_{i=p+1,\dotsc,d},\{j_e\}_{e\not\in \partial f}\ra \,.
\end{multline}
Here, two shift operators act on each site for $i=p+1,\cdots,d+1$ and we denote $k_i = j_i - \frac{\tilde{\epsilon}_i}{2} + \frac{\epsilon_i}{2}$. The first shift operator $\E_{e_i e_{i-1}}^{-\tilde{\epsilon}_i, -\tilde{\epsilon}_{i-1}}$ (in the bracket) acts on the spin network state and shifts $j_i$ and $j_{i-1}$ to $j_i-\tilde{\epsilon}_i/2$ and $j_{i-1}-\tilde{\epsilon}_{i-1}/2$ respectively. 
It also gives the first $q$-$6j$ symbol in the third line and the term $(-1)^{k_i - \frac{\epsilon_i}{2} + k_{i-1} - \frac{\epsilon_{i-1}}{2} + l_i}\sqrt{[d_{j_i}][d_{j_{i-1}}][d_{j_i-\f{\tilde{\epsilon}_i}{2}}][d_{j_{i-1}-\f{\tilde{\epsilon}_{i-1}}{2}}]} $. The result picks up a factor $1/[d_{j_i}]$ by the action of $1/\mathbf{N}_{e_i v_i}$ before the shift operator. In addition, $1/\mathbf{N}_{e_p v_{p+1}}$ acts on the spin network state after $\E_{e_{p+1} e_{p}}^{-\tilde{\epsilon}_{p+1}, -\tilde{\epsilon}_{p}}$ and thus brings a factor $1/[d_{j_p-\f{\tilde{\epsilon}_p}{2}}]$. The action of the second shift operator $\E_{e_i e_{i-1}}^{\epsilon_i, \epsilon_{i-1}}$ shifts the spins $j_i-\tilde{\epsilon}_i/2$ and $j_{i-1}-\tilde{\epsilon}_{i-1}/2$ to $k_i$ and $k_{i-1}$ respectively and brings the second $q$-$6j$ symbol in the third line as well as the term $(-1)^{k_i + k_{i-1} + l_i} \sqrt{[d_{k_i}][d_{k_{i-1}}][d_{k_i-\f{\epsilon_i}{2}}][d_{k_{i-1}-\f{\epsilon_{i-1}}{2}}]}$. Note that the spin $j_1=k_1$ and $j_p=k_p$ are kept unchanged in the result as $\epsilon_1=\tilde{\epsilon}_1$ and $\epsilon_p=\tilde{\epsilon}_p$.
The last equality is the rearrangement of the result. 

Putting them together, using the orthogonality of the spin network states, $\langle \{k_e\} | \{j_e\}\rangle \propto \prod_e \delta_{k_e, j_e}$ and eliminating the common terms $\f{1}{[d_{k_1}]}\prod_{i=p+1}^{d+1}
[d_{k_i-\f{\epsilon_i}{2}}][d_{k_i}] (-1)^{k_i+k_{i-1}+l_i} 
\begin{Bmatrix} 
k_i & k_i-\frac{\epsilon_i}{2} & \frac12 \\
 k_{i-1}-\frac{\epsilon_{i-1}}{2} &k_{i-1} &l_i
  \end{Bmatrix}_q$ leads to the expected difference equations.  
  \qed

The dependence of $\psi$ on the orientations is given by the following lemma.
\begin{lemma} \label{lemma:Orientation}
If $|\psi\rangle$, with spin network coefficients $\psi(\{j_e\})$, satisfies all the constraints \eqref{QuantumConstraint} for given edge orientations $\{o_e\}$, then $(-1)^{2j^*} \psi(\{j_e\})$ satisfies all the constraints on the same graph with reversed orientation $-o_{e^*}$ on the edge $e^*$.
\end{lemma}

{\bf Proof.} Consider the constraint \eqref{QuantumConstraint} on the fixed face $f$. If $e^*\not\in\partial f$, multiplication by $(-1)^{2j^*}$ does not change anything. If $e^*\equiv e_s \in\{e_2,\dotsc,e_{p-1}\}$, then the coefficient $A^{\tilde{\epsilon}_s, \tilde{\epsilon}_{s-1}}_{o_{s}}(k_s, k_{s-1},l_s)$ changes sign. 
Moreover, it is the only one that depends on $o_{s}$. The state coefficient on the first line of \eqref{QuantumConstraint} changes from $\psi(k_1-\epsilon_1, \dotsc, k_{e_p}-\tilde{\epsilon}_p,\dotsc)$ to $(-1)^{2k_{s}+1}\psi(k_1-\epsilon_1, \dotsc, k_p-\tilde{\epsilon}_p,\dotsc)$ since $(-1)^{\tilde{\epsilon}_{s}} = -1$. Moreover, the  coefficients $B$'s are independent of the orientation $o_{s}$ and the state coefficient on the second line changes from $\psi(k_1,\dotsc, k_p, \dotsc)$ to $(-1)^{2k_{s}}\psi(k_1,\dotsc, k_p, \dotsc)$. Factorizing $(-1)^{2k_{s}}$ from the equation reveals that only the first line is modified, by $-o_{s} \times (-1) = o_{s}$. The constraint therefore still holds.
If $e^*\equiv e_s \in\{e_{p+1},\cdots,e_{d+1}\}$, the coefficient $B^{\tilde{\epsilon}_s, \tilde{\epsilon}_{s-1}}_{o_{s}}(k_s-\f{\epsilon_s}{2}, k_{s-1}-\f{\epsilon_{s-1}}{2},l_s)$ changes sign while the coefficients $A$'s remain unchanged. The same analysis leads to the same conclusion.

The argument is the same for all edges in the boundary of $f$, since the orientation of any of those edges appears in a single coefficient of the equation.
\qed

In this section, we have quantized the four deformed spinors on each ribbon to $q$-deformed quantum spinors as given explicitly in \eqref{eq:def_quantum_spinor} and constructed the quantum Hamiltonian constraint of the $q$-deformed LQG model purely in terms of (the scalar products of) these $q$-deformed quantum spinors. \ref{thm:QuantumConstraint} is the main result of the current paper. It gives rise to the difference equations that the physical states satisfy. To verify that the Hamiltonian constraint we construct is the correct one, one can justify the topological invariance of the solutions to the constraint. That is, the solutions to the Hamiltonian constraints for graphs related by a series of Pachner moves are the same (up to normalization). This is what we will illustrate in the next section. The difference equations we derived in \ref{thm:QuantumConstraint} will turn out to play a key role in the analysis.  

%%%%%%%%%%%

%

%%%%%%%%%%%%%
\section{Pachner moves}
\label{sec:Pachner}

We now show how to relate the physical states on triangulations which are related by Pachner moves. This is an extension of \cite{Noui:2004iy} to $q$ real (using Hamiltonian constraints instead of projection on flat connections). In two dimensions, there are two types of Pachner moves, the $3-1$ moves and the $2-2$ moves (as well as their inverses).
In this section, we will first analyze the case of the $2-2$ moves. The $3-1$ moves result naturally follows the analysis of removing an edge of a triangle since, in this case, two out of the three vertices of the triangle are removed due to gauge invariance on the bi-valent vertices. 

%%%%%%%%%%%%%
\subsection{2-2 Pachner move} 
\label{sec:2-2}

The 2-2 Pachner move changes a portion of the graph into another one as follows,
\begin{equation}
\begin{array}{c} 
\begin{tikzpicture}[scale=0.8]
	\coordinate (L) at (0,0);
	\coordinate (R) at (2,0);
	\coordinate (Lu) at ([shift=(120:1)]L);
	\coordinate (Ld) at ([shift=(-120:1)]L);
	\coordinate (Ru) at ([shift=(60:1)]R);
	\coordinate (Rd) at ([shift=(-60:1)]R);
	
	\draw[thick] (Lu) -- node[pos=0.4, right]{$e_1$}(L);
	\draw[thick] (Ld) -- node[pos=0.4, right]{$e_4$}(L);
	\draw[thick] (Ru) -- node[pos=0.4, left]{$e_2$}(R);
	\draw[thick] (Rd) -- node[pos=0.4, left]{$e_3$}(R);
	\draw[thick](L) -- node[above,pos=.5]{$e_5$}  (R);

	\draw[thick,->] (4,0) -- (5,0);
	
	\coordinate (U) at (7,0.7);
	\coordinate (D) at (7,-0.7);
	\coordinate (Ul) at ([shift=(150:1)]U);
	\coordinate (Ur) at ([shift=(30:1)]U);
	\coordinate (Dl) at ([shift=(-150:1)]D);
	\coordinate (Dr) at ([shift=(-30:1)]D);
	
	\draw[thick] (Ul) -- node[pos=0.4, below]{$e_1$}(U);
	\draw[thick] (Ur) -- node[pos=0.4, below]{$e_2$}(U);
	\draw[thick] (Dl) -- node[pos=0.4, above]{$e_4$}(D);
	\draw[thick] (Dr) -- node[pos=0.4, above]{$e_3$}(D);

	\draw[thick] (D) -- node[right,pos=.5]{$e_0$}(U);
	
\end{tikzpicture}\,.
\end{array}
\label{pic:2to2}
\end{equation}
We denote the initial graph which contains the left-hand side as $\Gamma_i$, and the final graph which contains the right-hand side as $\Gamma_f$. The orientations of all edges are left arbitrary.

\begin{theorem} \label{thm:2to2}
Let $|\psi_f\rangle$ on $\Gamma_f$ be defined in the spin network basis by
\be
\psi_{f}(j_1, j_2, j_3, j_4, j_0, \dotsc) =  (-1)^{j_1+j_2+j_3+j_4} [d_{j_0}] \sum_{j_5} (-1)^{(1-o_5)j_5+(1-o_0)j_0}
\begin{Bmatrix} 
j_1 & j_2 & j_0\\ j_3 & j_4 & j_5
\end{Bmatrix}_q
\,\psi_{i}(j_1, j_2, j_3, j_4, j_5, \dotsc)\,.
\label{eq:2to2}
\ee
where the ellipses denote spins which are the same on both sides (for edges which are not affected by the move). Then $|\psi_{i}\rangle$ is a state which satisfies all the constraints on $\Gamma_i$ if and only if $|\psi_{f}\rangle$ satisfies all the constraints on  $\Gamma_f$.
\end{theorem}

Since the 2-2 move is its own inverse, there is symmetry between both sides of the move. This must translate into a symmetry which exchanges the role of $|\psi_i\rangle$ and $|\psi_f\rangle$ in \eqref{eq:2to2}. This is indeed true thanks to the orthonormality of the $q$-$6j$ symbols, 
\be
\sum_{j_5} [d_{j_5}][d_{j_0}] 
\begin{Bmatrix} 
j_1 & j_2 & j_0\\ j_3 & j_4 & j_5
\end{Bmatrix}_q
\begin{Bmatrix} 
j_1 & j_2 & j'_0\\ j_3 & j_4 & j_5
\end{Bmatrix}_q
=\delta_{j_0,j'_0}\,,
\ee
which transforms \eqref{eq:2to2} into
\be
\psi_{i}(j_1, j_2, j_3, j_4, j_5, \dotsc) =  (-1)^{j_1+j_2+j_3+j_4} [d_{j_5}] \sum_{j'_0} (-1)^{(1-o_5)j_5+(1-o_0)j'_0}
\begin{Bmatrix} 
j_1 & j_2 & j'_0\\ j_3 & j_4 & j_5
\end{Bmatrix}_q
\,\psi_{f}(j_1, j_2, j_3, j_4, j'_0, \dotsc)\,.
\label{eq:2to2_re}
\ee

{\bf Proof of Theorem \ref{thm:2to2}.} 
There are four faces involved in the move on each side. Clearly, $|\psi_i\rangle$ and $|\psi_f\rangle$ satisfy the same constraints associated to faces which are not among those four. Therefore, we can focus on the four faces involved in the move, and for symmetry reasons, we can simply look at the constraints on two faces: the face $f_{12}$ which has $e_1, e_2$ in its boundary, and the face $f_{14}$ which has $e_1, e_4$ in its boundary.

\paragraph{Face $f_{12}$.} It has a different boundary on $\Gamma_f$ and $\Gamma_i$, due to the disappearance of $e_5$. On $\Gamma_i$, there are constraints where $\E^{\epsilon_5,\epsilon_1}_{e_5 e_1}$ and $\E^{\epsilon_2,\epsilon_5}_{e_2 e_5}$ are both among the $A$-terms of the constraint \eqref{QuantumConstraint}. Let us denote the two reference edges ($e_1$ and $e_k$ in \eqref{QuantumConstraint}) as $e$ and $e'$, which may be $e_1$ and/or $e_2$. Then the difference equations \eqref{QuantumConstraint} read
\begin{multline} \label{Face12Initial}
\sum_{\{\tilde{\epsilon}\}} (\prod_{e\underset{\text{c.c.}}{\to} e'}  A) \sum_{\tilde{\epsilon}_5=\pm} A_{o_5}^{\tilde{\epsilon}_5, \tilde{\epsilon}_1}(k_5, k_1, k_4) A_{o_2}^{\tilde{\epsilon}_2, \tilde{\epsilon}_5}(k_2, k_5, k_3) \psi_i\Bigl(k_1-\frac{\tilde{\epsilon}_1}{2}, k_2-\frac{\tilde{\epsilon}_2}{2}, k_3, k_4, k_5-\frac{\tilde{\epsilon}_5}{2},\dotsc\Bigr) \\
 +(-1)^{d_{12,i}-d_{ee'}} \alpha^{\epsilon_e,\epsilon_{e'}}(k_e,k_{e'})\sum_{\{\tilde{\epsilon}\}} (\prod_{e'\underset{\text{c.c.}}{\to} e} B) \psi_i(k_1, k_2, k_3, k_4, k_5,\dotsc) = 0,
\end{multline}
where $d_{12,i}$ denotes the number of boundary edges of $f_{12}$ in $\Gamma_i$ and $d_{ee'}$ the number of edges from $e$ to $e'$ counter-clockwise. Notice that $\tilde{\epsilon}_1$ (\resp $\tilde{\epsilon}_2$) is fixed if $e=e_1$ (\resp if $e'=e_2$) and summed over otherwise. We have indicated in $\psi_i$ only the spins which are involved in the move.

\begin{figure}
\includegraphics[scale=.55]{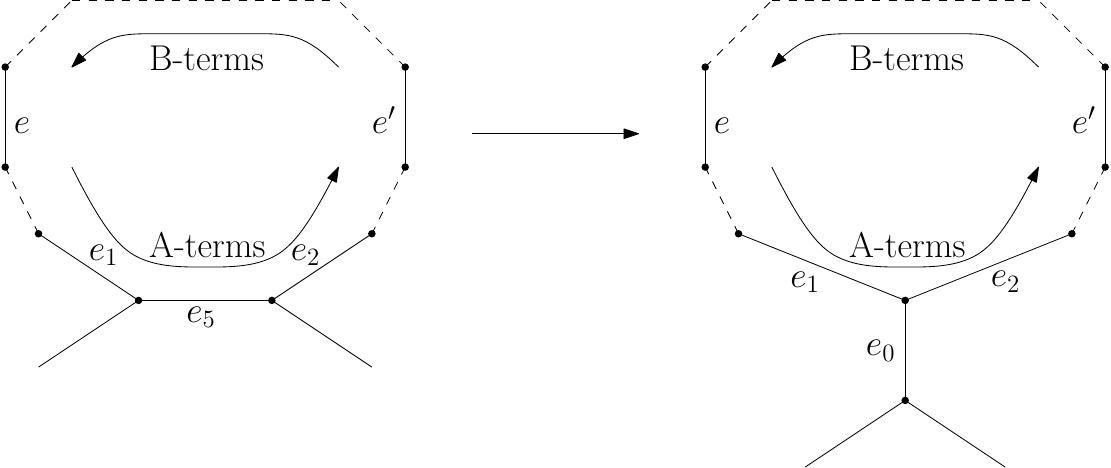}
\caption{\label{fig:2-2MoveF12} A graphical representation of Equations \eqref{Face12Initial} on the LHS and \eqref{Face12Final} on the RHS.}
\end{figure}

We have written $\sum_{\{\tilde{\epsilon}\}}(\prod_{e\underset{\text{c.c.}}{\to} e'} A)$ and $\sum_{\{\tilde{\epsilon}\}} (\prod_{e'\underset{\text{c.c.}}{\to} e} B)$ schematically the coefficients of the equation which are associated to corners \emph{not} involved in the move. Here $\sum_{\{\tilde{\epsilon}\}}(\prod_{e\underset{\text{c.c.}}{\to} e'} A)$ is the product of the $A$-terms over the corners from $e$ to $e'$ going counter-clockwise, except for the two corners with $e_5$, whose $A$-terms are distinguished. Then $\sum_{\{\tilde{\epsilon}\}} (\prod_{e'\underset{\text{c.c.}}{\to} e} B)$ is the product of the $B$-terms over the corners from $e'$ to $e$ counter-clockwise. This is depicted in the Figure \ref{fig:2-2MoveF12}.

On the other hand, a state on $\Gamma_f$ must also satisfy a constraint along the face $f_{12}$ with the two reference edges $e$ and $e'$. It reads
\begin{multline} \label{Face12Final}
\langle \{k_e\}_{\Gamma_f}| \H_{f_{12}, e, e'}| \psi_f\rangle
\propto \sum_{\{\tilde{\epsilon}\}} (\prod_{e\underset{\text{c.c.}}{\to} e'}  A) A_{o_2}^{\epsilon_2, \epsilon_1}(k_2, k_1, k_0) \psi_f\Bigl(k_1-\frac{\tilde{\epsilon}_1}{2}, k_2-\frac{\tilde{\epsilon}_2}{2}, k_3, k_4, k_0,\dotsc\Bigr) \\
 +(-1)^{d_{12,i}-d_{ee'}+1} \alpha^{\epsilon_e,\epsilon_{e'}}(k_e,k_{e'})\sum_{\{\tilde{\epsilon}\}} (\prod_{e'\underset{\text{c.c.}}{\to} e} B) \psi_f(k_1, k_2, k_3, k_4, k_0,\dotsc).
\end{multline}
There is also a constraint where $\mathbf{E}_{e_1e_2}^{\tilde{\epsilon_1}, \tilde{\epsilon}_2}$ gives rise to a $B$-term, but as we have shown this is equivalent to the above constraint. Here it is important that the products of the $A$-terms and $B$-terms over all corners except the one where $e_1$ and $e_2$ meet are the same as in \eqref{Face12Initial}. The reason is obviously that those terms are local and the 2-2 move does not involve their corners. As in \eqref{Face12Initial}, $\tilde{\epsilon}_1$ and $\tilde{\epsilon}_2$ may be fixed or summed over.

We now plug \eqref{eq:2to2} into \eqref{Face12Final} to check that it vanishes, provided the constraint \eqref{Face12Initial} holds. First compute, with $j_{1,2} := k_{1,2} - \frac{\tilde{\epsilon}_{1,2}}{2}$,
\begin{multline}
A_{o_2}^{\tilde{\epsilon}_2, \tilde{\epsilon}_1}(k_2, k_1, k_0) \psi_f(j_1, j_2, k_3, k_4, k_0,\dotsc) 
= 
o_2\tilde{\epsilon}_2 [d_{k_2}] [d_{k_0}] 
\sum_{j_5} (-1)^{(1-o_5)j_5+(1-o_0)k_0} (-1)^{j_1 + j_2 + k_3 + k_4} (-1)^{k_0 + k_1 + k_2} \\
\begin{Bmatrix} 
k_1 & j_1 & \frac12\\ j_2 & k_2 & k_0
\end{Bmatrix}_q 
\begin{Bmatrix} 
j_1 & j_2 & k_0\\ k_3 & k_4 & j_5
\end{Bmatrix}_q
\psi_{i}(j_1, j_2, k_3, k_4, j_5,\dotsc)\,.
\end{multline}
The Biedenharn-Elliott identity on $q$-$6j$ symbols gives precisely
\begin{multline}
(-1)^{j_1 + j_2 + k_3 + k_4} (-1)^{k_0 + k_1 + k_2} 
\begin{Bmatrix} 
k_1 & j_1 & \frac12\\ j_2 & k_2 & k_0
\end{Bmatrix}_q 
\begin{Bmatrix} 
j_1 & j_2 & k_0\\ k_3 & k_4 & j_5
\end{Bmatrix}_q \\
= \sum_{k_5} [d_{k_5}] (-1)^{k_5 + j_5 + \frac12} 
\begin{Bmatrix} 
k_1 & k_2 & k_0\\ k_3 & k_4 & k_5
\end{Bmatrix}_q 
\begin{Bmatrix} 
k_1 & j_1 & \frac12\\ j_5 & k_5 & k_4
\end{Bmatrix}_q 
\begin{Bmatrix} 
k_5 & j_5 & \frac12\\ j_2 & k_2 & k_3
\end{Bmatrix}_q\,.
\end{multline}
Setting $j_5 = k_5 - \frac{\tilde{\epsilon}_5}{2}$ to change the summation over $j_5$ to one over $\tilde{\epsilon}_5$ (there are no other values of $j_5$ allowed by the triangular inequalities on the $q$-$6j$ symbol), we get
\begin{multline}
A_{o_2}^{\epsilon_2, \epsilon_1}(k_2, k_1, k_0) \psi_f(j_1, j_2, k_3, k_4, k_0,\dotsc)
= 
o_2\epsilon_2 [d_{k_2}] [d_{k_0}]
\sum_{k_5, \tilde{\epsilon}_5} (-1)^{(1-o_5)j_5+(1-o_0)k_0} (-1)^{k_5 + j_5 + \frac12}\\
 [d_{k_5}]
\begin{Bmatrix} 
k_1 & k_2 & k_0\\ k_3 & k_4 & k_5
\end{Bmatrix}_q 
\begin{Bmatrix} 
k_1 & j_1 & \frac12\\ j_5 & k_5 & k_4
\end{Bmatrix}_q 
\begin{Bmatrix} 
k_5 & j_5 & \frac12\\ j_2 & k_2 & k_3
\end{Bmatrix}_q 
\psi_i(j_1, j_2, k_3, k_4, j_5,\dotsc)\,.
\end{multline}
Using $\epsilon_5 = (-1)^{\frac12 + j_5 - k_5}$, we find $(-1)^{k_5 + j_5 + \frac12} = \epsilon_5 (-1)^{2k_5}$. We also use $(-1)^{(1-o_5)j_5} = o_5(-1)^{(1-o_5)k_5}$ and notice that a $q$-$6j$ symbol can be factored. Thus,
\begin{multline}
\langle \{k_e\}_{\Gamma_f}| \H_{f_{12}, e, e'}| \psi_f\rangle
\propto  \sum_{k_5, \tilde{\epsilon}_5} (-1)^{(1-o_5)k_5+(1-o_0)k_0} (-1)^{k_1 + k_2 + k_3 + k_4} 
\begin{Bmatrix} 
k_1 & k_2 & k_0\\ k_3 & k_4 & k_5
\end{Bmatrix}_q [d_{k_0}]\\
\Biggl( \sum_{\{\tilde{\epsilon}\}} (\prod_{e\underset{\text{c.c.}}{\to} e'}  A) 
o_5 \epsilon_5 [d_{k_5}]
(-1)^{2k_5} (-1)^{k_1 + k_2 + k_3 + k_4}  
o_2\epsilon_2 [d_{k_2}]
\begin{Bmatrix} 
k_1 & j_1 & \frac12\\ j_5 & k_5 & k_4
\end{Bmatrix}_q 
\begin{Bmatrix} 
k_5 & j_5 & \frac12\\ j_2 & k_2 & k_3
\end{Bmatrix}_q 
\psi_i(j_1, j_2, k_3, k_4, j_5,\dotsc) \\
+(-1)^{d_{12,i}-d_{ee'}} \alpha^{\epsilon_e,\epsilon_{e'}}(k_e,k_{e'})
\sum_{\{\tilde{\epsilon}\}} (\prod_{e'\underset{\text{c.c.}}{\to} e}  B) \psi_i(k_1, k_2, k_3, k_4, k_5,\dotsc)\Biggr)\,.
\end{multline}
We now recognize the coefficients in \eqref{Face12Initial},
\begin{multline}
\langle \{k_e\}_{\Gamma_f}| \H_{f_{12}, e, e'}| \psi_f\rangle
\propto \sum_{k_5} (-1)^{(1-o_5)k_5+(1-o_0)k_0} (-1)^{k_1 + k_2 + k_3 + k_4} 
\begin{Bmatrix} 
k_1 & k_2 & k_0\\ k_3 & k_4 & k_5
\end{Bmatrix}_q  [d_{k_0}]\\
\Biggl(\sum_{\{\tilde{\epsilon}\}} (\prod_{e\underset{\text{c.c.}}{\to} e'}  A) \sum_{\tilde{\epsilon}_5=\pm} A_{o_5}^{\tilde{\epsilon}_5, \tilde{\epsilon}_1}(k_5, k_1, k_4) 
A_{o_2}^{\tilde{\epsilon}_2, \tilde{\epsilon}_5}(k_2, k_5, k_3) \psi_i(j_1, j_2, k_3, k_4, j_5,\dotsc) \\
+(-1)^{d_{12,i}-d_{ee'}}\alpha^{\epsilon_e,\epsilon_{e'}}(k_e,k_{e'})
\sum_{\{\tilde{\epsilon}\}} (\prod_{e'\underset{\text{c.c.}}{\to} e}  B) \psi_i(k_1, k_2, k_3, k_4, k_5,\dotsc)\Biggr),
\end{multline}
and conclude that \eqref{Face12Final} vanished provided \eqref{Face12Initial} and \eqref{eq:2to2}.

\paragraph{Face $f_{14}$.} We now perform the same analysis on the constraints which act on the face $f_{14}$. We use the same notation as for the face $f_{12}$, \ie let $e$ and $e'$ be two reference edges around $f_{14}$ and consider the Hamiltonian constraints associated to them on $\Gamma_i$ and $\Gamma_f$. On $\Gamma_i$, the Hamiltonians contain the operator $\E_{e_1 e_4}^{\epsilon_1, \epsilon_4}$ which, without loss of generality, can be considered to give rise to an $A$-term. The constraints on the spin network coefficients of $|\psi_i\rangle$ read
\begin{multline} \label{ConstraintF14}
\sum_{\{\tilde{\epsilon}\}} (\prod_{e\underset{\text{c.c.}}{\to} e'} A) A_{o_1}^{\tilde{\epsilon}_1, \tilde{\epsilon}_4}(k_1, k_4, k_5) \psi_i(j_1, k_2, k_3, j_4, k_5,\dotsc) \\
+(-1)^{d_{14,i}-d_{ee'}}
\alpha^{\epsilon_e,\epsilon_{e'}}(k_e,k_{e'})\sum_{\{\tilde{\epsilon}\}} (\prod_{e'\underset{\text{c.c.}}{\to} e} B) \psi_i(k_1, k_2, k_3, k_4, k_5,\dotsc) = 0,
\end{multline}
with $j_1 = k_1-\tilde{\epsilon_1}/2, j_4 = k_4 - \tilde{\epsilon}_4/2$. The sign $\tilde{\epsilon}_1$ (\resp $\tilde{\epsilon}_4$) is fixed if $e=e_1$ (\resp if $e'=e_4$) and summed over otherwise. Here, $\prod_{e\underset{\text{c.c.}}{\to} e'} A$ is the product of the $A$-terms from $e$ to $e'$ counter-clockwise, except for the one on the corner of $e_1, e_4$ which has been singled out. As for $\prod_{e'\underset{\text{c.c.}}{\to} e} B$, it is the product of the $B$-terms going counter-clockwise from $e'$ to $e$.

On $\Gamma_f$, we need to look at two types of constraints. Either the operators $\E^{\tilde{\epsilon}_0,\tilde{\epsilon}_4}_{e_0 e_4}$ and $\E^{\tilde{\epsilon}_1,\tilde{\epsilon}_0}_{e_1 e_0}$ which enter $\H_{f_{14}, e, e'}$ on $\Gamma_f$ both contribute to $A$-terms of the constraint (or both to $B$-terms but this is the same), or one gives rise to an $A$-term and the other one to a $B$-term.

In the case that  they both give rise to $A$-terms, we are in the same situation as in our previous analysis on the face $f_{12}$, with the role of $\Gamma_i$ and $\Gamma_f$ exchanged. Since the relation \eqref{eq:2to2} between $|\psi_f\rangle$ and $|\psi_i\rangle$ can be inverted with the same form, we have nothing to prove.

If $\E^{\tilde{\epsilon}_0,\tilde{\epsilon}_4}_{e_0 e_4}$ contributes to a $B$-term, and $\E^{\tilde{\epsilon}_1,\tilde{\epsilon}_0}_{e_1 e_0}$ contributes to an $A$-term, this means that $e_0=e$ is a reference edge chosen for the constraint. The Hamiltonians of this type on $\Gamma_f$ are $\H_{f_{14}, e_0, e'}^{\epsilon_0, \epsilon_{e'},\dotsc,\epsilon_4}$ and they are labelled by signs for all the edges from $e'$ to $e_0$ counter-clockwise. The matrix elements read 
\begin{multline}
\langle \{k_e\}_{\Gamma_f}| \H^{\epsilon_0, \epsilon_{e'},\dotsc,\epsilon_4}_{f_{14}, e_0, e'}| \psi_f\rangle \propto 
\sum_{\{\tilde{\epsilon}\}} (\prod_{e_1\underset{\text{c.c.}}{\to} e'} A) A_{o_1}^{\tilde{\epsilon}_1, \epsilon_0}(k_1, k_0, k_2) 
\psi_f(j_1, k_2, k_3, j_4, j_0,\dotsc) \\
+(-1)^{d_{14,f}-d_{e_0 e'}}\alpha^{\epsilon_0,\epsilon_{e'}}(k_0,k_{e'}) 
\sum_{\{\tilde{\epsilon}\}} (\prod_{e'\underset{\text{c.c.}}{\to} e_4}B) B_{o_0}^{\epsilon_0, \tilde{\epsilon}_4}(j_0, j_4, k_3) 
 \psi_f(k_1, k_2, k_3, l_4 , k_0,\dotsc)\,,
\end{multline}
where $\epsilon_0$ is fixed (but $\tilde{\epsilon}_4$ only is if $e'=e_4$) and $j_{0,4} = k_{0,4} - \epsilon_{0,4}/2$, and $l_4 = j_4 + \tilde{\epsilon}_4/2$.

We now plug \eqref{eq:2to2} into the above matrix elements. We first look at the $A$-term,
\begin{multline}
A_{o_1}^{\epsilon_1, \epsilon_0}(k_1, k_0, k_2) \psi_f(j_1, k_2, k_3, j_4, j_0,\dotsc) 
=[d_{j_0}] \sum_{k_5} 
o_1 \epsilon_1 [d_{k_1}]
 (-1)^{k_0 + k_1 + k_2} 
 (-1)^{(1-o_5)k_5+(1-o_0)j_0} (-1)^{j_1 + k_2 + k_3 + j_4} \\
\begin{Bmatrix} 
j_1 & k_2 & j_0\\ k_3 & j_4 & k_5
\end{Bmatrix}_q 
\begin{Bmatrix} 
k_1 & j_1 & \frac12\\ j_0 & k_0 & k_2
\end{Bmatrix}_q 
\psi_i(j_1, k_2, k_3, j_4, k_5,\dotsc).
\end{multline}
The relevant Biedenharn-Elliott identity is
\be
(-1)^{k_0 + k_1 + k_2 + j_0 + j_1 + k_3 + k_5} 
\begin{Bmatrix} 
j_1 & k_2 & j_0\\ k_3 & j_4 & k_5
\end{Bmatrix}_q 
\begin{Bmatrix} 
k_1 & j_1 & \frac12\\ j_0 & k_0 & k_2
\end{Bmatrix}_q 
= \sum_{l_4} [d_{l_4}] (-1)^{j_4 + l_4 + \frac12} 
\begin{Bmatrix} 
j_4 & l_4 & \frac12\\ k_1 & j_1 & k_5
\end{Bmatrix}_q 
\begin{Bmatrix} 
j_4 & l_4 & \frac12\\ k_0 & j_0 & k_3
\end{Bmatrix}_q 
\begin{Bmatrix} 
k_1 & k_2 & k_0\\ k_3 & l_4 & k_5
\end{Bmatrix}_q\,.
\ee
As for the $B$-term, 
\begin{multline}
\sum_{l_4} B_{o_0}^{\epsilon_0, \tilde{\epsilon}_4}(j_0, j_4, k_3) 
\psi_f(k_1, k_2, k_3, l_4, k_0,\dotsc) 
= \sum_{k_5, l_4} 
o_0 \epsilon_0 [d_{j_0}][d_{k_0}]
(-1)^{j_0 + k_3 + j_4} (-1)^{(1-o_5)k_5+(1-o_0)k_0} (-1)^{k_1 + k_2 + k_3 + l_4} \\
\begin{Bmatrix} 
j_4 & l_4 & \frac12\\ k_0 & j_0 & k_3
\end{Bmatrix}_q 
\begin{Bmatrix} 
k_1 & k_2 & k_0\\ k_3 & l_4 & k_5
\end{Bmatrix}_q 
\psi_i(k_1, k_2, k_3, l_4, k_5,\dotsc).
\end{multline}
We recognize the two same $q$-$6j$ symbols as in the Biendenharn-Elliott identity above. We can thus factor them out, so that the matrix elements of the Hamiltonian are proportional to
\begin{multline}
\langle \{k_e\}_{\Gamma_f}| \H^{\epsilon_0, \epsilon_{e'},\dotsc,\epsilon_4}_{f_{14}, e_0, e'}| \psi_f\rangle \propto 
\sum_{k_5, l_4} (-1)^{j_0 + k_3 + j_4} 
(-1)^{(1-o_5)k_5+(1-o_0)k_0} (-1)^{k_1 + k_2 + k_3 + l_4} 
\begin{Bmatrix} 
j_4 & l_4 & \frac12\\ k_0 & j_0 & k_3
\end{Bmatrix}_q 
\begin{Bmatrix} 
k_1 & k_2 & k_0\\ k_3 & l_4 & k_5
\end{Bmatrix}_q 
[d_{j_0}]
\\
\Biggl( \sum_{\{\tilde{\epsilon}\}} (\prod_{e_1\underset{\text{c.c.}}{\to} e'} A)  
o_0 o_1 \epsilon_1 [d_{k_1}][d_{l_4}]  
(-1)^{k_2+j_4-j_0-k_5}(-1)^{j_4+l_4+\f12} (-1)^{j_0+k_3+j_4} (-1)^{k_1+k_2+k_3+l_4}
\begin{Bmatrix} 
j_4 & l_4 & \frac12\\ k_1 & j_1 & k_5
\end{Bmatrix}_q \\
\times \psi_i(j_1, k_2, k_3, j_4, k_5,\dotsc)
+(-1)^{d_{14,f}-d_{e_0e'}}
\alpha^{\epsilon_0,\epsilon_{e'}}(k_0,k_{e'}) 
\sum_{\{\tilde{\epsilon}\}} (\prod_{e'\underset{\text{c.c.}}{\to} e_4}B)
o_0 \epsilon_0 [d_{k_0}]
 \psi_i(k_1, k_2, k_3, l_4, k_5,\dotsc)\Biggr)
\end{multline}
It now suffices to show that the expression into brackets vanishes thanks to \eqref{ConstraintF14}. Let us take care of the signs:  
$(-1)^{k_2+j_4-j_0-k_5}(-1)^{j_4+l_4+\f12} (-1)^{j_0+k_3+j_4} (-1)^{k_1+k_2+k_3+l_4} =(-1)^{2(k_2-k_5+k_3)}(-1)^{4j_4}(-1)^{\f{1+\tilde{\epsilon}_4}{2}}(-1)^{k_1+l_4+k_5}=-\tilde{\epsilon}_4 (-1)^{k_1+l_4+k_5}$.
Replacing the sign factor in the bracket, we get
\begin{multline}
-\sum_{\{\tilde{\epsilon}\}} (\prod_{e_1\underset{\text{c.c.}}{\to} e'} A)
 o_0 o_1 \epsilon_1 [d_{k_1}][d_{l_4}] 
(-1)^{j_1 - k_1 + j_4 - l_4} (-1)^{j_1 + l_4 + k_5 + \frac12} 
\begin{Bmatrix} 
j_4 & l_4 & \frac12\\ k_1 & j_1 & k_5
\end{Bmatrix}_q 
\psi_i(j_1, k_2, k_3, j_4, k_5,\dotsc) \\
+(-1)^{d_{14,f}-d_{e_0 e'}}\alpha^{\epsilon_0,\epsilon_{e'}}(k_0,k_{e'}) 
\sum_{\{\tilde{\epsilon}\}} (\prod_{e'\underset{\text{c.c.}}{\to} e_4}B)
o_0\epsilon_0 [d_{k_0}]
\psi_i(k_1, k_2, k_3, l_4, k_5,\dotsc)\\
= -o_0 \tilde{\epsilon}_4[d_{l_4}]\Bigl(
\sum_{\{\tilde{\epsilon}\}} (\prod_{e_1\underset{\text{c.c.}}{\to} e'} A) A_{o_1}^{\epsilon_1, \tilde{\epsilon}_4}(k_1, l_4, k_5)
 \psi_i(j_1, k_2, k_3, j_4, k_5,\dotsc) \\
+(-1)^{d_{14,i}-d_{e_4 e'}}\alpha^{\tilde{\epsilon}_4,\epsilon_{e'}}(l_4,k_{e'}) 
\sum_{\{\tilde{\epsilon}\}} (\prod_{e'\underset{\text{c.c.}}{\to} e_4}B)
\psi_i(k_1, k_2, k_3, l_4, k_5,\dotsc)\Bigr).
\end{multline}
The expression into brackets on the RHS is exactly the constraint \eqref{ConstraintF14} on $\Gamma_i$ with the choice $e=e_4$ of reference edge and arbitrary $\tilde{\epsilon}_4$ fixed.
\qed

%%%%%%%%%%%%%
\subsection{Removing an edge}

Consider two adjacent faces $F$ and $f$, separated by an edge $e_0$. We consider the move which consists in removing $e_0$ (as well as its two end vertices). By performing a series of 2-2 Pachner moves (described in Section \ref{sec:2-2}), we can always assume that $f$ is triangular,
\begin{equation} \label{fig:linkRemoval}
\begin{array}{c}  
\begin{tikzpicture}
	\coordinate (A) at (0,0);
	\coordinate (B) at ([shift=(150:1.5)]A);
	\coordinate (C) at ([shift=(-150:1.5)]A);
	\coordinate (Bl) at ([shift=(165:1)]B);
	\coordinate (Cl) at ([shift=(-165:1)]C);
	\coordinate (D) at (0.5,0);
	
	\draw[thick] (A) -- node[midway, above]{$e_2$}(B);
	\draw[thick] (A) -- node[midway, below]{$e_1$}(C);
	\draw[thick] (C) -- node[midway, left]{$e_0$}(B);
	\draw[thick] (B) -- node[midway, above]{$e_3$}(Bl);
	\draw[thick] (C) -- node[midway, below]{$e_4$}(Cl);
	\draw[thick,dashed] (Bl) to [out=-150,in=150] (Cl);
	\draw[thick] (A) -- (D);
	
	\draw (-0.8,0) node{$f$};
	\draw (-2.3,0) node{$F$};
	
	\draw[thick,->] (1,0) -- (2,0);
	
	\coordinate (a) at (5.5,0);
	\coordinate (b) at ([shift=(150:1.5)]a);
	\coordinate (c) at ([shift=(-150:1.5)]a);
	\coordinate (bl) at ([shift=(165:1)]b);
	\coordinate (cl) at ([shift=(-165:1)]c);
	\coordinate (d) at (6,0);
	
	\draw[thick] (a) -- node[midway, above]{$e_2$}(bl);
	\draw[thick] (a) -- node[midway, below]{$e_1$}(cl);
	\draw[thick] (a) -- (d);
	\draw[thick,dashed] (bl) to [out=-150,in=150] (cl);
	\draw (4,0) node{$F\cup f$};

\end{tikzpicture}
\end{array}
\end{equation}

If $|\psi_i\rangle$ is a state which satisfies all the constraints before the edge removal, we want to describe how it transforms through the move.

\begin{theorem} \label{thm:linkRemoval}
$|\psi_f\rangle$ with spin network coefficients
\begin{equation} \label{LinkRemoval}
\psi_f(j_1, j_2, \dotsc) = (-1)^{(1+o_1)j_1 + (1+o_2)j_2} \sqrt{[d_{j_1}]\,[d_{j_2}]}\,\psi_i(0, j_1, j_2, j_2, j_1, \dotsc)
\end{equation}
is a solution of the constraints on the graph after the edge removal. Here $o_1, o_2$ are the orientations of the edges $e_1, e_2$ with respect to $f$ (counter-clockwise oriented) and $\psi(j_0, j_1, j_2, j_3, j_4, \dotsc)$ is the spin network coefficient of $|\psi_i\rangle$.
\end{theorem}
In other words, $|\psi_i\rangle$ gives rise to a solution of the constraints on $\Gamma_f$, obtained by keeping only its $j_0=0$ components. We will use this relation to study the 3-1 Pachner move.

{\bf Proof.} Consider two reference edges $e, e'$ in $F\cup f$, and the associated constraint such that $\mathbf{E}_{e_1 e_2}^{\tilde{\epsilon}_1, \tilde{\epsilon}_2}$ is an $A$-term (without loss of generality since $A$- and $B$-terms can be exchanged). Its matrix elements $\langle \{k_e\}| \H_{F\cup f, e, e'}| \psi_f\rangle$ read
\begin{equation} \label{ConstraintMergedFace}
\langle \{k_e\}| \H_{F\cup f, e, e'}| \psi_f\rangle \propto \sum_{\{\tilde{\epsilon}\}} (\prod_{e\underset{\text{c.c.}}{\to} e'} A) A_{o_2}^{\tilde{\epsilon}_2, \tilde{\epsilon}_1}(k_2, k_1, l_2) \psi_f(j_1, j_2, \dotsc)
+(-1)^{d_f - d_{ee'}} \alpha^{\epsilon_e,\epsilon_{e'}}(k_e,k_{e'}) \sum_{\{\tilde{\epsilon}\}} (\prod_{e'\underset{\text{c.c.}}{\to} e}  B) \psi_f(k_1, k_2, \dotsc),
\end{equation}
with $j_{1,2} = k_{1,2} - \tilde{\epsilon}_{1,2}/2$, and $d_f$ denotes the number of boundary edges surrounding $F\cup f$.  Here, $\prod_{e\underset{\text{c.c.}}{\to} e'} A$ is the product of the $A$-terms from $e$ to $e'$ counter-clockwise, except for the one on the corner of $e_1, e_2$ which has been singled out. We will show those matrix elements vanish as soon as the constraints on $f$ and on $F$ are both satisfied on $|\psi_i\rangle$, given \eqref{LinkRemoval}.

On $f$, we have the constraint, for fixed $\epsilon_1, \epsilon_2$, and $j_{1,2} = k_{1,2} - \epsilon_{1,2}/2$,
\begin{multline}
B_{o_2}^{-\epsilon_2, -\epsilon_1}(k_2, k_1, l_2) \psi_i(k_0, j_1, j_2, k_3, k_4, \dotsc) \\
= 
\alpha^{\epsilon_1,\epsilon_2}(k_1,k_2) 
\sum_{\epsilon_0=\pm} A_{o_{0,f}}^{\epsilon_0, -\epsilon_2}(k_0, j_2, k_3) 
A_{o_1}^{-\epsilon_1, \epsilon_0}(j_1, k_0, k_4) \psi_i(k_0-\frac{\epsilon_0}{2}, k_1, k_2, k_3, k_4,\dotsc),
\label{eq:Constraintf}
\end{multline}
where $o_{0,f}$ is the orientation of $e_0$ relative to $f$. On $F$ there is a constraint similar to \eqref{ConstraintMergedFace}, from the Hamiltonians $\mathbf{H}_{F, e, e'}$ with the same signs $\epsilon$s. It reads,
\begin{multline} \label{eq:ConstraintF}
\sum_{\{\tilde{\epsilon}\}} (\prod_{e\underset{\text{c.c.}}{\to} e'} A) \sum_{\tilde{\epsilon}_0=\pm} A_{o_{0,F}}^{\tilde{\epsilon}_0, \tilde{\epsilon}_4}(k_0, k_4 + \frac{\tilde{\epsilon}_4}{2}, k_1) A_{o_3}^{\tilde{\epsilon}_3, \tilde{\epsilon}_0}(k_3 + \frac{\tilde{\epsilon}_3}{2}, k_0, k_2) \psi_i(k_0-\frac{\tilde{\epsilon}_0}{2}, k_1, k_2, k_3-\frac{\tilde{\epsilon}_3}{2}, k_4 -  \frac{\tilde{\epsilon}_4}{2},\dotsc) \\
+(-1)^{d_f-d_{ee'}}\alpha^{\epsilon_e,\epsilon_{e'}}(k_e,k_{e'}) \sum_{\{\tilde{\epsilon}\}} (\prod_{e'\underset{\text{c.c.}}{\to} e}  B)  \psi_i(k_0, k_1, k_2, k_3, k_4,\dotsc) = 0\,,
\end{multline}
where $o_{0,F}$ is the orientation of $e_0$ as the boundary of $F$, which is opposite to $o_{0,f}$. Here $\tilde{\epsilon}_4$ (respectively $\tilde{\epsilon}_3$) is fixed if $e=e_4$ (respectively if $e'=e_3$) and summed over otherwise.

We now specialize \eqref{eq:Constraintf} and \eqref{eq:ConstraintF} to $k_0=0$, where they simplify a lot. First, that enforces $\epsilon_0 =-$ in \eqref{eq:Constraintf} and $\tilde{\epsilon}_0 = -$ in \eqref{eq:ConstraintF}, so that those sums reduce to a single term. In \eqref{eq:ConstraintF} we further take $k_3=k_2$ and $k_4=k_1$. All $q$-$6j$-symbols with a spin equal to 0 can be evaluated as $\{\begin{smallmatrix} j_1 & k_1 & \frac12\\ \frac12 & 0 & k_4\end{smallmatrix}\}_q = \delta_{j_1, k_4} (-1)^{j_1 + k_1 + \frac12}/\sqrt{[2] [d_{j_1}]}$.

As a consequence, \eqref{eq:Constraintf} gives
\be \label{Triangulark_0=0}
 B_{o_2}^{-\epsilon_2, -\epsilon_1}(k_2, k_1, l_2) \psi_i(0, j_1, j_2, j_2, j_1, \dotsc) 
= - \alpha^{\epsilon_1,\epsilon_2}(k_1,k_2)  
o_{0,f} o_1 \epsilon_2
\f{[d_{j_1}]}{[2]\sqrt{[d_{j_1}] [d_{j_2}]}}
 \psi_i(\frac{1}{2}, k_1, k_2, j_2, j_1,\dotsc ),
\ee
where $k_3=j_2$ and $k_4=j_1$ on the last term are enforced by the special evaluations of the $q$-$6j$-symbols with a spin 0. Equation \eqref{eq:ConstraintF} gives
\begin{multline} \label{eq:ConstraintF2}
\sum_{\{\tilde{\epsilon}\}} (\prod_{e\underset{\text{c.c.}}{\to} e'} A) o_{0,f} o_3 \epsilon_1
\f{[d_{k_2}]}{[2]\sqrt{[d_{k_1}] [d_{k_2}]}} \psi_i(\frac{1}{2}, k_1, k_2, j_2, j_1,\dotsc )
\\
+(-1)^{d_f-d_{ee'}}\alpha^{\epsilon_e,\epsilon_{e'}}(k_e,k_{e'}) \sum_{\{\tilde{\epsilon}\}} (\prod_{e'\underset{\text{c.c.}}{\to} e}  B)  \psi_i(0, k_1, k_2, k_2, k_1,\dotsc) = 0\,.
\end{multline}
The term $\psi_i(\frac{1}{2}, k_1, k_2, j_2, j_1,\dotsc )$ can be eliminated using \eqref{Triangulark_0=0}. Moreover we turn the $B$-coefficient of this equation into an $A$-coefficient using $- A_{o_2}^{\epsilon_2, \epsilon_1}(k_2, k_1, l_2) = B_{o_2}^{-\epsilon_2, -\epsilon_1}(k_2, k_1, l_2)$. It is then enough to recognize $\psi_f$ as given in \eqref{LinkRemoval} to obtain that \eqref{ConstraintMergedFace} vanishes.
\qed

%%%%%%%%%%%%%%%%%%%%%
\subsection{3-1 Pachner move}

The 3-1 Pachner move removes a triangular face from the graph and replaces it with a vertex. The edges incident to the face become incident to the vertex,
\begin{equation}
\begin{array}{c} 
\begin{tikzpicture}[scale=0.8]
	\coordinate (A) at (0,0);
	\coordinate (B) at (2,0);
	\coordinate (C) at ([shift=(-60:2cm)]A);
	\coordinate (A1) at ([shift=(150:0.8cm)]A);
	\coordinate (B1) at ([shift=(30:0.8cm)]B);
	\coordinate (C1) at ([shift=(-90:0.8cm)]C);
	
	\draw[thick] (A) -- node[above,pos=.5]{$e_2$}(B);
	\draw[thick] (C) -- node[right,pos=.5]{$e_6$}(B);
	\draw[thick] (A) -- node[left,pos=.5]{$e_1$}(C);

	\draw[thick] (A) -- node[left,pos=.4]{$e_3$}(A1);
	\draw[thick] (B) -- node[right,pos=.3]{$e_4$}(B1);
	\draw[thick] (C1) -- node[left,pos=.5]{$e_5$}(C);
 
	\draw[->,thick] (3.5,-0.866) -- (4.5,-0.866);
	
	\coordinate (A) at (0+6,0);
	\coordinate (B) at (2+6,0);
	\coordinate (C) at ([shift=(-60:2cm)]A);
	\coordinate (A1) at ([shift=(150:0.8cm)]A);
	\coordinate (B1) at ([shift=(30:0.8cm)]B);
	\coordinate (C1) at ([shift=(-90:0.8cm)]C);
	\coordinate (O) at (1+6,-0.866);
		
	\draw[thick] (O) -- node[left,pos=.5]{$e_3$}(A1);
	\draw[thick] (O) -- node[below,pos=.5]{$e_4$}(B1);
	\draw[thick] (C1) -- node[left,pos=.5]{$e_5$}(O);
\end{tikzpicture}
 \end{array}
\end{equation}
The orientations of all the edges are left arbitrary.

\begin{theorem}
If $|\psi_i\rangle$ is a state on the initial graph $\Gamma_i$ which satisfy all the constraints, then its spin network coefficients can be written
\be
\psi_i(j_1, j_2, j_3, j_4, j_5, j_6, \dotsc)
=(-1)^{(1+o_1)j_1 + (1+o_2) j_2 + (1+o_6) j_6}  
(-1)^{j_3+j_4+j_5}
\begin{Bmatrix} 
 j_1 & j_2 & j_3 \\ j_4 & j_5 & j_6
 \end{Bmatrix}_q
\psi_f(j_3, j_4, j_5, \dotsc)   \, .
 \label{eq:3to1}
\ee
where $\psi_f(j_3, j_4, j_5, \dotsc)$ are the spin network coefficients of a state $|\psi_f\rangle$ which satisfies all the constraints on the final graph $\Gamma_f$.
\end{theorem}

{\bf Proof.} 
Let us write the constraints on the triangular face. There is one constraint for each pair of edges of the boundary. For the pair $(e_2, e_6)$, for instance, one gets
\begin{multline}
\sum_{\epsilon_1=\pm} A_{o_1}^{\epsilon_1, \epsilon_2}(k_1, k_2, k_3) A_{o_6}^{\epsilon_6, \epsilon_1}(k_6, k_1, k_5) \psi_i(j_1, j_2, k_3, k_4, k_5, j_6, \dotsc) \\
+ \alpha^{\epsilon_2,\epsilon_6}(k_2,k_6) B_{o_2}^{\epsilon_2, \epsilon_6}(j_2, j_6, k_4) \psi_i(k_1, k_2, k_3, k_4, k_5, k_6, \dotsc) = 0\,. 
\end{multline}
Here $j_i = k_i - \epsilon_i/2$, for $i=1, 2, 6$.  
The coefficients are
\be
\begin{aligned}
&A_{o_1}^{\epsilon_1, \epsilon_2}(k_1, k_2, k_3) 
= 
o_1 \epsilon_1 [d_{k_1}]
(-1)^{k_1 + k_2 + k_3} 
\begin{Bmatrix} 
k_1 & j_1 & \frac12\\ j_2 & k_2 & k_3
\end{Bmatrix}_q,\\
&A_{o_6}^{\epsilon_6, \epsilon_1}(k_6, k_1, k_5) 
= 
o_6 \epsilon_6 [d_{k_6}]
 (-1)^{k_1 + k_5 + k_6} 
\begin{Bmatrix} 
k_6 & j_6 & \frac12\\ j_1 & k_1 & k_5
\end{Bmatrix}_q,\\
&B_{o_2}^{\epsilon_2, \epsilon_6}(j_2, j_6, k_4) 
=
o_2 \epsilon_2 [d_{j_2}]
 (-1)^{j_2 + k_4 + j_6} 
\begin{Bmatrix} 
k_6 & j_6 & \frac12\\ j_2 & k_2 & k_4
\end{Bmatrix}_q\,.
\end{aligned}
\ee
We thus have the recursion
\begin{multline}
\sum_{\epsilon_1=\pm}
o_1o_2o_6 [d_{k_1}]
(-1)^{2k_1 + k_2 +k_3 + k_5 + k_6 + \frac{1-\epsilon_1}{2} + \frac{\epsilon_2}{2} + \frac{\epsilon_6}{2}} 
\begin{Bmatrix} 
k_1 & j_1 & \frac12\\ j_2 & k_2 & k_3
\end{Bmatrix}_q 
\begin{Bmatrix} 
k_6 & j_6 & \frac12\\ j_1 & k_1 & k_5
\end{Bmatrix}_q 
\psi_i(j_1, j_2, k_3, k_4, k_5, j_6, \dotsc) \\
+  (-1)^{k_2 + k_4 + k_6} 
\begin{Bmatrix} 
k_6 & j_6 & \frac12\\ j_2 & k_2 & k_4
\end{Bmatrix}_q 
\psi_i(k_1, k_2, k_3, k_4, k_5, k_6, \dotsc) = 0,
\end{multline}
and similarly for the pairs $(e_1, e_2)$, $(e_6, e_1)$. A similar result for the flat case was found in \cite{Bonzom:2011nv}, where $q$ is set to 1. Those recursions determine the dependence of $\psi_i$ on $j_1, j_2, j_6$ up to a single initial condition. As the recursion involves three terms, it may seem like several initial conditions are required. However, at $k_1=0$, only two terms are left in the recursion, as shown in Equation \eqref{Triangulark_0=0}. This means that from the initial condition $\psi_i(0, k_3,k_3, k_4, k_5, k_5)$, one gets $\psi_i(\frac12,  k_3 -\epsilon_2/2, k_3, k_4, k_5, k_5 - \epsilon_6/2)$. Then this determines $\psi_i$ for arbitrary $k_1, k_2, k_6$. The result is known to be
\be
\psi_i(k_1, k_2, k_3, k_4, k_5, k_6, \dotsc) = 
\begin{Bmatrix} 
k_1 & k_2 & k_3\\ k_4 & k_5 & k_6
\end{Bmatrix}_q\,
\phi(k_3, k_4, k_5, \dotsc),
\label{eq:TV_amplitude}
\ee
where $\phi(k_3, k_4, k_5, \dotsc)$ is independent of $k_1, k_2, k_6$. To determine $\phi$, we set $k_1 = 0$,
\be
\phi(k_3, k_4, k_5, \dotsc) = 
\begin{Bmatrix} 
0 &k_3 & k_3\\ k_4 & k_5 & k_5
\end{Bmatrix}_q^{-1} 
\psi_i(0, k_3, k_3, k_4, k_5, k_5, \dotsc) 
= (-1)^{k_3 + k_4 + k_5} \sqrt{[d_{k_3}] [d_{k_5}]} \psi_i(0,k_3, k_3, k_4, k_5, k_5, \dotsc).
\ee
We conclude with Theorem \ref{thm:linkRemoval}. 
\qed

The relation between the physical states before and after the $3-1$ Pachner move provides a way to relate the $q$-deformed LQG to the Turaev-Viro model with $q$ real. Consider the graphs on two adjacent time-slices in a spin foam different by a $3-1$ move. This part of the spin foam gives a Turaev-Viro vertex amplitude which is simply a $q$-$6j$ symbol. We have reproduced this vertex amplitude in \eqref{eq:TV_amplitude} by relating the coefficients of the physical states before and after the Pachner move. This is also consistent with the method to relate LQG to the spin foam model by considering the physical scalar product of states introduced in \cite{Noui:2004iy}.

In this section, we have proved that the physical states for graphs related by a Pachner move, either $2-2$ move or $3-1$ move, are equivalent hence the physical states are topological states. The equivalence is shown by the exact relation between the coefficients of the spin network basis for physical states before and after the Pachner move as shown in \eqref{eq:2to2} and \eqref{eq:3to1}. This also justifies the validity of the Hamiltonian expression \eqref{eq:Hamiltonian} from the direct quantization of the scalar products of deformed spinors in the classical Hamiltonian \eqref{eq:ClassicalHamiltonian}.

%%%%%%%%%%%%%%%%%%
\section*{Conclusion}

In this paper, we have given a realization of the interplay between the cosmological constant, curved geometries and quantum group structure in the 3D loop quantum gravity framework in Euclidean signature with a negative cosmological constant, which we call the $q$-deformed LQG model. In particular, the deformed constraints at the classical level represent discrete hyperbolic geometries, as shown in \cite{Bonzom:2014wva}. Upon the standard quantization procedure, these deformed constraints become quantum constraints with a quantum group structure. 

We have focused on the Hamiltonian constraints, obtained from the flatness constraints. We have written them with the deformed spinors and performed the quantization following the companion paper \cite{Bonzom:2022bpv}. The result is a generalization of the quantum Hamiltonian constraints derived in \cite{Bonzom:2011nv} for flat space. By studying the way the solutions to the quantum constraints change under Pachner moves, we provide a generalization of the Noui-Perez transition amplitudes \cite{Noui:2004iy} to $q\neq 1$ real: the transition amplitudes are the coefficients relating the physical states in the spin network basis under Pachner moves. Here, they clearly lead to a Turaev-Viro model for $q$ real. It is a topological model (with the same finiteness issues as the $q=1$ version, the Ponzano-Regge model).

Our method is radically different from \cite{Noui:2004iy}, however, and maybe more in the spirit of LQG. On its way to linking $q$-deformed LQG to spin foams, our method derives the Wheeler-DeWitt equations as difference equations on the spin network coefficients of the states, see Equation \eqref{QuantumConstraint}. In the flat case, the Hamiltonian constraint can be interpreted as displacements of the vertices of the triangulation \cite{BonzomDittrich}. Our difference equations \eqref{QuantumConstraint} are quantum implementations of those symmetries.

Although our constraints are in fact derived from the flatness constraints, we believe that this approach is promising to study both how to incorporate the cosmological constant in 4D and how to write interesting dynamics for curved 4D geometries. A first step in the continuous theory has been initiated in \cite{Girelli:2021pol}.

%%%%%%%%%%%%%%%%%
\section*{Acknowledgements}

The authors would like to thank Florian Girelli and Etera Livine for their early participation in this work. This research was supported in part by Perimeter Institute for Theoretical Physics. Research at Perimeter Institute is supported by the Government of Canada through the Department of Innovation, Science and Economic Development Canada and by the Province of Ontario through the Ministry of Research, Innovation and Science. QP is supported by an NSERC Discovery grant awarded to MD. VB is partially supported by the ANR-20-CE48-0018 "3DMaps" grant. The University of Waterloo and the Perimeter Institute for Theoretical Physics are located in the traditional territory of the Neutral, Anishnawbe and Haudenosaunee peoples. We thank them for allowing us to conduct this research on their land.

\bibliographystyle{bib-style} 
\bibliography{QH}

\end{document}